\documentclass[a4paper,12pt]{article}
\usepackage{indentfirst}
\usepackage{amsfonts,amssymb,bm,a4wide,subfigure,cite,makeidx,multicol}
\usepackage[usenames]{color}
\usepackage{fullpage}
\usepackage[centertags]{amsmath}
\allowdisplaybreaks[4]
\usepackage{mathrsfs}
\usepackage[dvips]{graphicx}
\usepackage{url}
\usepackage[dvips,hyperindex]{hyperref}
\graphicspath{{./figures/}}

\newcommand{\boss}[2]{\ensuremath{\rlap{\kern-2.5pt\ensuremath{\overset{\scriptscriptstyle(-)}{\phantom{#1}}}}{\ensuremath{{#1}_{#2}}}}}

\begin{document}

\title{NEUTRINO ELECTROMAGNETIC PROPERTIES}

\author{Carlo Giunti $^{1)}$\footnote{e-mail: giunti@to.infn.it,
$^{**}$e-mail: studenik@srd.sinp.msu.ru} \ , Alexander Studenikin
$^{2)**}$
   \\
   $^{1)}$ \small{\it INFN Section of Turin,
   University of Turin, Italy}
   \\
   $^{2)}$ \small {\it Department of Theoretical Physics,}
      \small {\it Moscow State University, Russia}}

\date{}
\maketitle

\sloppy



The main goal of the paper is to give a short review on neutrino
electromagnetic properties. In the introductory part of the paper a
summary on what we really know about neutrinos is given: we discuss
the basics  of neutrino mass and mixing as well as the phenomenology
of neutrino oscillations. This is important for the following
discussion on neutrino electromagnetic properties that starts with a
derivation of the neutrino electromagnetic vertex function in the
most general form, that follows from the requirement of Lorentz
invariance, for both the Dirac and Majorana cases. Then, the problem
of the neutrino form factors definition and calculation within gauge
models is considered. In particular, we discuss the neutrino electric
charge form factor and charge radius, dipole magnetic and electric
and anapole form factors. Available experimental constraints on
neutrino electromagnetic properties are also discussed, and the
recently obtained experimental limits on neutrino magnetic moments
are reviewed. The most important neutrino electromagnetic processes
involving a direct neutrino coupling with photons (such as neutrino
radiative decay, neutrino Cherenkov radiation, spin light of neutrino
and plasmon decay into neutrino-antineutrino pair in media) and
neutrino resonant spin-flavor precession in a magnetic field are
discussed at the end of the paper.


\section{Introduction}
\label{sec1}
The neutrino is a very fascinating particle which has remained under the focus
of intensive investigations, both theoretical an experimental, for a
couple of decades. These studies have given evidence of an ultimate
relation between the knowledge of neutrino properties and
the understanding of the fundamentals of particle physics. The birth of the
neutrino was due to
an attempt, by W. Pauli in 1930, to explain the continuous spectrum of beta-particles
through ``a way out for saving the law of
conservation of energy'' \cite{Pauli57}. This new particle, called at
first the ``neutron'' and then renamed the ``neutrino'', was an
essential part of the first model of weak interactions
(E. Fermi, 1934). Further important milestones of particle
physics, such as parity nonconservation (T.D. Lee, C.N. Yang and
L. Landau, 1956) and the $V-A$ model of local weak interactions
(E. Sudarshan, R. Marshak, 1956; R. Feynman, M. Gell-Mann, 1958), as well
as the structure of the Glashow-Weinberg-Salam {\it standard model},
were based on the clarification of the specific properties of the
neutrino.  It has happened more than once that a novel discovery in
neutrino physics stimulates far-reaching consequences in the theory
of particle interactions.

The neutrino plays a crucial role in particle physics because it is a
``tiny'' particle. Indeed, the scale of neutrino mass is much lower
than that of the charged fermions ($m_{\nu_{f}}<<m_f, \ \ f=e,\mu,
\tau$). The weak and electromagnetic interactions of neutrinos with
other particles are really very weak. That is a reason for the
neutrino to fall under the focus of researchers during the latest
stages of a particular particle physics evolution paradigm when all
of the ``principal'' phenomena have been already observed and
theoretically described.

Neutrino electromagnetic properties, that is the main subject of
this paper, are of particular importance because they provide a kind
of bridge to ``new physics'' beyond the standard model. In spite of
reasonable efforts in studies of neutrino electromagnetic
properties, up to now there is no experimental
confirmation in favour of nonvanishing neutrino electromagnetic characteristics.
The available experimental data in the field do
not rule out the possibility that neutrinos have ``zero''
electromagnetic properties. However, in the course of the recent
development of knowledge on neutrino mixing and oscillations,
supported by the discovery of flavor conversions of neutrinos from
different sources, non-trivial neutrino electromagnetic properties
seem to be very plausible.

The structure of the paper is as follows. In the first part of the
paper we summarize {\it what we really know about neutrinos}: the
basics of neutrinos mass and mixing are discussed, as well as the
phenomenology of neutrino oscillations. This introductory part is
important for understanding the second part of the paper that is
devoted to electromagnetic (in the sense mentioned above still
``unknown'') properties of neutrinos. We start discussing neutrino
electromagnetic properties deriving the neutrino electromagnetic
vertex function in the most general form for both the Dirac and
Majorana cases. Then, we consider the neutrino electric charge form
factor and charge radius, magnetic, electric and anapole form
factors. We discuss the relevant theoretical items as well as
available experimental constraints. In particular, the neutrino
magnetic and electric moments, in both theoretical and experimental
aspects, are discussed in detail. In Section~\ref{sec4}, the most
important neutrino electromagnetic processes involving the direct
neutrino couplings with photons (such as neutrino radiative decay,
neutrino Cherenkov radiation, spin light of neutrino and plasmon
decay into neutrino-antineutrino pair in media) and neutrino resonant
spin-flavor precession in a magnetic field are discussed.

\section{What we know about neutrino}
\label{sec2}
\begin{figure}[t!]
\begin{center}
\includegraphics*[bb=74 679 645 778, width=0.99\textwidth]{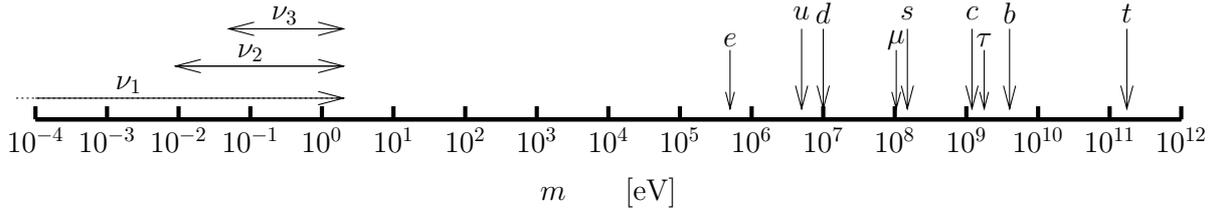}
\caption{ \label{002} Order of magnitude of the masses of leptons and quarks. }
\end{center}
\end{figure}

In the Standard Model of electroweak interaction, forged in the 60's by
Glashow, Weinberg and Salam \cite{Glashow:1961tr,Weinberg:1967tq,Salam:1968rm},
neutrinos are massless by construction (see Ref.~\cite{Giunti-Kim-2007}). This
requirement was motivated by the low experimental upper limit on the neutrino
mass (see Fig.~\ref{002}) and by the theoretical description of neutrinos
through massless left-handed Weyl spinors in the two-component theory of
Landau, Lee and Yang, and Salam \cite{Landau:1957tp,Lee:1957qr,Salam:1957st},
which prompted the $V-A$ theory of charged-current weak interactions of Feynman
and Gell-Mann, Sudarshan and Marshak, and Sakurai
\cite{Feynman:1958ty,Sudarshan:1958vf,Sakurai-NuovoCimento7-64-1958}.

The massless of neutrinos in the Standard Model is due to the absence of
right-handed neutrinos, without which it is not possible to have Dirac mass
terms, and to the absence of Higgs triplets, without which it is not possible
to have Majorana mass terms. In the following we will consider the extension of
the Standard Model with the introduction of three right-handed neutrinos. We
will see that this seemingly innocent addition has the very powerful effect of
allowing not only Dirac mass terms, but also Majorana mass terms for the
right-handed neutrinos which induce Majorana masses for the observable
neutrinos.

Table~\ref{003} shows the values of the weak isospin, hypercharge, and electric
charge of the lepton and Higgs doublets and singlets in the extended Standard
Model under consideration. For simplicity, we work in the flavor base in which
the mass matrix of the charged leptons is diagonal. Hence, $ e $, $ \mu $, $
\tau $ are the physical charged leptons with definite masses. The three singlet
neutrinos are often called \emph{sterile}, since they do not take part in weak
interactions, in contrast with the standard \emph{active} neutrinos $\nu_{e}$,
$\nu_{\mu}$, $\nu_{\tau}$.

\begin{table}[t!]
\caption{ \label{003} Eigenvalues of the weak isospin $I$, of its
third component $I_{3}$, of the hypercharge $Y$, and of the charge
$Q=I_{3}+Y/2$ of the lepton and Higgs doublets and singlets in the
extension of the Standard Model with the introduction of right-handed
neutrinos. }
\begin{center}
\renewcommand{\arraystretch}{1.45}
\setlength{\tabcolsep}{0.3cm}
\begin{tabular}{cccccc}
\hline ($ \alpha = e, \mu, \tau $ and $ s = s_{1}, s_{2}, s_{3} $) & & $I$ &
$I_{3}$ & $Y$ & $Q$
\\
\cline{3-6} left-handed lepton doublets & $ L_{\alpha L} \equiv
\begin{pmatrix}
\nu_{\alpha L} \\ \alpha_{L}
\end{pmatrix}
$ & $1/2$ & $
\begin{matrix}
1/2 \\ -1/2
\end{matrix}
$ & $-1$ & $
\begin{matrix}
0 \\ -1
\end{matrix}
$
\\
right-handed charged-lepton singlets & $\alpha_{R}$ & $0$ & $0$ & $-2$ & $-1$
\\
right-handed neutrino singlets & $\nu_{sR}$ & $0$ & $0$ & $0$ & $0$
\\
Higgs doublet & $ \Phi \equiv
\begin{pmatrix}
\phi^{+} \\ \phi^{0}
\end{pmatrix}
$ & $1/2$ & $
\begin{matrix}
1/2 \\ -1/2
\end{matrix}
$ & $+1$ & $
\begin{matrix}
1 \\ 0
\end{matrix}
$
\\
\hline
\end{tabular}
\end{center}

\end{table}

\subsection{Dirac mass term}
\label{004} \nopagebreak

The fields in Tab.~\ref{003} allow us to construct the Yukawa Lagrangian term
\begin{equation}
\mathscr{L}_{\text{Y}} = - \sum_{\alpha=e,\mu,\tau} \sum_{k=1}^{3} Y_{\alpha k}
\, \overline{L_{{\alpha}L}} \, \widetilde{\Phi} \, \nu_{s_{k}R} + \text{H.c.}
\,, \label{005}
\end{equation}
where $Y$ is a matrix of Yukawa couplings and $ \widetilde{\Phi} \equiv i
\sigma_{2} \Phi^{*} $. In the Standard Model, a nonzero vacuum expectation
value of the Higgs doublet,
\begin{equation}
\langle \Phi \rangle = \frac{ 1 }{ \sqrt{2} }
\begin{pmatrix}
0 \\ v
\end{pmatrix}
\,, \label{006}
\end{equation}
induces the spontaneous symmetry breaking of the Standard Model symmetries
$\text{SU(2)}_{L} \times \text{U(1)}_{Y} \to \text{U(1)}_{Q} $. In the unitary
gauge, the Higgs doublet is given by
\begin{equation}
\Phi(x) = \frac{1}{\sqrt{2}} \,
\begin{pmatrix}
0
\\
v + H(x)
\end{pmatrix}
\,, \label{007}
\end{equation}
where $H(x)$ is the physical Higgs field. From the Yukawa Lagrangian term in
Eq.~(\ref{005}) we obtain the neutrino Dirac mass term
\begin{equation}
\mathscr{L}_{\text{D}} = - \frac{v}{\sqrt{2}} \sum_{\alpha=e,\mu,\tau}
\sum_{k=1}^{3} Y_{\alpha k} \, \overline{\nu_{{\alpha}L}} \, \nu_{s_{k}R} +
\text{H.c.} \,. \label{008}
\end{equation}
Since the matrix $Y$ is, in general, a complex $3\times3$ matrix, the flavor
neutrino fields $\nu_{e}$, $\nu_{\mu}$, $\nu_{\tau}$ do not have a definite
mass. The massive neutrino fields are obtained through the diagonalization of
$\mathscr{L}_{\text{D}}$. This is achieved through the transformations
\begin{equation}
\nu_{{\alpha}L} = \sum_{k=1}^{3} U_{\alpha k} \, \nu_{kL} \,, \qquad
\nu_{s_{j}R} = \sum_{k=1}^{3} V_{jk} \, \nu_{kR} \,, \label{009}
\end{equation}
with unitary matrices $U$ and $V$ which perform the biunitary diagonalization
\begin{equation}
\frac{v}{\sqrt{2}} \left( U^{\dagger} \, Y \, V \right)_{kj} = m_k \,
\delta_{kj} \,, \label{010}
\end{equation}
with real and positive masses $m_k$. The resulting diagonal Dirac mass term is
\begin{equation}
\mathscr{L}_{\text{D}} = - \sum_{k=1}^{3} m_{k} \, \overline{\nu_{kL}} \,
\nu_{kR} + \text{H.c.} = - \sum_{k=1}^{3} m_{k} \, \overline{\nu_{k}} \,
\nu_{k} \,, \label{011}
\end{equation}
with the Dirac fields of massive neutrinos $ \nu_{k} = \nu_{kL} + \nu_{kR} $.

\subsection{Dirac--Majorana mass term}
\label{012} \nopagebreak

In the above derivation of Dirac neutrino masses we have implicitly assumed
that the total lepton number is conserved. If this assumption is lifted,
neutrino masses receive an important contribution from the Majorana mass term
of the right-handed singlet neutrinos,
\begin{equation}
\mathscr{L}_{R} = \frac{1}{2} \sum_{k,j=1}^{3} \nu^{T}_{s_{k}R} \,
\mathcal{C}^{\dagger} \, M^{R}_{\alpha\beta} \, \nu_{s_{j}R} + \text{H.c.} \,,
\label{013}
\end{equation}
where $\mathcal{C}$ is the charge-conjugation matrix. The mass matrix $M^{R}$
is complex and symmetric.

The Majorana mass term in Eq.~(\ref{013}) is allowed by the symmetries of the
Standard Model, since right-handed neutrino fields are invariant. On the other
hand, an analogous Majorana mass term of the left-handed neutrinos,
\begin{equation}
\mathscr{L}_{L} = \frac{1}{2} \sum_{\alpha,\beta=e,\mu,\tau}
\nu^{T}_{{\alpha}L} \, \mathcal{C}^{\dagger} \, M^{L}_{\alpha\beta} \,
\nu_{{\beta}L} + \text{H.c.} \,, \label{014}
\end{equation}
is forbidden, since it has $I_{3}=1$ and $Y=-2$. There is no Higgs triplet in
the Standard Model to compensate these quantum numbers.

In the extension of the Standard Model with the introduction of right-handed
neutrinos, the neutrino masses and mixing are given by the Dirac--Majorana mass
term
\begin{equation}
\mathscr{L}_{\text{D+M}} = \mathscr{L}_{\text{D}} + \mathscr{L}_{R} \,.
\label{015}
\end{equation}
The neutrino fields with definite masses are obtained through the
diagonalization of $\mathscr{L}_{\text{D+M}}$. It is convenient to define the
vector $N_{L}$ of 6 left-handed fields
\begin{equation}
N^{T}_{L} \equiv \left( \nu_{eL}, \nu_{\mu L}, \nu_{\tau L},
(\nu_{s_{1}R})^{C}, (\nu_{s_{2}R})^{C}, (\nu_{s_{3}R})^{C} \right) \,,
\label{016}
\end{equation}
with the charge-conjugated sterile neutrino fields $ (\nu_{sR})^{C} =
\mathcal{C} \overline{\nu_{sR}}^{T} $. The Dirac--Majorana mass term in
Eq.~(\ref{015}) can be written in the compact form
\begin{equation}
\mathscr{L}_{\text{D+M}} = \frac{1}{2} \, N^{T}_{L} \, \mathcal{C}^{\dagger} \,
M^{\text{D+M}} \, N_{L} + \text{H.c.} \,, \label{017}
\end{equation}
with the $N \times N$ symmetric mass matrix
\begin{equation}
M^{\text{D+M}} \equiv
\begin{pmatrix}
0 & {M^{\text{D}}}^{T}
\\
M^{\text{D}} & M^{R}
\end{pmatrix}
\,, \label{018}
\end{equation}
where
\begin{equation}
M^{\text{D}} = \frac{v}{\sqrt{2}} \, Y \,. \label{019}
\end{equation}
Notice that the Dirac--Majorana mass term in Eq.~(\ref{017}) has the structure
of a Majorana mass term. Therefore, it will not be a surprise to find in the
following that the 6 massive neutrinos obtained from the diagonalization of $
\mathscr{L}_{\text{D+M}} $ are Majorana particles.

Equation~(\ref{017}) is diagonalized through the unitary transformation
\begin{equation}
N_{L} = V \, n_{L} \,, \qquad \text{with} \qquad n_{L}^{T} = \left( \nu_{1L},
\ldots, \nu_{6L} \right) \,. \label{020}
\end{equation}
The unitary matrix $V$ is chosen in order to diagonalize the symmetric mass
matrix $M^{\text{D+M}}$:
\begin{equation}
V^{T} \, M^{\text{D+M}} \, V = M \,, \qquad \text{where} \qquad M_{kj} = m_{k}
\, \delta_{kj} \qquad (k,j=1,\ldots,6) \,, \label{021}
\end{equation}
with real and positive masses $m_{k}$. In this way, the Dirac--Majorana mass
term in Eq.~(\ref{017}) can be written in terms of the massive fields as
\begin{equation}
\mathscr{L}_{\text{D+M}} = \frac{1}{2} \, n_{L}^{T} \, \mathcal{C}^{\dagger} \,
M \, n_{L} + \text{H.c.} = \frac{1}{2} \sum_{k=1}^{6} m_{k} \, \nu_{kL}^{T} \,
\mathcal{C}^{\dagger} \, \nu_{kL} + \text{H.c.} = \frac{1}{2} \sum_{k=1}^{6}
m_{k} \, \nu_{k}^{T} \, \mathcal{C}^{\dagger} \, \nu_{k} \,, \label{022}
\end{equation}
where $ \nu_{k} = \nu_{kL} + \nu_{kL}^{C} $ are Majorana fields which satisfy
the constraint $ \nu_{k}^{C} = \nu_{k} $. Hence, a general result of the
diagonalization of a Dirac--Majorana mass term is that massive neutrinos are
Majorana particles.

Note that in the limit $M^{R}=0$ we recover the Dirac case, since there are
three pairs of degenerate mass eigenvalues. Each pair of massive Majorana
fields with the same mass corresponds to a Dirac field (see
Ref.~\cite{Giunti-Kim-2007}). For example, if the massive Majorana fields
$\nu_{1}$ and $\nu_{4}$ have the same mass $m_1$, they correspond to the Dirac
field $ ( \nu_{1} + i \nu_{4} ) / \sqrt{2} $.

\subsection{Weak interactions}
\label{023} \nopagebreak

The mixing of neutrinos in Eq.~(\ref{020}) is observable through its effect in
weak interactions. Let us first consider the leptonic weak charged current
\begin{equation}
j_{\text{CC}}^{\rho} = 2 \sum_{\alpha=e,\mu,\tau} \overline{\nu_{{\alpha}L}} \,
\gamma^{\rho} \, \alpha_{L} = 2 \sum_{\alpha=e,\mu,\tau} \sum_{k=1}^{6}
\overline{\nu_{kL}} \, V_{\alpha k}^{*} \, \gamma^{\rho} \, \alpha_{L} \,.
\label{024}
\end{equation}
Hence, only the rectangular submatrix of $V$ composed by the first three rows
is relevant for charged-current weak interactions. Each of the 6 massive
neutrinos partake in charged-current weak interactions if the elements the
first three rows and corresponding column of the mixing matrix are not
negligibly small. In Section~\ref{026} we will see that in the celebrated
see-saw mechanism the effective number of light Majorana massive neutrinos
which take part in weak interactions is reduced to three.

The neutrino weak neutral current is given by
\begin{equation}
j_{\text{NC}}^{\rho} = \sum_{\alpha=e,\mu,\tau} \overline{\nu_{{\alpha}L}} \,
\gamma^{\rho} \, \nu_{{\alpha}L} = \sum_{k,j=1}^{6} \overline{\nu_{kL}} \left(
\sum_{\alpha=e,\mu,\tau} V_{\alpha k}^{*} V_{\alpha j} \right) \gamma^{\rho} \,
\nu_{jL} \,. \label{025}
\end{equation}
Unless the mixing of 6 neutrinos reduces to an effective mixing of 3 neutrinos,
as in the see-saw mechanism discussed in Section~\ref{026}, $
\sum_{\alpha=e,\mu,\tau} V_{\alpha k}^{*} V_{\alpha j} \neq \delta_{kj} $ and
there can be neutral-current transitions among different massive neutrinos (no
GIM mechanism \cite{Glashow:1970gm}).

\subsection{See-saw mechanism}
\label{026} \nopagebreak

The order of magnitude of the elements of the Dirac mass matrix $M^{\text{D}}$
in Eq.~(\ref{019}) is expected to be smaller than $ v \sim 10^{2} \, \text{GeV}
$, since the Yukawa couplings are expected not to be unnaturally large. In
general, since a Dirac mass term is forbidden by the symmetries of the Standard
Model, it can arise only as a consequence of symmetry breaking and Dirac masses
are proportional to the symmetry-breaking scale. This fact is often summarized
by saying that Dirac masses are \emph{protected} by the symmetries of the
Standard Model. On the other hand, since the Majorana mass term in
Eq.~(\ref{013}) is a Standard Model singlet, the elements of the Majorana mass
matrix $M_{R}$ are not protected by the Standard Model symmetries. It is
plausible that the Majorana mass term $\mathscr{L}_{R}$ is generated by new
physics beyond the Standard Model and the right-handed chiral neutrino fields
$\nu_{sR}$ belong to nontrivial multiplets of the symmetries of the high-energy
theory. In this case, the elements of the mass matrix $M_{R}$ are protected by
the symmetries of the high-energy theory and their order of magnitude
corresponds to the breaking scale of these symmetries, which may be as large as
the grand unification scale, of the order of $10^{14}$--$10^{16} \,
\text{GeV}$. The mass matrix can be diagonalized by blocks, up to corrections
of the order $ (M^{R})^{-1} M^{\text{D}} $:
\begin{equation}
W^{T} \, M^{\text{D+M}} \, W \simeq
\begin{pmatrix}
M_{\text{light}} & 0
\\
0 & M_{\text{heavy}}
\end{pmatrix}
\,, \label{027}
\end{equation}
with
\begin{equation}
W \simeq
\begin{pmatrix}
1 - \frac{1}{2} \, {M^{\text{D}}}^{\dagger} ( M^{R} {M^{R}}^{\dagger} )^{-1}
M^{\text{D}} & [ (M^{R})^{-1} M^{\text{D}} ]^{\dagger}
\\
- (M^{R})^{-1} M^{\text{D}} & 1 - \frac{1}{2} \, ( M^{R} )^{-1} M^{\text{D}}
{M^{\text{D}}}^{\dagger} ( {M^{R}}^{\dagger} )^{-1}
\end{pmatrix}
\,. \label{028}
\end{equation}
The light $3\times3$ mass matrix $M_{\text{light}}$ and the heavy $3\times3$
mass matrix $M_{\text{heavy}}$ are given by
\begin{equation}
M_{\text{light}} \simeq - {M^{\text{D}}}^{T} \, ( M^{R} )^{-1} \, M^{\text{D}}
\,, \qquad M_{\text{heavy}} \simeq M^{R} \,. \label{029}
\end{equation}
The heavy masses are given by the eigenvalues of $M^{R}$, whereas the light
masses are given by the eigenvalues of $M_{\text{light}}$, whose elements are
suppressed with respect to the elements of the Dirac mass matrix $M^{\text{D}}$
by the very small matrix factor $ {M^{\text{D}}}^{T} ( M^{R} )^{-1} $. This is
the celebrated \emph{see-saw mechanism}, which explains naturally the smallness
of light neutrino masses (see Fig.~\ref{002}). Notice, however, that the values
of the light neutrino masses and their relative sizes can vary over wide
ranges, depending on the specific values of the elements of $M^{\text{D}}$ and
$M^{R}$.

Since the off-diagonal block elements of $W$ are very small, the three flavor
neutrinos are mainly composed by the three light neutrinos. Therefore, the
see-saw mechanism implies the effective low-energy mixing of three Majorana
neutrinos with an approximately unitary $3\times3$ mixing matrix $U$ composed
by the first 3 rows and the first 3 columns of $V$.

\subsection{Three-neutrino mixing}
\label{030} \nopagebreak

In the case of three-neutrino mixing, the three left-handed flavor neutrino
fields $\nu_{e L}$, $\nu_{\mu L}$, $\nu_{\tau L}$ which partake in weak
interactions are unitary linear combinations of three left-handed massive
neutrino fields $\nu_{1L}$, $\nu_{2L}$, $\nu_{3L}$:
\begin{equation}
\nu_{\alpha L} = \sum_{k=1}^{3} U_{{\alpha}k} \, \nu_{kL} \qquad
(\alpha=e,\mu,\tau) \,, \label{031}
\end{equation}
where $U$ is a $3\times3$ unitary mixing matrix. Motivated by the see-saw
mechanism, we consider Majorana massive neutrinos. The deviation from the
unitarity of $U$ in the see-saw mechanism is negligible. In this approximation,
the expression in Eq.~(\ref{024}) for the leptonic weak charged current reduces
to
\begin{equation}
j_{\text{CC}}^{\rho} = 2 \sum_{\alpha=e,\mu,\tau} \overline{\nu_{{\alpha}L}} \,
\gamma^{\rho} \, \alpha_{L} = 2 \sum_{\alpha=e,\mu,\tau} \sum_{k=1}^{3}
\overline{\nu_{kL}} \, U_{\alpha k}^{*} \, \gamma^{\rho} \, \alpha_{L} \,,
\label{032}
\end{equation}
and the expression in Eq.~(\ref{025}) for the leptonic weak neutral current
reduces to
\begin{equation}
j_{\text{NC}}^{\rho} = \sum_{\alpha=e,\mu,\tau} \overline{\nu_{{\alpha}L}} \,
\gamma^{\rho} \, \nu_{{\alpha}L} = \sum_{k,j=1}^{3} \overline{\nu_{kL}} \,
\gamma^{\rho} \, \nu_{jL} \,, \label{033}
\end{equation}
where the unitarity of $U$ implies the absence of neutral-current transitions
among different massive neutrinos (GIM mechanism \cite{Glashow:1970gm}).

Notice that in the case of three-neutrino mixing the mixing matrix $U$ enters
only in the leptonic weak charged current $j_{\text{CC}}^{\rho}$. Hence, it is
observable only through weak charged-current interactions.

The unitary matrix $U$ can be parameterized in terms of 3 mixing angles and 6
phases. However, 3 phases are unphysical, because they can be eliminated by
rephasing the three charged lepton fields in $j_{\text{CC}}^{\rho}$. In the
case of Majorana massive neutrinos, no additional phase can be eliminated,
because a Majorana mass term $\nu_{k}^{T} \, \mathcal{C}^{\dagger} \, \nu_{k}$
is not invariant under rephasing of $\nu_{k}$. On the other hand, in the case
of Dirac massive neutrinos, two additional phases can be eliminated by
rephasing the massive neutrino fields. Hence, the mixing matrix has 3 physical
phases in the case of Majorana massive neutrinos or 1 physical phase in the
case of Dirac massive neutrinos. In general, in the case of Majorana massive
neutrinos $U$ can be written as
\begin{equation}
U = U^{\text{D}} \, D^{\text{M}} \,, \label{034}
\end{equation}
where $U^{\text{D}}$ is a Dirac unitary mixing matrix which can be
parameterized in terms of three mixing angles and one physical phase, called
\emph{Dirac phase}, and $D^{\text{M}}$ is a diagonal unitary matrix with two
physical phases, usually called \emph{Majorana phases}.

The standard parameterization of $U^{\text{D}}$ is
\begin{equation}
U^{\text{D}} =
\begin{pmatrix}
c_{12} c_{13} & s_{12} c_{13} & s_{13} e^{-i\delta_{13}}
\\
- s_{12} c_{23} - c_{12} s_{23} s_{13} e^{i\delta_{13}} & c_{12} c_{23} -
s_{12} s_{23} s_{13} e^{i\delta_{13}} & s_{23} c_{13}
\\
s_{12} s_{23} - c_{12} c_{23} s_{13} e^{i\delta_{13}} & - c_{12} s_{23} -
s_{12} c_{23} s_{13} e^{i\delta_{13}} & c_{23} c_{13}
\end{pmatrix}
\,, \label{035}
\end{equation}
where $ c_{ab} \equiv \cos\vartheta_{ab} $ and $ s_{ab} \equiv
\sin\vartheta_{ab} $. $\vartheta_{12}$, $\vartheta_{13}$, $\vartheta_{23}$ are
the three mixing angles ($ 0 \leq \vartheta_{ab} \leq \pi/2 $) and
$\delta_{13}$ is the Dirac phase ($ 0 \leq \delta_{13} < 2 \pi $).

The diagonal unitary matrix $D^{\text{M}}$ can be written as
\begin{equation}
D^{\text{M}} = \text{diag}\!\left( e^{i\lambda_{1}} \, , \, e^{i\lambda_{2}} \,
, \, e^{i\lambda_{3}} \right) \,, \quad \text{with} \quad \lambda_{1}=0 \,.
\label{036}
\end{equation}
The phases $\lambda_{2}$ and $\lambda_{3}$ are the two physical Majorana
CP-violating phases. Since all measurable quantities depend only on the
differences of the three phases $\lambda_{1}$, $\lambda_{2}$, $\lambda_{3}$,
the choice $\lambda_{1}=0$ is a matter of convention and other choices are
equivalent from the physical point of view. In fact, rephasing all the charged
lepton fields in $j_{\text{CC}}^{\rho}$ by $e^{i\varphi}$, we have
$e^{i\lambda_{k}} \to e^{i(\lambda_{k}-\varphi)}$, whereas
$e^{i(\lambda_{k}-\lambda_{j})}$ remains constant.

All the phases in the mixing matrix violate the CP symmetry. Since the Majorana
phases are observable only in processes which are allowed only in the case of
Majorana neutrinos, as neutrinoless double-$\beta$ decay, in most observable
processes CP violation is generated by the Dirac phase. The size of this CP
violation can be quantified in a parameterization-invariant way by the Jarlskog
invariant
\cite{Jarlskog:1985ht,Jarlskog:1985cw,Greenberg:1985mr,Dunietz:1985uy,Wu:1985ea,KraPetPLB88}
\begin{equation}
J \equiv \text{Im}\!\left( U_{e 2} \, U_{\mu 3} \, U_{e 3}^{*} \, U_{\mu 2}^{*}
\right) = \text{Im}\!\left( U^{\text{D}}_{e 2} \, U^{\text{D}}_{\mu 3} \,
U^{\text{D}*}_{e 3} \, U^{\text{D}*}_{\mu 2} \right) \,. \label{037}
\end{equation}
The second equality is due to the invariance of $J$ under the rephasing $
U_{\alpha k} \to e^{-i\varphi_{\alpha}} U_{\alpha k} e^{i\psi_{k}} $, with
arbitrary phases $\varphi_{\alpha}$ and $\psi_{k}$, which implies that the
Majorana phases do not contribute. Using the unitarity of the mixing matrix, it
can be shown that
\begin{equation}
J_{\alpha\beta kj} \equiv \text{Im}\!\left( U_{\alpha k} \, U_{\beta j} \,
U_{\alpha j}^{*} \, U_{\beta k}^{*} \right) = \pm J \,. \label{038}
\end{equation}
Therefore, in the case of three-neutrino mixing $|J|$ quantifies Dirac CP
violation independently from the parameterization of the mixing matrix. In the
parameterization in Eq.~(\ref{035}), we have
\begin{equation}
J = c_{12} s_{12} c_{23} s_{23} c_{13}^{2} s_{13} \sin\delta_{13} = \frac{1}{8}
\, \sin2\vartheta_{12} \, \sin2\vartheta_{23} \, \cos\vartheta_{13} \,
\sin2\vartheta_{13} \, \sin\delta_{13} \,. \label{039}
\end{equation}

\subsection{Neutrino oscillations}
\label{040} \nopagebreak

Neutrino oscillations was proposed by B. Pontecorvo in the late 1950s in
analogy with $K^{0}$--$\bar{K}^{0}$ oscillations
\cite{Pontecorvo:1957cp,Pontecorvo:1958qd} The oscillations are generated by
the interference of the phases of different massive neutrinos, which are
produced and detected coherently because of their very small mass differences.

Let us consider a flavor neutrino state
\begin{equation}
| \nu_{\alpha} \rangle = \sum_{k} U_{\alpha k}^{*} \, | \nu_{k} \rangle \,,
\label{041}
\end{equation}
which describes a neutrino with flavor $\alpha$ created in a charged-current
weak interaction process from a charged lepton $\alpha^{-}$ or together with a
charged antilepton $\alpha^{+}$ ($ \alpha = e, \mu, \tau $). The presence of
the weight $U_{\alpha k}^{*}$ for $| \nu_{k} \rangle$ in the flavor state $|
\nu_{\alpha} \rangle$ is due to the decomposition in Eq.~(\ref{032}) of the
leptonic charged current $j_{\text{CC}}^{\rho}$ in terms of the massive
neutrino contributions, which contain the creation operators of massive
neutrinos. Additional coefficients due to different effects of neutrino masses
in the interaction process are negligible in neutrino oscillation experiments.

The massive neutrino states $| \nu_{k} \rangle$ are eigenstates of the free
Hamiltonian with energy eigenvalues
\begin{equation}
E_{k} = \sqrt{ |\bm {p}_{k}|^{2} + m_{k}^{2} } \,, \label{042}
\end{equation}
where $\bm {p}_{k}$ is the respective momentum. Since the massive neutrinos
evolve in space-time as plane waves, the space-time evolution of the flavor
state in Eq.~(\ref{041}) is
\begin{equation}
| \nu_{\alpha} (\bm {L},T) \rangle = \sum_{k} U_{\alpha k}^{*} \, e^{ - i E_{k}
T + i \bm {p}_{k} \cdot \bm {L} } \, | \nu_{k} \rangle =
\sum_{\beta=e,\mu,\tau} \left( \sum_{k} U_{\alpha k}^{*} \, e^{ - i E_{k} T + i
\bm {p}_{k} \cdot \bm {L} } \, U_{\beta k} \right) | \nu_{\beta} \rangle \,,
\label{043}
\end{equation}
where we have used the unitarity of the mixing matrix for inverting the
relation in Eq.~(\ref{041}). One can see that the phase differences of
different massive neutrinos generate flavor transitions with probability
\begin{equation}
P_{\nu_{\alpha}\to\nu_{\beta}}(\bm {L},T) = | \langle \nu_{\beta} |
\nu_{\alpha} (\bm {L},T) \rangle |^2 = \left| \sum_{k} U_{\alpha k}^{*} \, e^{
- i E_{k} T + i \bm {p}_{k} \cdot \bm {L} } \, U_{\beta k} \right|^2 \,.
\label{044}
\end{equation}

Since the source-detector distance $ L \equiv |\bm {L}| $ is macroscopic, we
can consider all massive neutrino momenta $\bm {p}_{k}$ aligned along $\bm
{L}$. Moreover, taking into account the smallness of neutrino masses, in
oscillation experiments in which the neutrino propagation time $T$ is not
measured it is possible to approximate $T = L$. With these approximations, the
phases in Eq.~(\ref{044}) reduce to
\begin{equation}
- E_{k} T + p_{k} L = - \left( E_{k} - p_{k} \right) L = - \frac{ E_{k}^{2} -
p_{k}^{2} }{ E_{k} + p_{k} } \, L = - \frac{ m_{k}^{2} }{ E_{k} + p_{k} } \, L
\simeq - \frac{ m_{k}^{2} }{ 2 E } \, L \,, \label{045}
\end{equation}
at lowest order in the neutrino masses. Here, $ p_{k} \equiv |\bf
{p}_{k}| $ and $E$ is the neutrino energy neglecting mass
contributions. Equation~(\ref{045}) shows that the phases of massive
neutrinos relevant for oscillations are independent of the values of
the energies and momenta of different massive neutrinos, because of
the relativistic dispersion relation in Eq.~(\ref{042}). The flavor
transition probabilities are
\begin{align}
P_{\nu_{\alpha}\to\nu_{\beta}}(L,E) = \delta_{\alpha\beta} \null &
\null - 4 \sum_{k>j} \text{Re}\!\left( U_{{\alpha}k}^{*} \,
U_{{\beta}k} \, U_{{\alpha}j} \, U_{{\beta}j}^{*} \right)
\sin^{2}\left( \frac{\Delta{m}^{2}_{kj} L}{4E} \right) - \nonumber
\\
\null & \null - 2 \sum_{k>j} J_{\alpha\beta kj} \, \sin\!\left(
\frac{\Delta{m}^{2}_{kj} L}{2E} \right) \,. \label{046}
\end{align}
The C and T conjugated flavor transition probabilities are given by
\begin{equation}
P_{\bar\nu_{\alpha}\to\bar\nu_{\beta}} = P_{\nu_{\beta}\to\nu_{\alpha}} =
\left. P_{\nu_{\alpha}\to\nu_{\beta}} \right|_{U \to U^{*}} \,. \label{047}
\end{equation}
The survival probabilities ($\alpha=\beta$) are CP-invariant (a consequence of
CPT symmetry),
\begin{equation}
P_{\bar\nu_{\alpha}\to\bar\nu_{\alpha}} = P_{\nu_{\alpha}\to\nu_{\alpha}} \,,
\label{048}
\end{equation}
whereas CP violation is observable in flavor transitions by measuring the
asymmetries
\begin{equation}
A^{\text{CP}}_{\alpha\beta} = P_{\nu_{\alpha}\to\nu_{\beta}} -
P_{\bar\nu_{\alpha}\to\bar\nu_{\beta}} \qquad (\alpha\neq\beta) \,. \label{049}
\end{equation}
CPT symmetry implies that the CP asymmetries are equal to the corresponding T
asymmetries: $ A^{\text{T}}_{\alpha\beta} = - \bar{A}^{\text{T}}_{\alpha\beta}
= A^{\text{CP}}_{\alpha\beta} $, with $ A^{\text{T}}_{\alpha\beta} =
P_{\nu_{\alpha}\to\nu_{\beta}} - P_{\nu_{\beta}\to\nu_{\alpha}} $ and $
\bar{A}^{\text{T}}_{\alpha\beta} = P_{\bar\nu_{\alpha}\to\bar\nu_{\beta}} -
P_{\bar\nu_{\beta}\to\bar\nu_{\alpha}} $.

In the approximation of two-neutrino mixing, in which one of the three massive
neutrino components of two flavor neutrinos is neglected, the mixing matrix
reduces to
\begin{equation}
U =
\begin{pmatrix}
\cos\vartheta & \sin\vartheta
\\
-\sin\vartheta & \cos\vartheta
\end{pmatrix}
\,, \label{050}
\end{equation}
where $\vartheta$ is the mixing angle ($0 \leq \vartheta \leq \pi/2$). In this
approximation, there is only one squared-mass difference $\Delta{m}^2$ and the
transition probability is given by
\begin{equation}
P_{\nu_{\alpha}\to\nu_{\beta}}(L,E) = \sin^{2} 2\vartheta \, \sin^{2}\!\left(
\frac{\Delta{m}^{2} L}{4E} \right) \qquad (\alpha\neq\beta) \,. \label{051}
\end{equation}
In the case $\alpha=\beta$, the \emph{survival probability} is
\begin{equation}
P_{\nu_{\alpha}\to\nu_{\alpha}}(L,E) = 1 - \sin^{2} 2\vartheta \,
\sin^{2}\!\left( \frac{\Delta{m}^{2} L}{4E} \right) \,. \label{052}
\end{equation}
These simple expressions are often used in the analysis of experimental data.

When neutrinos propagate in matter, the potential generated by the coherent
forward elastic scattering with the particles in the medium (electrons and
nucleons) modifies mixing and oscillations \cite{Wolfenstein:1978ue}. In a
medium with varying density it is possible to have resonant flavor transitions
\cite{Mikheev:1985gs}. This is the famous MSW effect.

\begin{figure}[t!]
\begin{center}
\includegraphics*[bb=143 663 478 760, width=0.90\textwidth]{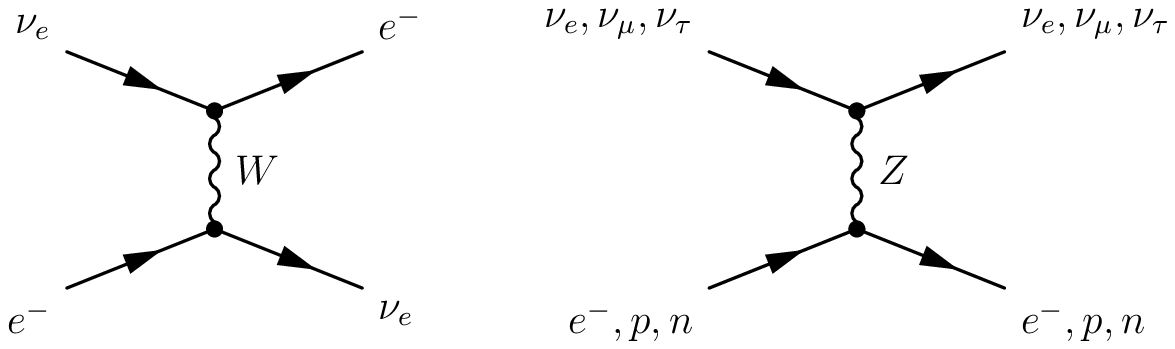}
\end{center}
\caption{ \label{i007} Feynman diagrams of the coherent forward elastic
scattering processes that generate the CC potential $V_{\text{CC}}$ through $W$
exchange and the NC potential $V_{\text{NC}}$ through $Z$ exchange. }
\end{figure}

The effective potentials for $\nu_{\alpha}$ and $\bar\nu_{\alpha}$ are,
respectively,
\begin{equation}
V_{\alpha} = V_{\text{CC}} \, \delta_{\alpha e} + V_{\text{NC}} \,, \qquad
\overline{V}_{\alpha} = - V_{\alpha} \,, \label{i025}
\end{equation}
with the charged-current and neutral-current potentials
\begin{equation}
V_{\text{CC}} = \sqrt{2} \, G_{\text{F}} \, N_{e} \,, \qquad V_{\text{NC}} = -
\frac{1}{2} \, \sqrt{2} \, G_{\text{F}} \, N_{n} \,, \label{i017}
\end{equation}
generated by the Feynman diagrams in Fig.~\ref{i007}. Here $N_{e}$ and $N_{n}$
are the electron and neutron number densities in the medium (in an electrically
neutral medium the neutral-current potentials of protons and electrons cancel
each other). In normal matter, these potentials are very small, because
\begin{equation}
\sqrt{2} \, G_{\text{F}} \simeq 7.63 \times 10^{-14} \, \frac{ \text{eV} \,
\text{cm}^{3} }{ N_{\text{A}} } \,, \label{i026}
\end{equation}
where $N_{\text{A}}$ is Avogadro's number.

Let us consider, for simplicity, two-neutrino $\nu_{e}$--$\nu_{\mu}$ mixing. In
general, a neutrino produced at $x=0$ is described at a distance $x$ by a state
\begin{equation}
| \nu(x) \rangle = \varphi_{e}(x) \, | \nu_{e} \rangle + \varphi_{\mu}(x) \, |
\nu_{\mu} \rangle \,. \label{state}
\end{equation}
The evolution of the flavor amplitudes $\varphi_{e}(x)$ and $\varphi_{\mu}(x)$
with the distance $x$ is given by the differential equation
\cite{Wolfenstein:1978ue}
\begin{equation}
i \frac{ \text{d} }{ \text{d}x }
\begin{pmatrix}
\varphi_{e}(x)
\\
\varphi_{\mu}(x)
\end{pmatrix}
=
\begin{pmatrix}
\frac{ \Delta{m}^{2} }{ 2 E } \sin^2\vartheta + V_{e} & \frac{ \Delta{m}^{2} }{
4 E } \sin{2\vartheta}
\\
\frac{ \Delta{m}^{2} }{ 4 E } \sin{2\vartheta} & \frac{ \Delta{m}^{2} }{ 2 E }
\cos^2\vartheta + V_{\mu}
\end{pmatrix}
\begin{pmatrix}
\varphi_{e}(x)
\\
\varphi_{\mu}(x)
\end{pmatrix}
\,. \label{i064}
\end{equation}
For an initial $\nu_{e}$, the boundary condition for the solution of the
differential equation is
\begin{equation}
\begin{pmatrix}
\varphi_{e}(0)
\\
\varphi_{\mu}(0)
\end{pmatrix}
=
\begin{pmatrix}
1
\\
0
\end{pmatrix}
\,, \label{i066}
\end{equation}
and the probabilities of $\nu_{e}\to\nu_{\mu}$ transitions and $\nu_{e}$
survival are, respectively,
\begin{equation}
P_{\nu_{e}\to\nu_{\mu}}(x) = |\varphi_{\mu}(x)|^{2} \,, \qquad
P_{\nu_{e}\to\nu_{e}}(x) = |\varphi_{e}(x)|^{2} = 1 -
P_{\nu_{e}\to\nu_{\mu}}(x) \,. \label{i067}
\end{equation}

The evolution equation~(\ref{i064}) has the structure of a Schr\"odinger
equation with the effective Hamiltonian matrix
\begin{equation}
\mathcal{H} = \frac{ \Delta{m}^{2} }{ 4 E } + \frac{1}{2} \, V_{\text{CC}} +
V_{\text{NC}} + \frac{1}{4E}
\begin{pmatrix}
- \Delta{m}^{2} \cos{2\vartheta} + 2 E V_{\text{CC}} & \Delta{m}^{2}
\sin{2\vartheta}
\\
\Delta{m}^{2} \sin{2\vartheta} & \Delta{m}^{2} \cos{2\vartheta} - 2 E
V_{\text{CC}}
\end{pmatrix}
\,. \label{i068}
\end{equation}
This matrix can be diagonalized by the orthogonal transformation
\begin{equation}
U_{\text{M}}^{T} \, \mathcal{H} \, U_{\text{M}} = \frac{ \Delta{m}^{2} }{ 4 E }
+ \frac{1}{2} \, V_{\text{CC}} + V_{\text{NC}} + \frac{1}{4E} \,
\text{diag}(-\Delta{m}^{2}_{\text{M}},\Delta{m}^{2}_{\text{M}}) \,.
\label{i069}
\end{equation}
The orthogonal matrix
\begin{equation}
U_{\text{M}} =
\begin{pmatrix}
\cos\vartheta_{\text{M}} & \sin\vartheta_{\text{M}}
\\
-\sin\vartheta_{\text{M}} & \cos\vartheta_{\text{M}}
\end{pmatrix}
\label{i071}
\end{equation}
is the effective mixing matrix in matter, and
\begin{equation}
\Delta{m}^{2}_{\text{M}} = \sqrt{ \left( \Delta{m}^{2}\cos2\vartheta - 2 E
V_{\text{CC}} \right)^{2} + \left( \Delta{m}^{2}\sin2\vartheta \right)^{2} }
\label{i072}
\end{equation}
is the effective squared-mass difference. The effective mixing angle in matter
$\vartheta_{\text{M}}$ is given by
\begin{equation}
\tan2\vartheta_{\text{M}} = \dfrac{\tan2\vartheta}{1-\dfrac{2 E
V_{\text{CC}}}{\Delta{m}^{2}\cos2\vartheta}} \,. \label{i073}
\end{equation}
The most interesting characteristic of this expression is that there is a
resonance \cite{Mikheev:1985gs} when
\begin{equation}
V_{\text{CC}} = \frac{ \Delta{m}^{2} }{ 2 E } \, \cos2\vartheta \,,
\label{i074}
\end{equation}
which corresponds to the electron number density
\begin{equation}
N_{e}^{\text{R}} = \frac{ \Delta{m}^{2} \cos2\vartheta }{ 2 \sqrt{2} E
G_{\text{F}} } \,. \label{i075}
\end{equation}
At the resonance the effective mixing angle is equal to $\pi/4$, i.e.\ the
mixing is maximal, leading to the possibility of total transitions between the
two flavors if the resonance region is wide enough.

In general, the evolution equation~(\ref{i064}) must be solved numerically or
with appropriate approximations. In a constant matter density, it is easy to
derive an analytic solution, leading to the transition probability
\begin{equation}
P_{\nu_{e}\to\nu_{\mu}}(x) = \sin^{2} 2\vartheta_{\text{M}} \, \sin^{2}\left(
\frac{\Delta{m}^{2}_{\text{M}} x}{4E} \right) \,, \label{i082}
\end{equation}
which has the same structure as the two-neutrino transition probability in
vacuum in Eq.~(\ref{051}), with the mixing angle and the squared-mass
difference replaced by their effective values in matter.

\subsection{Phenomenology}
\label{053} \nopagebreak

\begin{figure}[t!]
\begin{center}
\includegraphics*[bb=93 42 571 695, height=0.8\linewidth, angle=-90]{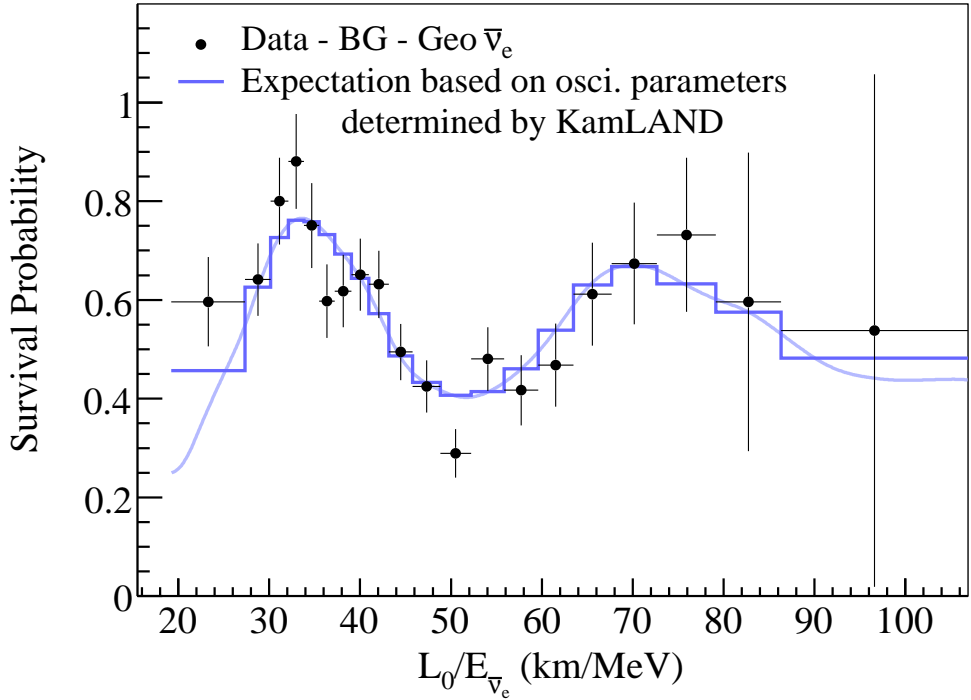}
\end{center}
\caption{ \cite{0801.4589} \ \label{055} Ratio of the background and
geoneutrino-subtracted $\bar\nu_{e}$ spectrum to the expectation for
no-oscillation as a function of $L_{0}/E$. $L_{0}$ is the effective
baseline taken as a flux-weighted average ($L_{0}$\,=\,180\,km). }
\end{figure}

\begin{figure}[t!]
\begin{center}
\includegraphics*[bb=36 35 539 521, width=0.8\linewidth]{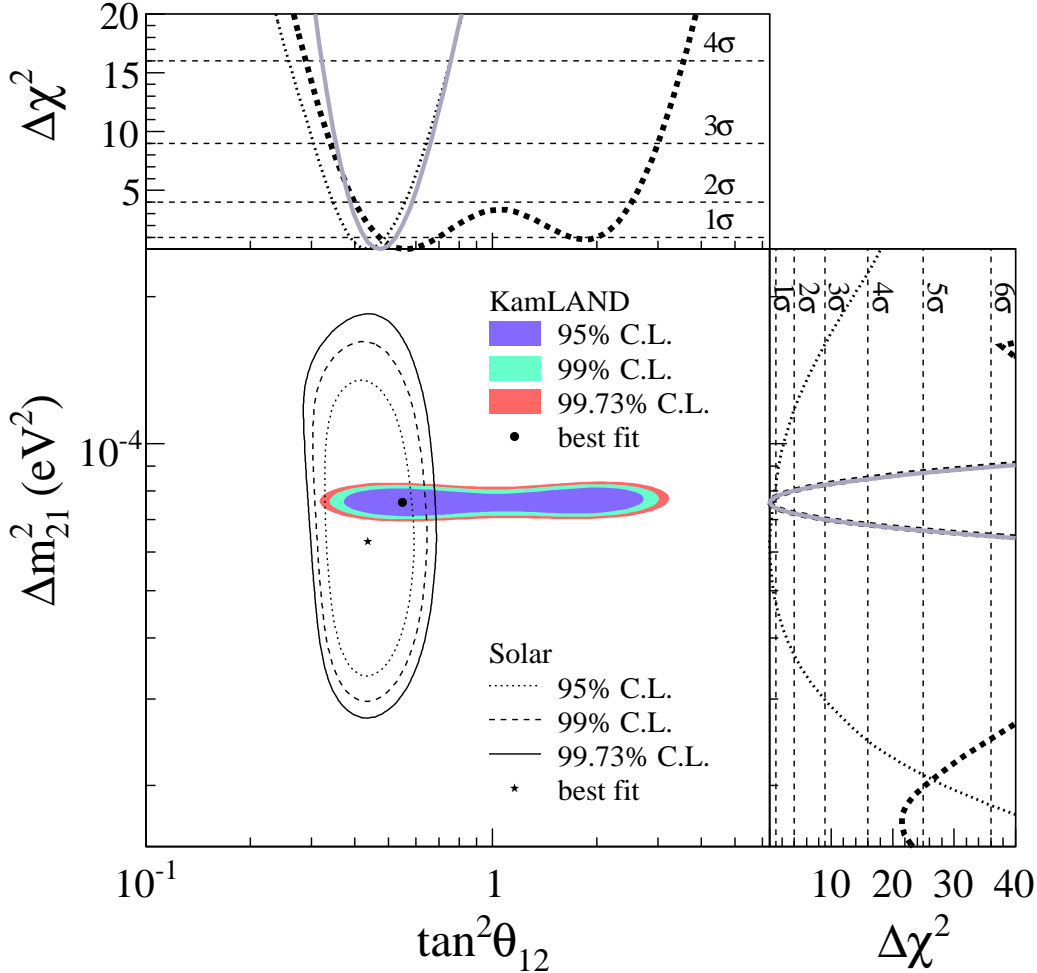}
\end{center}
\caption{ \cite{0801.4589}\ \label{056} Allowed regions for
$\nu_{e}\to\nu_{\mu},\nu_{\tau}$ oscillation parameters from KamLAND
and solar neutrino data. The side-panels show the $\Delta
\chi^{2}$-profiles for KamLAND (dashed) and solar experiments
(dotted) individually, as well as the combination of the two
(solid).}
\end{figure}

\begin{figure}[t!]
\begin{center}
\includegraphics*[bb=5 2 505 268, width=0.8\linewidth]{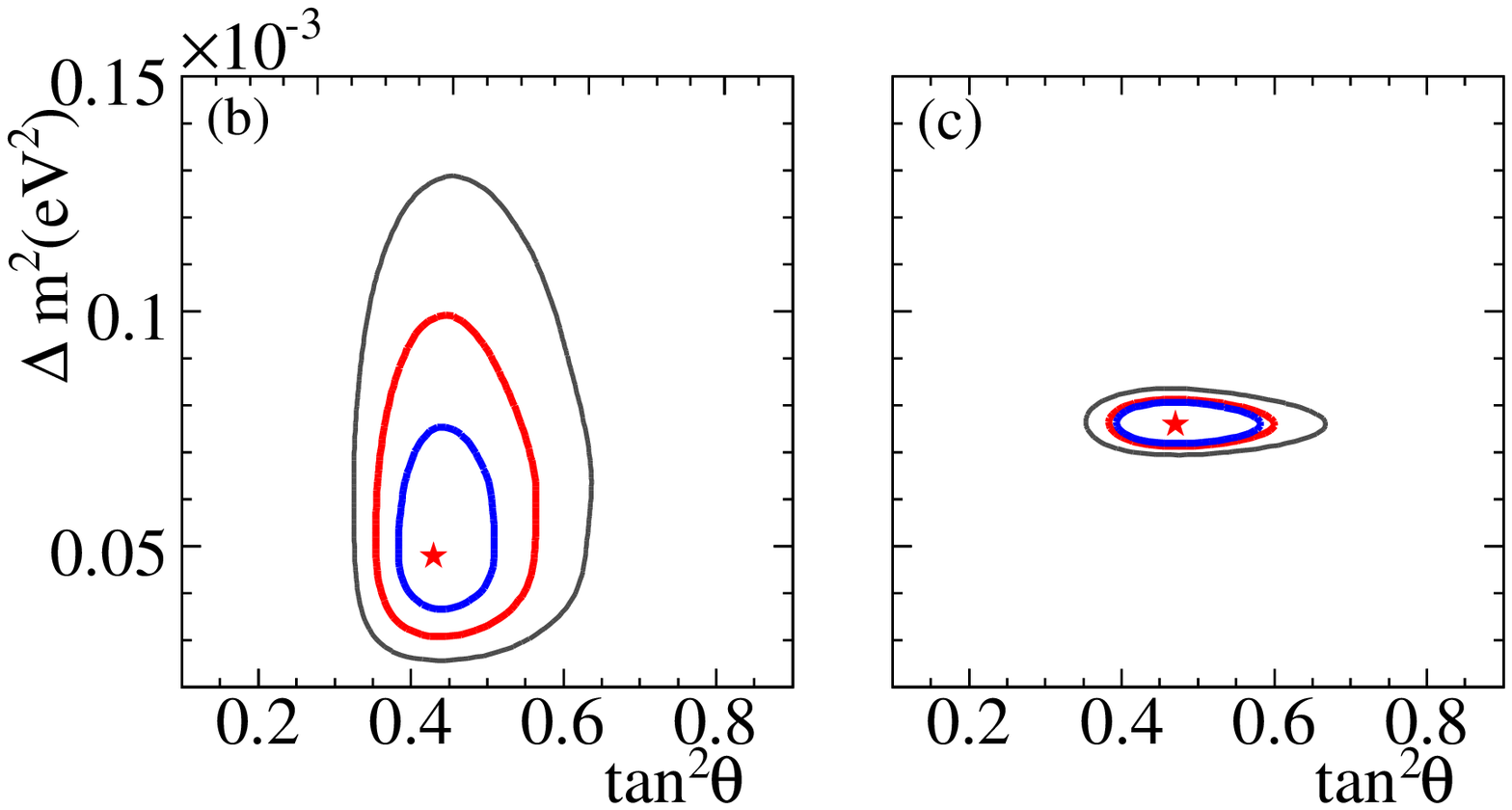}
\end{center}
\caption{ \cite{0806.0989}\label{057} Allowed regions for
$\nu_{e}\to\nu_{\mu},\nu_{\tau}$ oscillation parameters from solar
(b) and solar+KamLAND (c) data. The best-fit points are: $\Delta
m^2=4.90\times10^{-5}$ eV$^2$, $\tan^{2}\theta=0.437$ from solar data
(b) and $\Delta m^2=7.59\times10^{-5}$ eV$^2$, $\tan^{2}\theta=0.468$
from solar+KamLAND data (c).}
\end{figure}

Since the 1960s it has been known, mainly through the insight of Pontecorvo
\cite{Pontecorvo:1968fh}, that neutrino oscillations can be revealed not only
in terrestrial neutrino experiments, but also in experiments which are
sensitive to neutrinos coming from astrophysical sources. The largest
astrophysical neutrino flux on Earth coming from the Sun has been measured by
several experiments, starting with the pioneering Homestake experiment
\cite{Cleveland:1998nv}, which first observed the deficit of electron solar
neutrinos with respect to the Standard Solar Model prediction (see
Ref.~\cite{Bahcall:1989ks}). In 2002 the SNO experiment \cite{nucl-ex/0204008}
has shown that the solar neutrino problem is due to
$\nu_{e}\to\nu_{\mu},\nu_{\tau}$ transitions. The reactor long-baseline KamLAND
experiment established at the end of 2002 \cite{hep-ex/0212021} that these
$\nu_{e}\to\nu_{\mu},\nu_{\tau}$ transitions are due to neutrino oscillations.
The evidence for oscillations obtained by the KamLAND experiment is illustrated
in Fig.~\ref{055}. The current solar and KamLAND data are fitted well by
effective two-neutrino oscillations, including MSW
\cite{Wolfenstein:1978ue,Mikheev:1985gs} effects of neutrino propagation in
matter, with the solar squared-mass difference and mixing angle
\cite{0801.4589}
\begin{equation}
\Delta{m}^2_{\text{SUN}} = \left( 7.59 \pm 0.21 \right) \times 10^{-5} \,
\text{eV}^2 \,, \qquad \tan^2\vartheta_{\text{SUN}} = 0.47 {}^{+0.06}_{-0.05}
\,. \label{054}
\end{equation}
The allowed regions in the
$\tan^2\vartheta_{\text{SUN}}$--$\Delta{m}^2_{\text{SUN}}$ plane obtained from
KamLAND data and the data of solar neutrino experiments (Homestake
\cite{Cleveland:1998nv}, GALLEX/GNO \cite{hep-ex/0504037}, SAGE
\cite{astro-ph/0204245}, Super-Kamiokande \cite{0803.4312}, SNO
\cite{0806.0989}, Borexino \cite{0808.2868}) are shown in Fig.~\ref{056}.
Figure~\ref{057} shows the results of the analysis of solar neutrino data alone
and in combination with KamLAND data.

\begin{figure}[t!]
\includegraphics*[bb=16 25 453 443, width=0.45\linewidth]{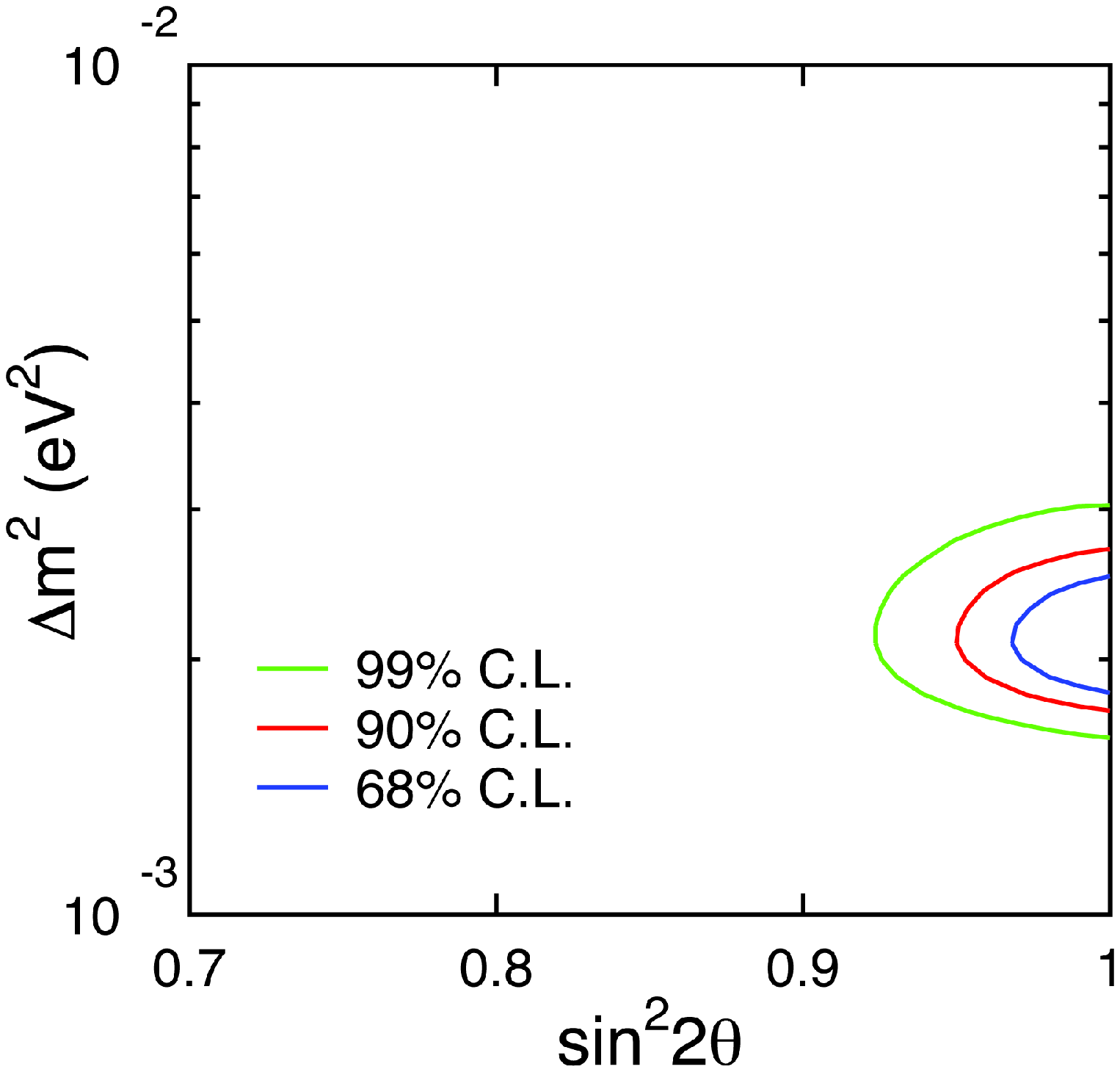}
\hfill
\includegraphics*[bb=16 25 453 443, width=0.45\linewidth]{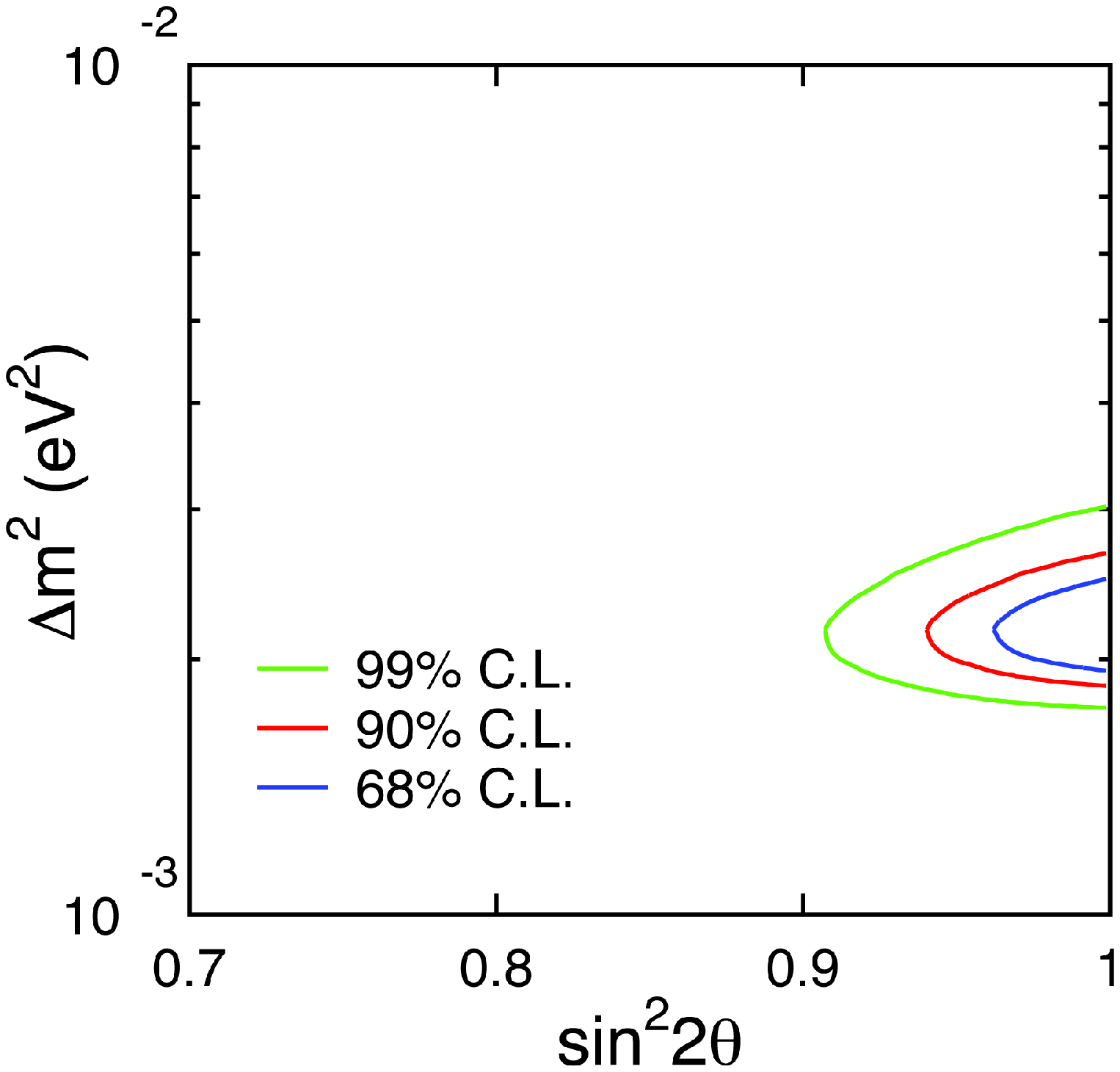}
\caption{ \cite{0810.0595} \label{060} Allowed regions for $
\protect\boss{\nu}{\mu} \to \protect\boss{\nu}{\tau} $ oscillation
parameters from Super-Kamiokande atmospheric neutrino data obtained
with a zenith angle analysis (left) and a $L/E$ analysis (right). The
best-fit values of the oscillation parameters are:
$\sin^22\vartheta=1.02$, $\Delta{m}^2=2.1\times10^{-3}\,\text{eV}^2$
(zenith) and $\sin^22\vartheta=1.04$,
$\Delta{m}^2=2.2\times10^{-3}\,\text{eV}^2$ ($L/E$).}
\end{figure}

\begin{figure}[t!]
\begin{center}
\includegraphics*[bb=10 30 490 490, width=0.6\linewidth]{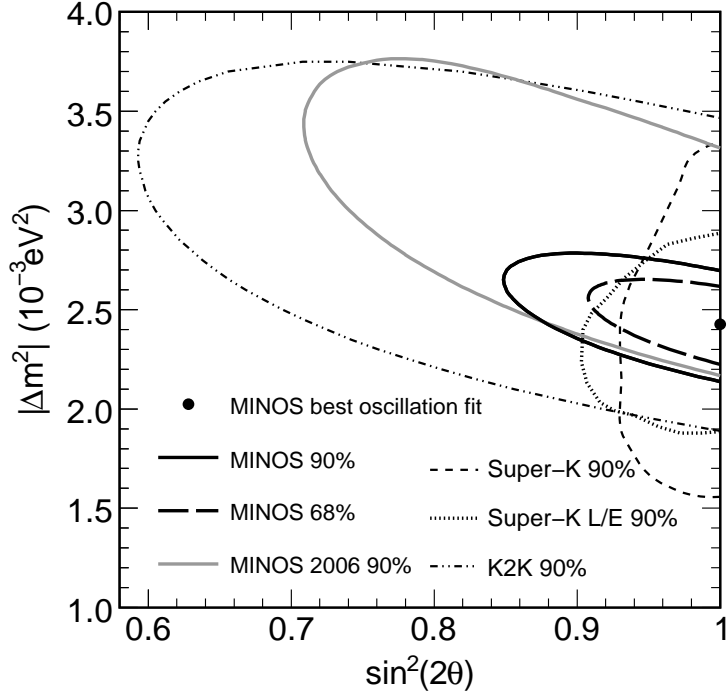}
\end{center}
\caption{ \cite{0806.2237} \ \label{061} Allowed regions for $
\protect\boss{\nu}{\mu} $-disappearance oscillation parameters from
K2K and MINOS data.}
\end{figure}

The atmospheric neutrino anomaly was discovered in the late 1980s in
the Kamiokande \cite{Fukuda:1994mc} and IMB
\cite{Becker-Szendy:1992ym} experiments. In 1998 the Super-Kamiokande
experiment found a model independent evidence of muon (anti)neutrino
disappearance in atmospheric neutrino data \cite{Fukuda:1998mi}.
Atmospheric neutrinos are produced by the decay of pions and muons
created by the interactions of cosmic rays with the nuclei in the
atmosphere. Since at energies higher than about 1~GeV the flux of
atmospheric neutrinos is approximately isotropic, the corresponding
number of events generated in a detector by atmospheric neutrinos
must be the same in any direction. The Super-Kamiokande collaboration
measured the up-down asymmetry of high-energy
$\boss{\nu}{\mu}$-induced events \cite{Fukuda:1998mi}
\begin{equation}
A_{\mu}^{\text{up-down}} \equiv \left( \frac { U - D } { U + D } \right)_{\mu}
= - 0.296 \pm 0.049 \,, \label{058}
\end{equation}
where $ U $ and $ D $ are, respectively, the neutrino fluxes  integrated in the
ranges $ -1 < \cos\theta_{z} < -0.2 $ and $ 0.2 < \cos\theta_{z} < 1 $
($\theta_{z}$ is the angle between the zenith and the neutrino arrival
direction). Since the measured asymmetry deviates from zero by about $6\sigma$,
the model-independent evidence of an atmospheric neutrino anomaly is
indisputable. The negative value of $A_{\mu}^{\text{up-down}}$ indicates that
muon (anti)neutrinos coming from the opposite hemisphere disappear, most likely
because of $ \boss{\nu}{\mu} \to \boss{\nu}{\tau} $ oscillations with the
mixing parameters in Fig.~\ref{060}, since atmospheric electron (anti)neutrinos
do not show any anomalous behavior. This interpretation has been confirmed by
the independent observations of $\boss{\nu}{\mu}$ disappearance in the
accelerator long-baseline experiments K2K \cite{hep-ex/0606032} and MINOS
\cite{0806.2237} which are generated by the same values of the mixing
parameters, as shown in Fig.~\ref{061}. From MINOS data \cite{0806.2237},
\begin{equation}
\Delta{m}^2_{\text{ATM}} = \left( 2.43 \pm 0.13 \right) \times 10^{-3} \,
\text{eV}^2 \,, \qquad \sin^22\vartheta_{\text{ATM}}
>
0.90 \, (\text{90\% C.L.}) \,. \label{059}
\end{equation}

\begin{figure}[t!]
\underline{NORMAL} \quad
\begin{minipage}[c]{0.25\textwidth}
\includegraphics*[bb=175 469 415 779, width=\linewidth]{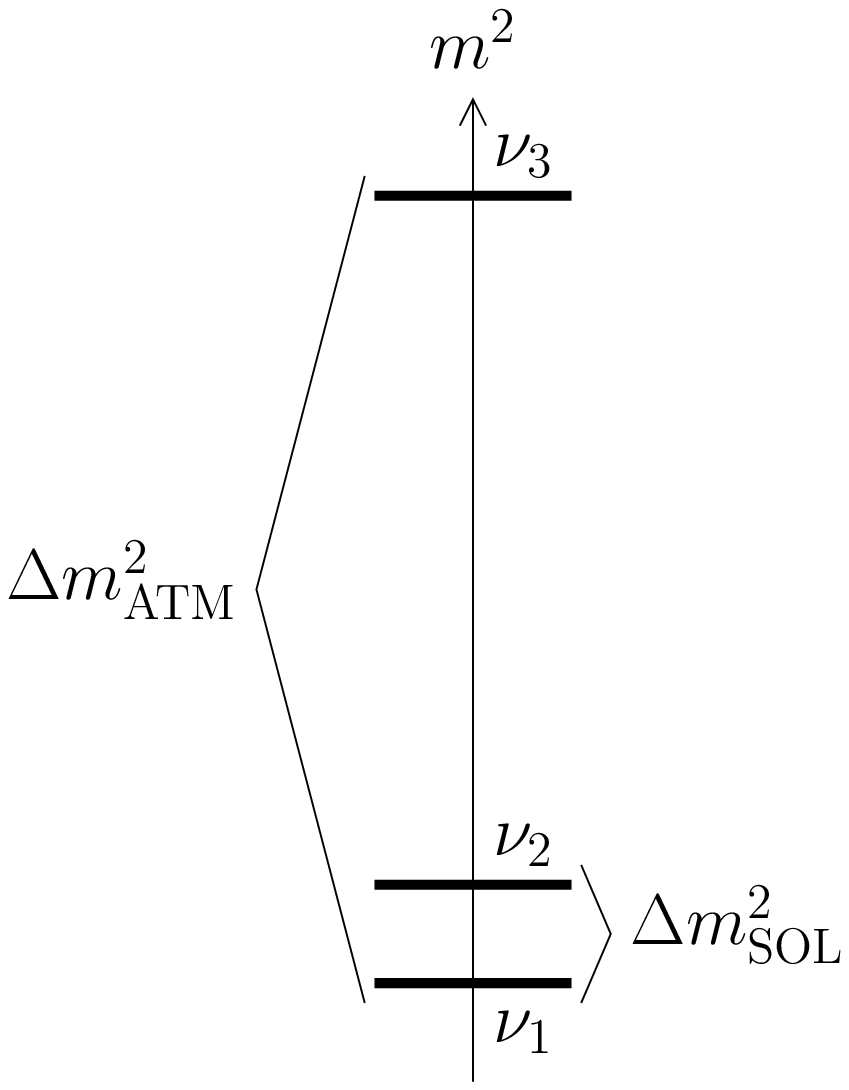}
\end{minipage}
\hfill
\begin{minipage}[c]{0.25\textwidth}
\includegraphics*[bb=180 469 420 779, width=\linewidth]{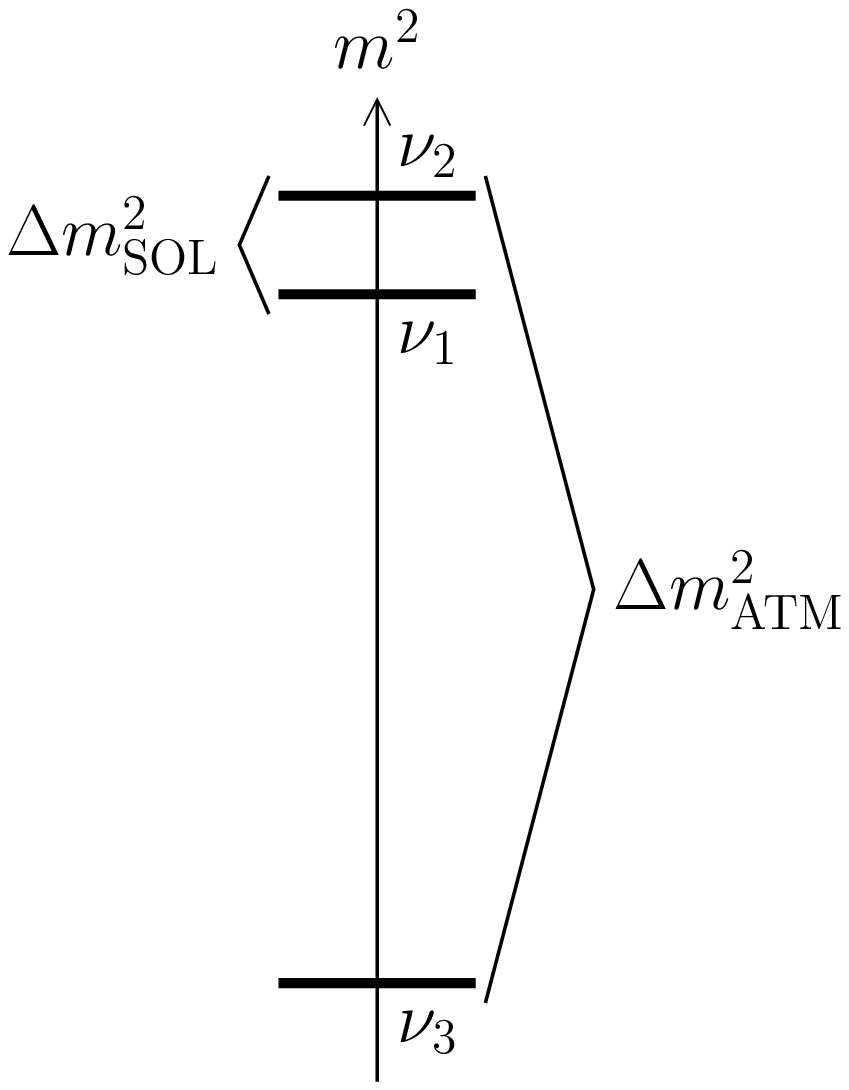}
\end{minipage}
\quad \underline{INVERTED} \caption{ \label{062} The two three-neutrino schemes
allowed by the hierarchy $\Delta{m}^{2}_{\text{SOL}} \ll
\Delta{m}^{2}_{\text{ATM}}$. }
\end{figure}

The different values of solar and atmospheric squared-mass differences in
Eqs.~(\ref{054}) and (\ref{059}) imply that two-neutrino mixing, with one
squared-mass difference, is not sufficient for the description of all
oscillation data. Moreover, all three neutrino flavors are involved in the
observed oscillations (solar and reactor
$\boss{\nu}{e}\to\boss{\nu}{\mu},\boss{\nu}{\tau}$; atmospheric and accelerator
$ \boss{\nu}{\mu} \to \boss{\nu}{\tau} $). Therefore we must consider the
mixing of three neutrinos in Eq.~(\ref{031}). The observed hierarchy $
\Delta{m}^{2}_{\text{SOL}} \ll \Delta{m}^{2}_{\text{ATM}} $ can be accommodated
in the \emph{normal} and \emph{inverted} three-neutrino mixing schemes shown
schematically in Fig.~\ref{062}. We choose the arbitrary labeling numbers of
the massive neutrinos in order to have $ \Delta{m}^{2}_{\text{SOL}} =
\Delta{m}^{2}_{21} $ and $ \Delta{m}^{2}_{\text{ATM}} = |\Delta{m}^{2}_{31}| $,
with $ \Delta{m}^{2}_{21} \ll \Delta{m}^{2}_{31} \simeq \Delta{m}^{2}_{32} $.

\begin{figure}[t!]
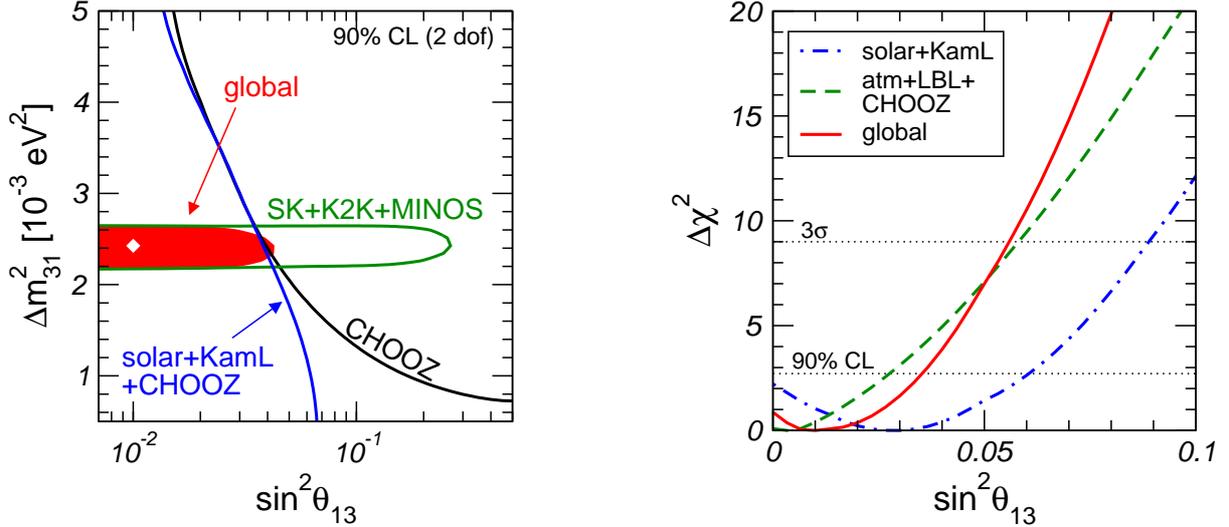

\includegraphics*[bb=268 9 556 310, height=7cm]{fig/Schwetz-Tortola-Valle-0808.2016v1-f3a.eps}
\hfill
\includegraphics*[bb=6 480 322 787, height=7cm]{fig/Schwetz-Tortola-Valle-0808.2016v1-f3b.eps}
\caption{\cite{0808.2016} \label{065} Constraints on
$\sin^2\vartheta_{13}$ from a global analysis of neutrino oscillation
data.}
\end{figure}

In principle, the analysis of neutrino oscillation data in a three-neutrino
mixing framework could yield results which are different from those obtained in
a two-neutrino mixing approximation. However, in practice the two-neutrino
mixing approximation is quite accurate, because the only element of the mixing
matrix which affects both solar and atmospheric neutrino oscillations,
$U_{e3}$, is small. This information comes from the results of the CHOOZ
\cite{hep-ex/0301017} and Palo Verde \cite{hep-ex/0107009} reactor
long-baseline experiments, which excluded $\bar\nu_{e}$ disappearance generated
by $\Delta{m}^{2}_{\text{ATM}}$ with the effective mixing angle $ \sin^{2}
2\vartheta_{ee}^{\text{eff}} = 4 \, |U_{e3}|^{2} \left( 1 - |U_{e3}|^{2}
\right) = \sin^{2} 2\vartheta_{13} $ \cite{hep-ph/9802201}. The results of a
global analysis of neutrino oscillation data are shown in Fig.~\ref{065}. The
90\%~C.L. ($3\sigma$) bounds for $\sin^2\vartheta_{13}$ are \cite{0808.2016}
\begin{equation}\label{063}
\sin^2\vartheta_{13} \le \left\lbrace \begin{array}{l@{\qquad}l}
0.060~(0.089) & \text{(solar+KamLAND)} \,, \\
0.027~(0.058) & \text{(CHOOZ+atm+K2K+MINOS)} \,, \\
0.035~(0.056) & \text{(global data)} \,.
\end{array} \right.
\end{equation}
Therefore, in practice we have
\begin{equation}
\vartheta_{\text{SOL}} \simeq \vartheta_{12} \,, \qquad \vartheta_{\text{ATM}}
\simeq \vartheta_{23} \,, \label{064}
\end{equation}
and the results in Eqs.(\ref{054}) and (\ref{059}) apply to three-neutrino
mixing.

\begin{figure}[t!]
\begin{center}
\setlength{\tabcolsep}{0cm}
\begin{tabular*}{\textwidth}{l@{\extracolsep{\fill}}r}
\includegraphics*[bb=102 430 445 754, width=0.49\textwidth]{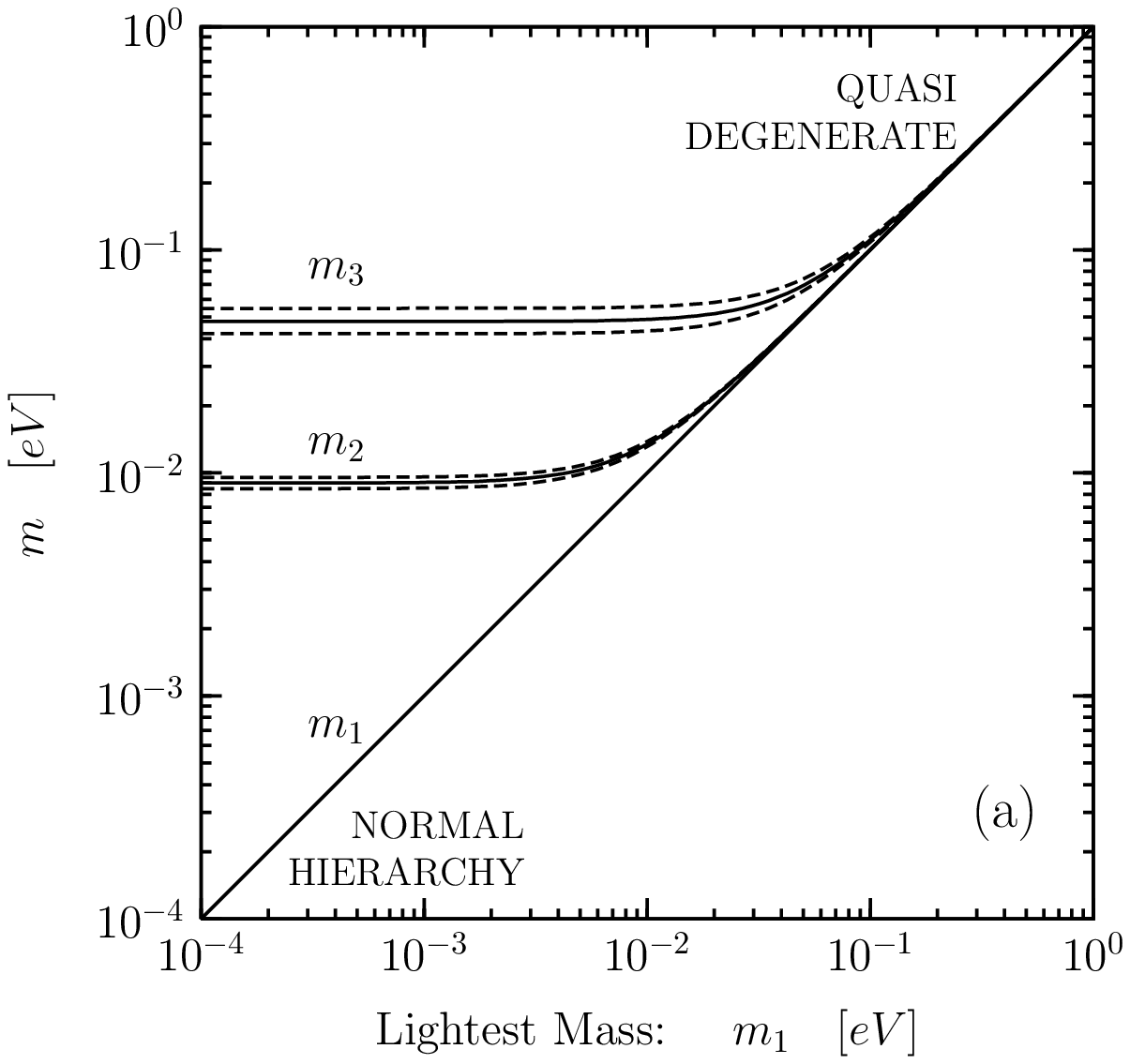}
&
\includegraphics*[bb=102 430 445 754, width=0.49\textwidth]{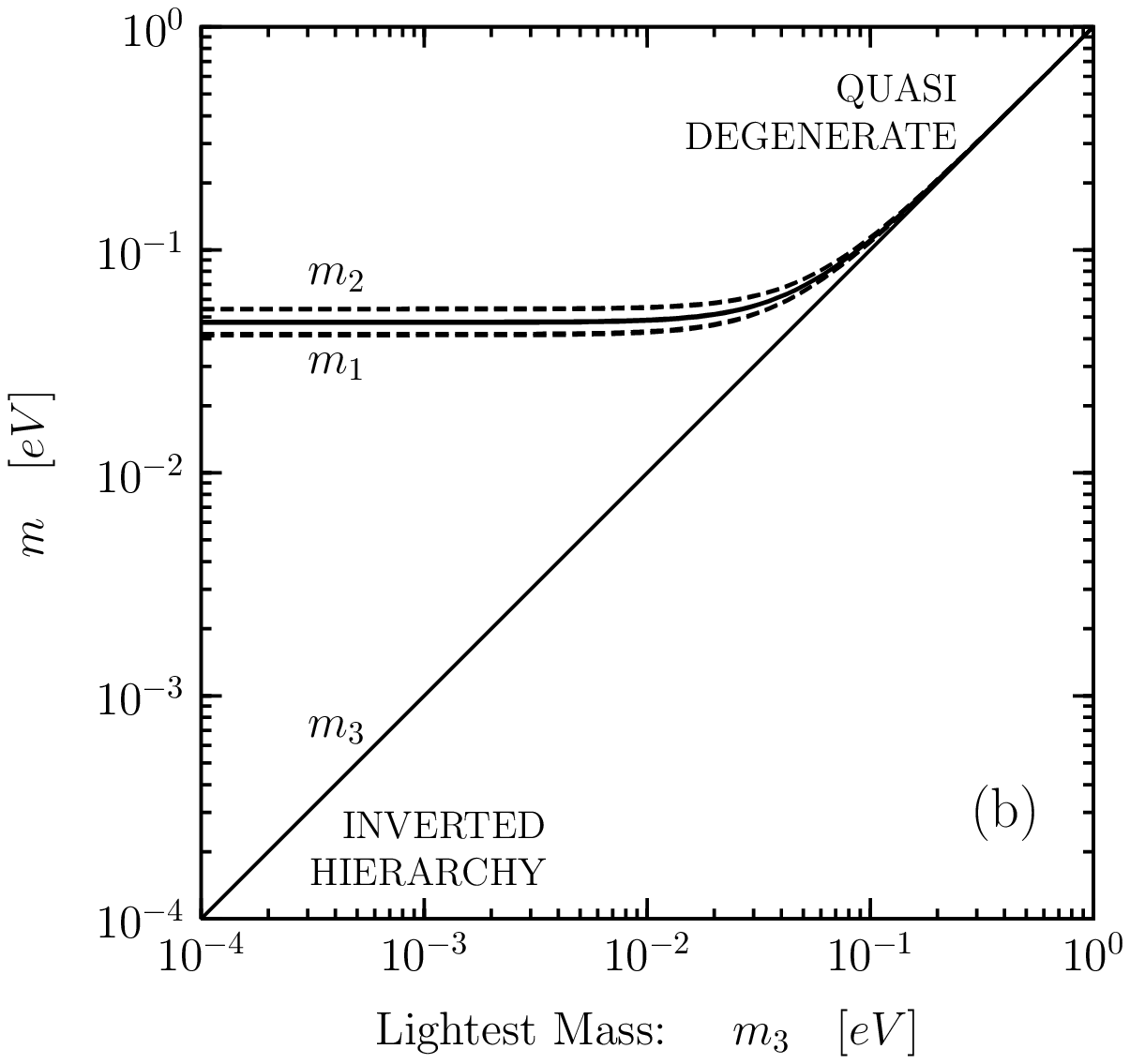}
\end{tabular*}
\end{center}
\caption{ \label{066} Values of neutrino masses as functions of the lightest
mass $m_{1}$ in the normal scheme and $m_{3}$ in the inverted scheme. Solid
lines correspond to the best-fit values of $\Delta{m}^2_{\text{SUN}}$ and
$\Delta{m}^2_{\text{ATM}}$. Dashed lines enclose $3\sigma$ ranges. }
\end{figure}

So far we have considered only neutrino oscillation data, which give
information on neutrino mixing and the differences of neutrino squared masses.
The absolute scale of neutrino masses must be determined with other means.
However, since we know the squared-mass differences from Eqs.~(\ref{054}) and
(\ref{059}), it is possible to express the neutrino masses as functions of only
one unknown parameter representing the absolute mass scale. Figure~\ref{066}
shows the values of the three neutrino masses as functions of the lightest
mass, which is $m_{1}$ in the normal scheme and $m_{3}$ in the inverted scheme.
One can see that in both schemes there is quasidegeneracy of the three masses
when $ m_{1} \simeq m_{2} \simeq m_{3} \gg \sqrt{\Delta{m}^{2}_{\text{ATM}}}
\simeq 5 \times 10^{-2} \, \text{eV} $. In this case, it is very difficult to
distinguish the two schemes. On the other hand, the two schemes have very
different features if the lightest mass is much smaller than
$\sqrt{\Delta{m}^{2}_{\text{ATM}}}$. In this case, in the normal scheme there
is a \emph{hierarchy} of masses: $ m_{1} \ll m_{2} \ll m_{3} $. In the inverted
scheme there is a so-called \emph{inverted hierarchy} $ m_{3} \ll m_{1} \simeq
m_{2} $ in which $m_{1}$ and $m_{2}$ are quasidegenerate. In fact, in the
inverted scheme $m_{1}$ and $m_{2}$ are always quasidegenerate, because their
separation is due to the small solar squared-mass difference
$\Delta{m}^{2}_{\text{SOL}}$. Let us note that, independently of the mass
scheme, at least two neutrinos are massive, with masses larger than about $ 8
\times 10^{-3} \, \text{eV} $.

\begin{figure}[t!]
\begin{center}
\setlength{\tabcolsep}{0cm}
\begin{tabular*}{\textwidth}{l@{\extracolsep{\fill}}r}
\includegraphics*[bb=102 430 445 754, width=0.49\textwidth]{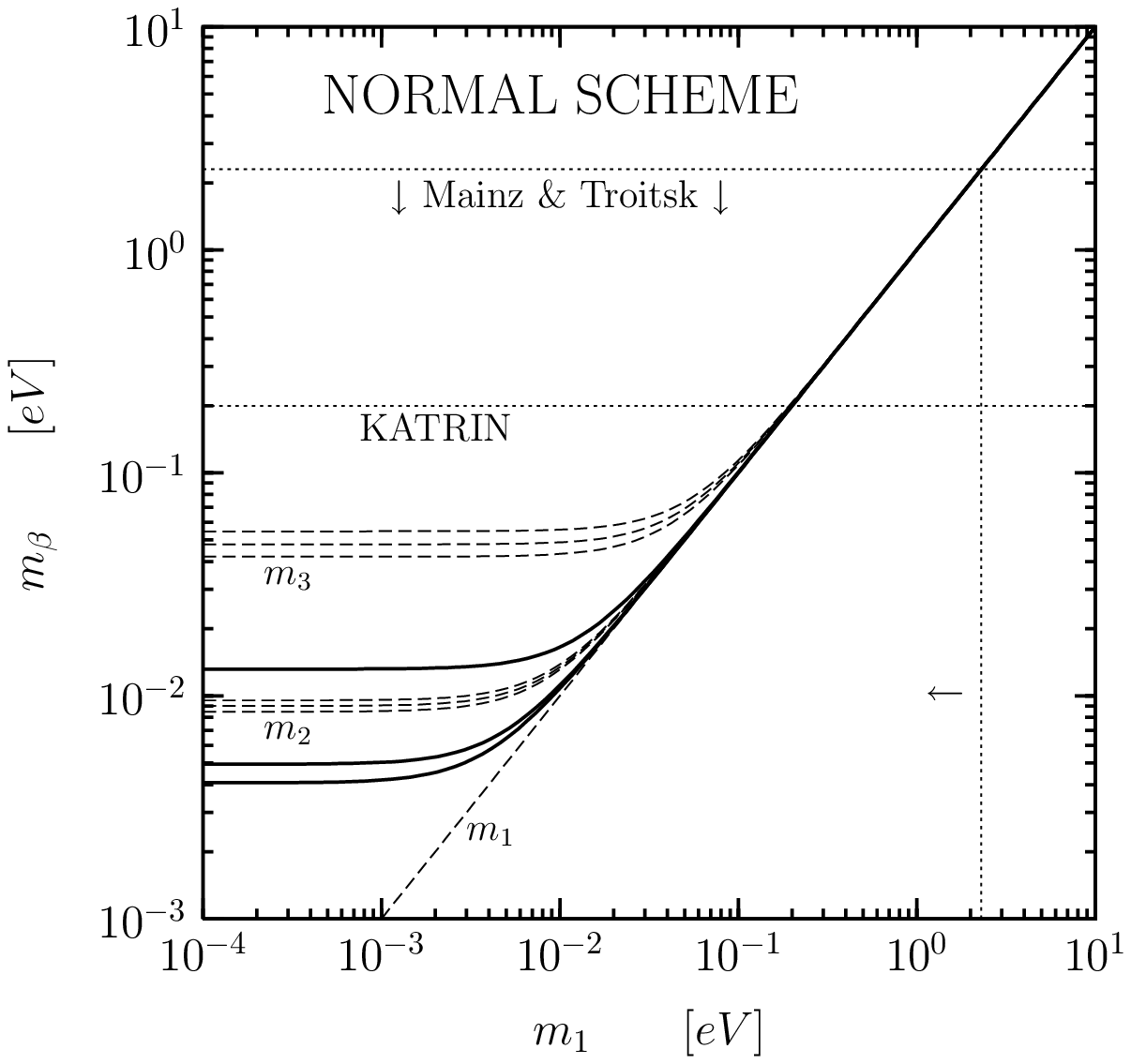}
&
\includegraphics*[bb=102 430 445 754, width=0.49\textwidth]{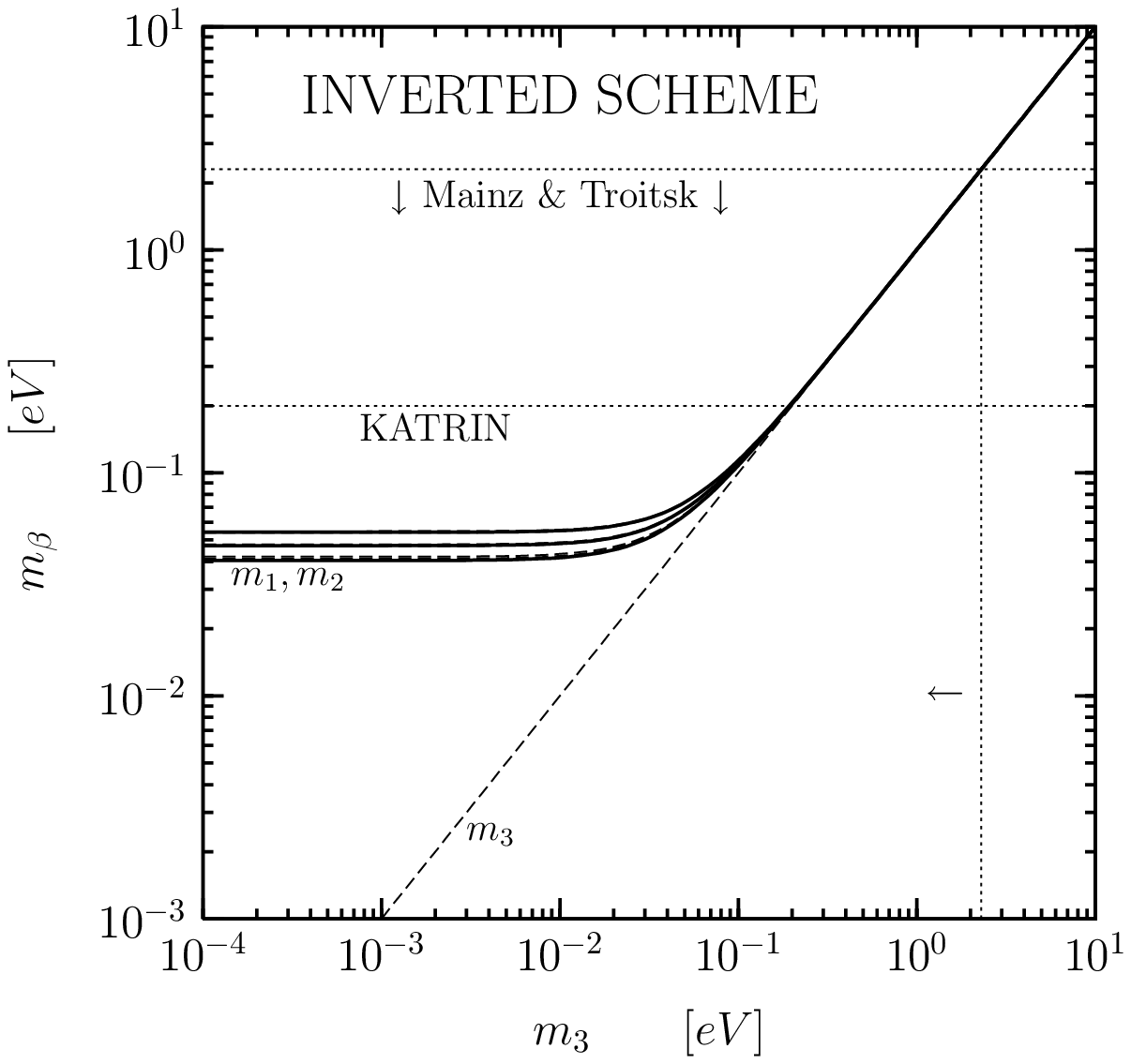}
\end{tabular*}
\end{center}
\caption{ \label{069} Effective neutrino mass $m_{\beta}$ in tritium
$\beta$-decay experiments as a function of the lightest mass ($m_{1}$ in the
normal scheme and $m_{3}$ in the inverted scheme; see Fig.~\ref{062}). Middle
solid lines correspond to the best-fit values of $\Delta{m}^2_{\text{SUN}}$ and
$\Delta{m}^2_{\text{ATM}}$. Extreme solid lines enclose $3\sigma$ ranges.
Dashed lines show the best-fit values and $3\sigma$ ranges of individual
masses. In the inverted scheme, the best-fit values and $3\sigma$ ranges of
$m_{1}$ and $m_{2}$ are practically the same and coincide with the best-fit
value and $3\sigma$ range of $m_{\beta}$. }
\end{figure}

The most reliable method for the determination of the absolute value of
neutrino masses is the kinematic measurement of neutrino masses in
interactions. Currently, the best limit is obtained in tritium $\beta$-decay
experiments, which are sensitive to the effective mass
\begin{equation}
m_{\beta} = \sqrt{ \sum_{k} |U_{ek}|^{2} m_{k}^{2} } \,. \label{067}
\end{equation}
The current bound on $m_{\beta}$ was obtained in the Mainz
\cite{hep-ex/0412056} and Troitzk \cite{Lobashev:1999tp} experiments:
\begin{equation}
m_{\beta} < 2.3 \, \text{eV} \quad (95\% \, \text{C.L.}) \,. \label{068}
\end{equation}
Figure~\ref{069} shows the comparison of this bound with the possible value of
$m_{\beta}$ in the normal and inverted schemes as a function of the lightest
mass.

\begin{figure}[t!]
\begin{center}
\setlength{\tabcolsep}{0cm}
\begin{tabular*}{\textwidth}{l@{\extracolsep{\fill}}r}
\includegraphics*[bb=106 429 441 749, width=0.49\textwidth]{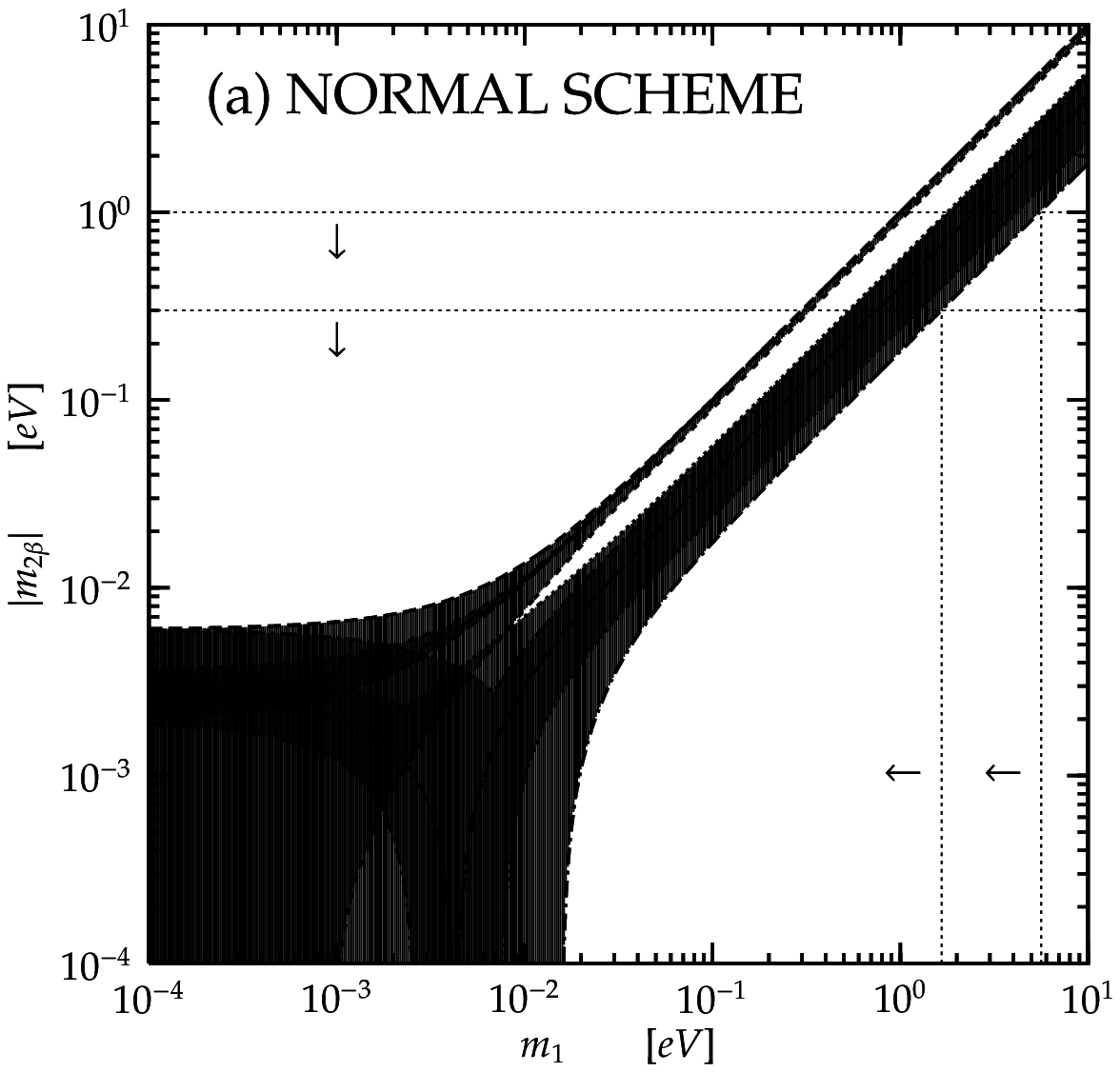}
&
\includegraphics*[bb=106 429 441 749, width=0.49\textwidth]{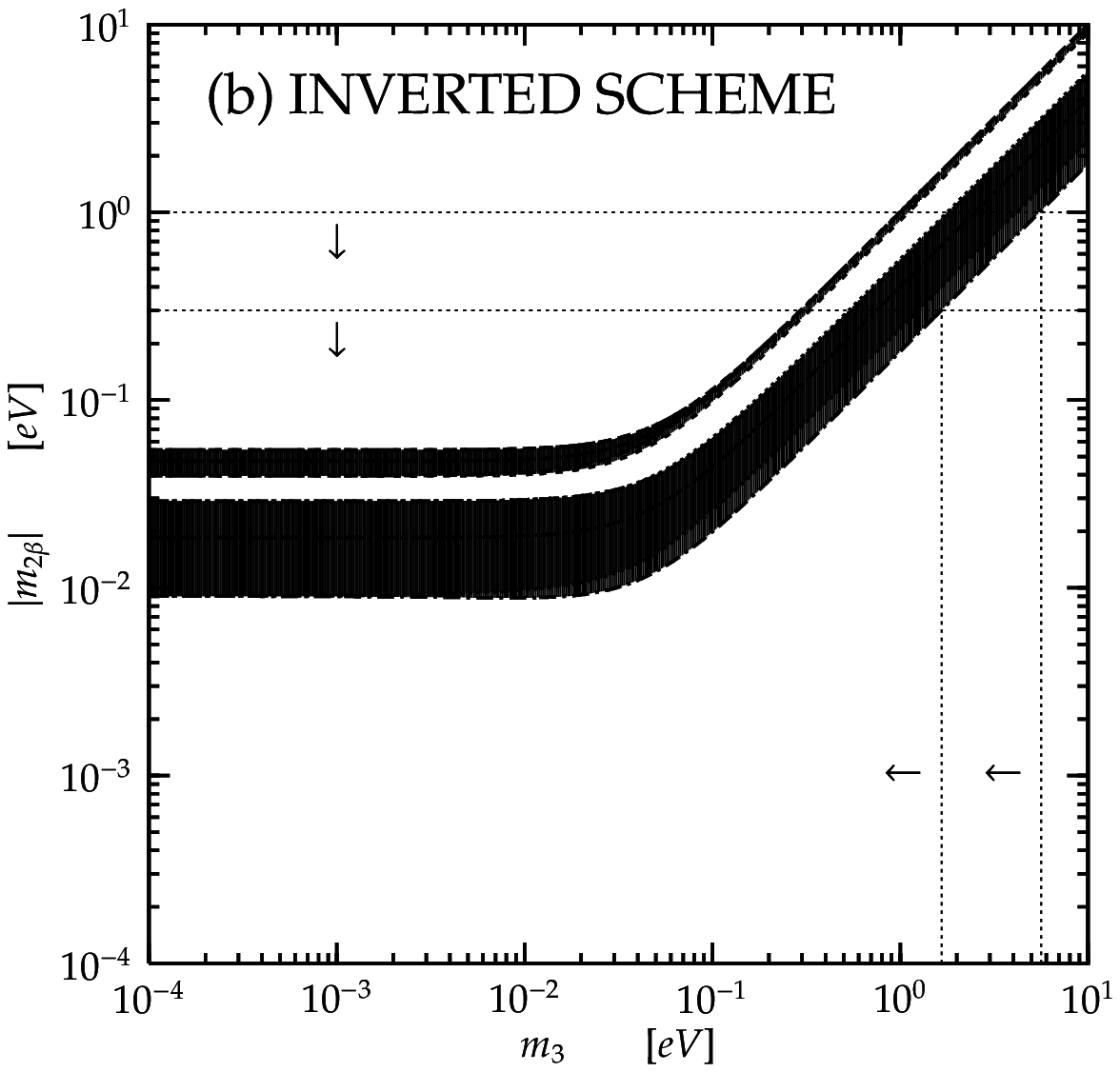}
\end{tabular*}
\end{center}
\caption{ \label{072} Absolute value $|m_{2\beta}|$ of the effective Majorana
neutrino mass in $2\beta_{0\nu}$-decay as a function of the lightest mass
$m_{1}$ in the normal scheme (a) and $m_{3}$ in the inverted scheme (b). The
two horizontal dotted lines correspond to the extremes of the upper bound range
in Eq.~(\ref{071}). The two vertical dotted lines show the corresponding upper
bounds for $m_{1}$ (a) and $m_{3}$ (b). }
\end{figure}

Another very important process which is sensitive to the absolute scale of
neutrino masses is neutrinoless double-$\beta$-decay, which occurs only if
massive neutrinos are Majorana particles. Neutrinoless double-$\beta$-decay
depends on the effective Majorana mass
\begin{equation}
m_{2\beta} = \sum_{k=1}^{3} U_{ek}^{2} \, m_{k} \,. \label{070}
\end{equation}
The best limit on $m_{2\beta}$, obtained in the Heidelberg--Moscow
$^{76}\text{Ge}$ experiment, \cite{Klapdor-Kleingrothaus:2001yx} is
\begin{equation}
|m_{2\beta}| \lesssim 0.3 - 1.0 \, \text{eV} \,, \label{071}
\end{equation}
where the large uncertainty is of theoretical nuclear physics origin.
Figure~\ref{072} shows the comparison of this bound with the possible value of
$m_{2\beta}$ in the normal and inverted scheme as a function of the lightest
mass. The unshaded strip within the shadowed bands can be obtained only in the
case of CP violation.

Let us finally mention that the evolution of the Universe depends on the values
of neutrino masses (see Ref.~\cite{Giunti-Kim-2007}). Current cosmological data
limit the sum of neutrino masses by \cite{0805.2517}
\begin{equation}
\sum_{k=1}^{3} m_{k} \lesssim 0.2 - 0.7 \, \text{eV} \,, \label{073}
\end{equation}
in the framework of the very successful flat $\Lambda$CDM model.

\section{Neutrino electromagnetic properties}
\label{sec4} The importance of neutrino electromagnetic properties
was first mentioned  by Pauli just in 1930 when he postulated the
existence of this particle and discussed the possibility that the
neutrino might have a magnetic moment.   Systematic theoretical
studies of neutrino electromagnetic properties have started after it
was shown that in the extended Standard Model  with right-handed
neutrinos the magnetic moment of a massive neutrino is, in general,
nonvanishing  and that its value is determined by the neutrino mass
\cite{MarSanPLB77,LeeShrPRD77,FujShrPRL80,ShrNP82,PetSNP77,
BilPetRMP87}.

Neutrino electromagnetic properties are of particular importance
because they are directly connected to fundamentals of particle
physics. For example, neutrino electromagnetic properties can be used
to distinguish Dirac and Majorana  neutrinos (see
\cite{SchValPRD81,ShrNP82,PalWolPRD82,NiePRD82,KayPRD82_KayPRD84} for
the correspondent discussion) and also as a probe of new physics that
might exist beyond the Standard Model  (see, for instance,
\cite{BelCirRamVogWisPRL05_BelGorRamVogWanPLB06_BelIJMPA07,NovRosSaTos08054177}).
\begin{figure}
  \centering
  \includegraphics[scale=.35]{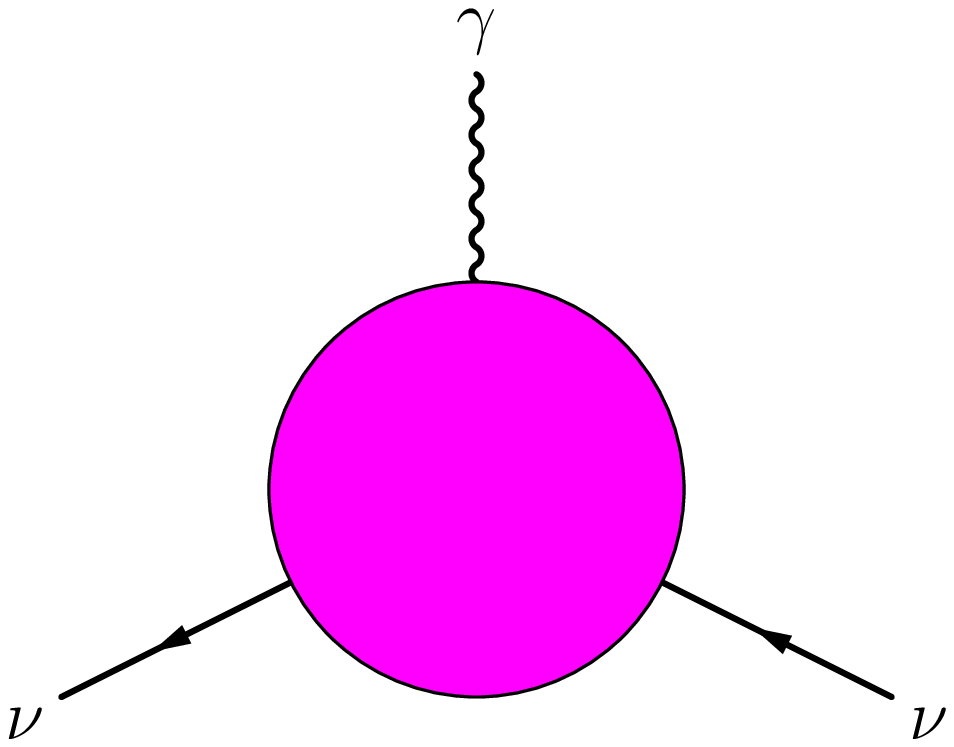}
  \caption{\label{fig0}Neutrino electromagnetic vertex function.}
  \label{nuvert}
\end{figure}

 Consider the matrix element of the electromagnetic current
between the fermion initial state $\psi (p)$ and final state $\psi (p')$
can be presented in the form
\begin{equation}\label{matr_elem}
<{\psi}(p^{\prime})|J_{\mu}^{EM}|\psi(p)>= {\bar
u}(p^{\prime})\Lambda_{\mu}(q,l)u(p),
\end{equation}
where $q_{\mu}=p^{\prime}_{\mu}-p_{\mu}$,
$l_{\mu}=p^{\prime}_{\mu}+p_{\mu}$. The matrix element between the
spinors of the electromagnetic vertex function $\Lambda_{\mu}(q,l)$
(Fig.~\ref{fig0}) should be a Lorentz vector (requirement of
Lorentz-covariance). In constructing the covariant operator
$\Lambda_{\mu}(q,l)$ we recall \footnote{A rather pedagogical
discussion on the electromagnetic form factors of spin-$\frac{1}{2}$
particles is given in \cite{NowPasRodEJP05_phys0402058}.} that there
are 16 linearly independent traceless (with the exception of the unit
matrix) matrices,
\begin{equation}\label{mat_1}
{\bf {\hat 1}}, \ \ \gamma _5,  \ \ \gamma _{\mu},  \ \ \gamma _5 \gamma
_{\mu},  \ \ \sigma_{\mu\nu},
\end{equation}
$\sigma_{\mu\nu}=\frac{i}{2}[\gamma_{\mu},\gamma_{\nu}]$. There are
in addition also the metric tensor $g_{\mu \nu} $, two vectors
$q_{\mu}$ and $l_{\mu}$, and the anti-symmetric tensor $\epsilon
_{\mu  \nu  \sigma  \gamma } $ that can be used.

There are three sets of operators from which $\Lambda_{\mu}(q,l)$ can be
formed. In the first set the Lorentz index is carried by the vectors $q_{\mu}$
and $l_\mu $,
\begin{equation}\label{set_1}
{\bf {\hat 1}}q_\mu , \ \ {\bf {\hat 1}}l_\mu, \ \ \gamma _5 q_\mu , \ \ \gamma
_5 l_\mu .
\end{equation}
There is another set of the same type,
\begin{equation}\label{set_2}
{\not q}q_\mu, \ \ {\not l} q_\mu, \ \ \gamma _5 q_\mu , \ \ \gamma_5 {\not
q}q_\mu , \ \ \gamma_5 {\not l}q_\mu ,\ \ \sigma_{\alpha \beta} q^\alpha
l^\beta q_\mu,
\end{equation}
and the correspondent terms obtained from (\ref{set_2}) by the substitution
$q_\mu \leftrightarrow \l_\mu$.

The second type of possible contributions to $\Lambda_{\mu}(q,l)$ can be
obtained from (\ref{mat_1}) with the demand that the Lorentz index is carried
by a matrix itself,
\begin{equation}\label{set_3}
\gamma_\mu , \ \ \gamma_5 \gamma_\mu ,  \ \ \sigma_{\mu \nu}q^\nu ,\ \
\sigma_{\mu \nu}l^\nu .
\end{equation}

The third type of terms from which the vertex $\Lambda_{\mu}(q,l)$ can be
constructed contains the tensor $\epsilon _{\mu  \nu  \sigma  \gamma } $,
\begin{equation}\label{set_4}
\epsilon _{\mu  \nu  \sigma  \gamma }\sigma^{\alpha \beta} q^{\nu}, \ \
\epsilon _{\mu  \nu  \sigma  \gamma }\sigma^{\alpha \beta} l^{\nu}, \ \
\epsilon _{\mu  \nu  \sigma  \gamma }\sigma^{\nu \beta}
q_{\beta}q^{\sigma}l^{\gamma}, \ \ \epsilon _{\mu  \nu  \sigma  \gamma
}\sigma^{\nu \beta} l_{\beta}q^{\sigma}l^{\gamma}, \ \ \epsilon _{\mu  \nu
\sigma  \gamma }\gamma^{\nu} q^{\sigma}l^{\gamma}{\bf {\hat 1}} , \ \ \epsilon
_{\mu  \nu \sigma  \gamma }\gamma^{\nu} q^{\sigma}l^{\gamma}\gamma _5.
\end{equation}

Taking all terms (\ref{set_1}), (\ref{set_2}), (\ref{set_3}) and (\ref{set_4})
together and using some $\gamma_\mu $ algebra (for details see
\cite{NowPasRodEJP05_phys0402058}), it is possible to arrive to the most
general expression for the vertex $\Lambda_{\mu}(q,l)$,
\begin{equation}
\Lambda_{\mu}(q,l)= f_1(q^2)q_\mu + f_2(q^2)q_\mu \gamma _5+ f_3(q^2)\gamma_\mu
+f_4(q^2)\gamma_\mu \gamma _5 + f_5 (q^2)\sigma_{\mu \nu}q^{\nu} + f_6
(q^2)\epsilon _{\mu  \nu \rho  \gamma }\sigma^{\rho \gamma}q^{\nu},
\end{equation}
where the only dependence on $q^2$ remains (because $p^2=p'^2=m^2$
where $m$ is the fermion mass and $l^2=4m^2-q^2)$.

From the natural requirement of current conservation (electromagnetic
gauge invariance) $\partial_{\mu}j^{\mu}=0$ it follows, that
\begin{equation}
f_1(q^2)q^2+f_2(q^2)q^2\gamma _5 + 2m f_4 (q^2) \gamma _5 =0,
\end{equation}
from which one gets
\begin{equation}
f_1(q^2)=0, \ \ \ \ f_2(q^2)q^2+ 2m f_4 (q^2) =0.
\end{equation}

 Therefore, in the most general case consistent with Lorentz and
electromagnetic gauge invariance, the vertex function is defined in terms of
four form factors \cite{NiePRD82,KayPRD82_KayPRD84},
\begin{multline}
\label{vert_func}\Lambda_{\mu}(q)=
f_{Q}(q^{2})\gamma_{\mu}+f_{M}(q^{2})i\sigma_{\mu\nu}q^{\nu}
+f_{E}(q^{2})\sigma_{\mu\nu}q^{\nu}\gamma_{5}+
f_{A}(q^{2})(q^{2}\gamma_{\mu}-q_{\mu}{\not q})\gamma_{5},
\end{multline}
where $f_{Q}(q^{2})$, $f_{M}(q^{2})$, $f_{E}(q^{2})$ and $f_{A}(q^{2})$ are
charge, dipole magnetic and electric, and anapole neutrino form factors.

Note that the form factors are Lorentz invariant and they depend only
on $q^2$, which is the only independent dynamical quantity which is
Lorentz invariant.

\begin{figure*}

  \centering
  \includegraphics{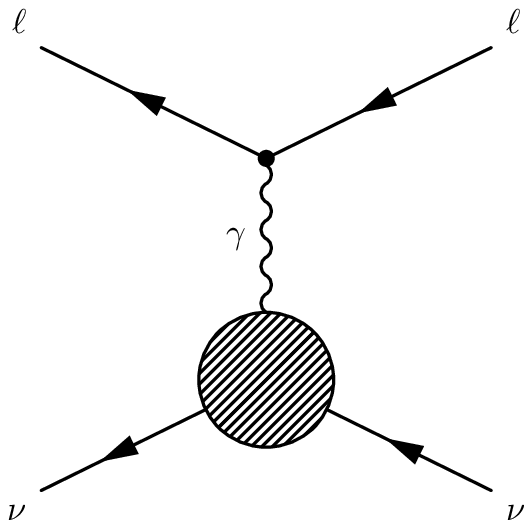}
  \caption{ \label{fig1} Contribution of the neutrino vertex function to neutrino elastic
  scattering on a charged lepton.}
  \label{examp}
\end{figure*}
The hermiticity of the electromagnetic current and the assumption of its
invariance under discreet symmetries transformations put certain
constraints on neutrino form factors, which are in general different
for the Dirac and Majorana cases. In the case of Dirac neutrinos, the
assumption of $CP$ invariance combined with the hermiticity of the
electromagnetic current $J_{\mu}^{EM}$ implies that the electric
dipole form factor vanishes. At zero momentum transfer only
$f_{Q}(0)$ and $f_{M}(0)$, which are called electric charge and
magnetic moments, contribute to the Hamiltonian, $H_{int}\sim
J_{\mu}^{EM}A^{\mu}$, which describes the neutrino interaction with
external electromagnetic field $A^{\mu}$. It is also possible to show
\cite{KayPRD82_KayPRD84,NiePRD82} that hermiticity by itself implies
that $f_Q$, $f_M$, and $f_A$ are real,
\begin{equation}
Im f_Q=Im f_M = Im f_A=0.
\end{equation}

In the case of Majorana neutrinos, regardless of whether $CP$-invariance is
violated or not, the charge, dipole magnetic and electric form factors vanish
\cite{NiePRD82,SchValPRD81},
\begin{equation}
f_Q=f_M =f_E=0.
\end{equation}
This means that in the case of  Majorana neutrinos only the anapole
moment can be non-vanishing among the electromagnetic moments (see
also \cite{KobOkuProThePhy72}). Note that it is possible to prove
\cite{NiePRD82} that the existence of a non vanishing magnetic moment
for a Majorana neutrino would bring a clear indication of $CPT$
nonconservation.

In general the matrix element of the electromagnetic current
(\ref{matr_elem}) can be considered between different neutrino
initial $\psi_{i} (p)$ and final $\psi_{j} (p')$ states  of different
masses, $p^2=m_i^2, \ p'^2=m_j^2$:
\begin{equation}
<{\psi}_j(p^{\prime})|J_{\mu}^{EM}|\psi_i(p)>= {\bar
u_j}(p^{\prime})\Lambda_{\mu}(q)u_i(p),
\end{equation}
and the correspondent vertex function is defined in the most general form
\begin{multline}
\label{Lambda} \Lambda_{\mu}(q)=
\Big(f_{Q}(q^{2})_{ij}+f_{A}(q^{2})_{ij}\gamma_{5}\Big)
(q^{2}\gamma_{\mu}-q_{\mu}{\not q})+f_{M}(q^{2})_{ij}i\sigma_{\mu\nu}q^{\nu}
+f_{E}(q^{2})_{ij}\sigma_{\mu\nu}q^{\nu}\gamma_{5}.
\end{multline}
The form factors are matrices in the space of neutrino mass
eigenstates \cite{ShrNP82}. General properties of the form factors in
the diagonal case ($i=j$) have been already discussed. In the
off-diagonal case ($i\neq j$) the hermiticity by itself does not
imply restrictions on the form factors of Dirac neutrinos. It is
possible to show \cite{NiePRD82} that if the assumption of $CP$
invariance is added, the form factors $f_{Q}(q^{2})$,
$f_{M}(q^{2})$, $f_{E}(q^{2})$ and $f_{A}(q^{2})$ should be
relatively real to each other (no relative phases exist). For the
Majorana neutrino, if $CP$ invariance holds, there could be either a
transition magnetic or a transition electric moment but not both.
The anapole form factor of a Majorana neutrino can be nonzero.

\subsection{Neutrino form factors in gauge models}
\label{form_fac_gauge_mod} \nopagebreak

 From the demand that the form factors at
zero momentum transfer, $q^2=0$, are elements of the scattering
matrix, it follows that  in any consistent theoretical model the form
factors in the matrix element (\ref{matr_elem}) should be gauge
independent and finite. Then, the form factors values at $q^{2}=0$
determine the static electromagnetic properties of the neutrino that
can be probed or measured in the direct interaction with external
electromagnetic fields. This is the case for charge, dipole magnetic
and electric neutrino form factors in the minimally extended Standard
Model . The neutrino anapole form factor is an exceptional case (see,
for instance, \cite{DegMarSirPRD89, DubKuzIJMPA98,RosPRD99}) and will
be discussed later in Section \ref{anap_mom}.

In non-Abelian gauge theories, the form factors in the matrix element
(\ref{matr_elem}) at nonzero momentum transfer, $q^2\neq 0$, can be not
invariant under the gauge transformation. This happens because in general the
off-shell photon's propagator is gauge dependent. Therefore, the one-photon
approximation is not enough to get physical quantities. In this case the form
factors in the matrix element (\ref{matr_elem}) cannot be directly measured in
an experiment with an external electromagnetic field, however they can
contribute to high order diagrams describing some processes that are accessible
for experimental observation (for a discussion on this item see, for instance,
\cite{BarGasLauNP72}). As an example, a diagram for a neutrino elastic
scattering on a charged lepton is shown in Fig.~\ref{fig1} where the hatched
plaque represents the neutrino electromagnetic vertex function that includes
contributions from the form factors.

It should be noted that  there is an important difference between the electromagnetic vertex function of massive and
massless neutrinos \cite{DvoStuPRD04_DvoStuJETP04}. For the case of a massless neutrino, the matrix element of the
electromagnetic current (\ref{matr_elem}) can be expressed in terms of only one Dirac form factor $f_D (q^2)$ (see, for
example, also \cite{NovRosSaTos08054177}),
\begin{equation}
  {\bar u}(p^{\prime})\Lambda_{\mu}(q)u(p)=
  f_{D}(q^{2}){\bar u}(p^{\prime})\gamma_{\mu}(1+\gamma_{5})u(p).
\end{equation}
It follows that  the electric charge and anapole form factors for a massless
neutrino are related to the Dirac form factor $f_{D}(q^{2})$ and hence to each
other
\begin{equation}
  f_{Q}(q^{2})=f_{D}(q^{2}), \quad f_{A}(q^{2})=f_{D}(q^{2})/q^{2}.
\end{equation}
\begin{figure}
\begin{center}
  \subfigure[]
  {\label{prverta}
    \includegraphics{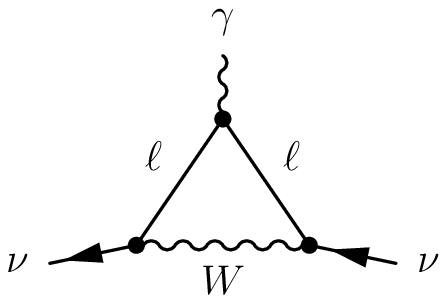}}
    \hspace{2cm}
  \subfigure[]
  {\label{prvertb}
  \includegraphics{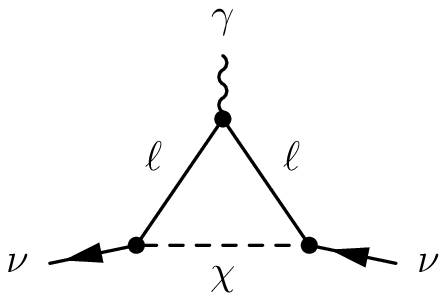}}
    \\
  \subfigure[]
  {\label{prvertc}
  \includegraphics{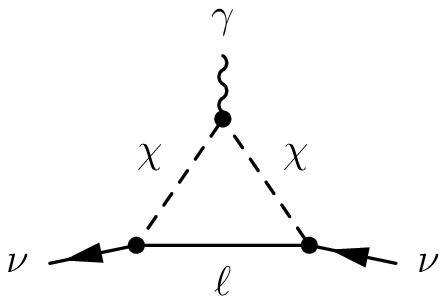}}
    \hspace{2cm}
  \subfigure[]
  {\label{prvertd}
  \includegraphics{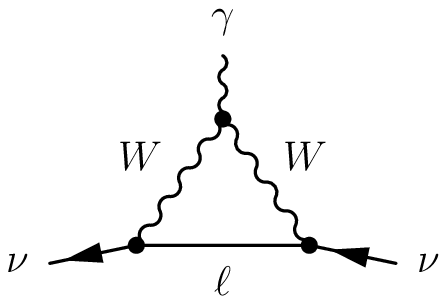}}
    \\
  \subfigure[]
  {\label{prverte}
  \includegraphics{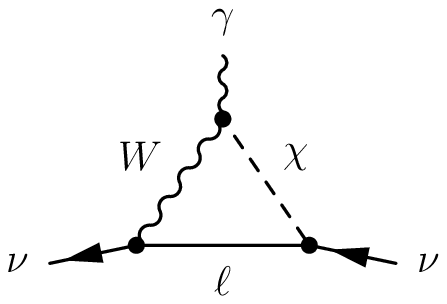}}
    \hspace{2cm}
  \subfigure[]
  {\label{prvertf}
  \includegraphics{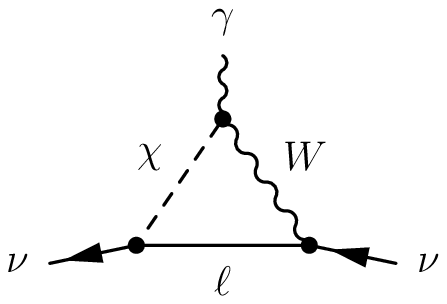}}
    \caption{\label{fig1f} \subref{prverta}-\subref{prvertf} Contributions to the neutrino vertex function
    from proper vertices ($\chi$ is the unphysical would-be charged scalar boson; the correspondent
    Feynman rules necessary for the massive neutrino electromagnetic vertex
    calculations can be found in  \cite{DvoStuPRD04_DvoStuJETP04}).}
\end{center}
\end{figure}

In the case of a massive neutrino, there is no such simple relation between
electric charge and anapole form factors since the $q_{\mu}{\not
q}\gamma_{5}$ term in the anapole part of the vertex function
(\ref{vert_func}) cannot be neglected.
\begin{figure}
  \centering
  \includegraphics[scale=.7]{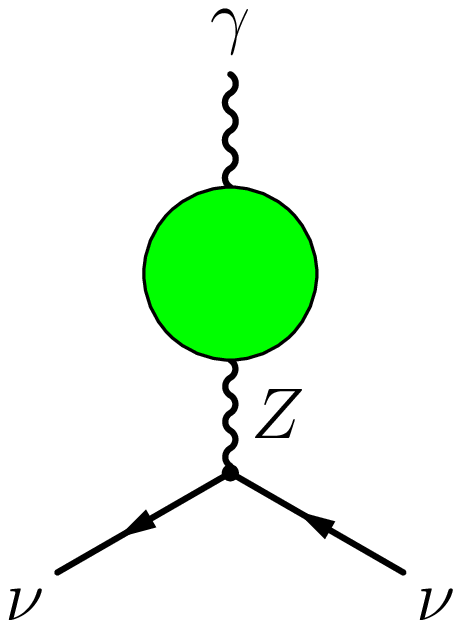}
  \caption{\label{fig4_0} Contributions to the neutrino vertex function of $\gamma-Z$ self-energy
  diagrams.}
  \label{gZcontr}
\end{figure}

\begin{figure}
\begin{center}
  \subfigure[]
  {\label{gZverta}
    \includegraphics{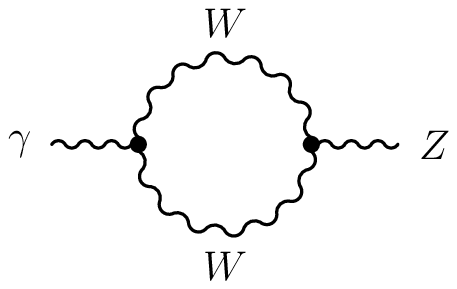}}
    \hspace{2cm}
  \subfigure[]
  {\label{gZvertb}
  \includegraphics{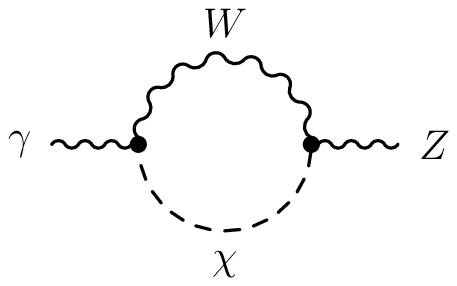}}
    \\
  \subfigure[]
  {\label{gZvertc}
  \includegraphics{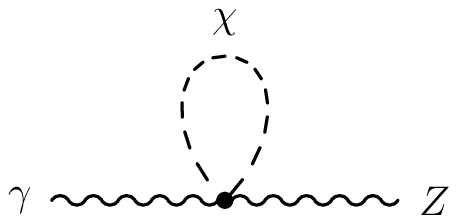}}
    \hspace{2cm}
  \subfigure[]
  {\label{gZvertd}
  \includegraphics{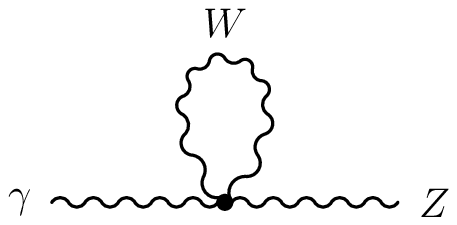}}
    \\
  \subfigure[]
  {\label{gZverte}
  \includegraphics{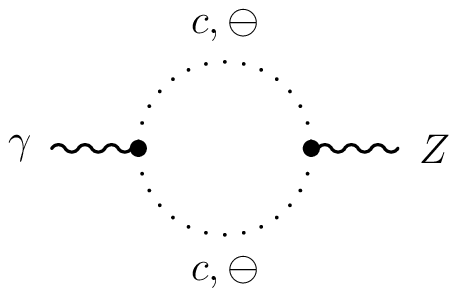}}
    \hspace{2cm}
  \subfigure[]
  {\label{gZvertf}
  \includegraphics{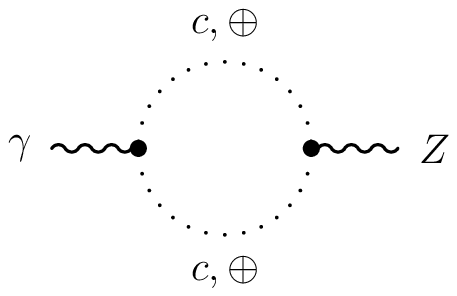}}
    \\
  \subfigure[]
  {\label{gZvertg}
  \includegraphics{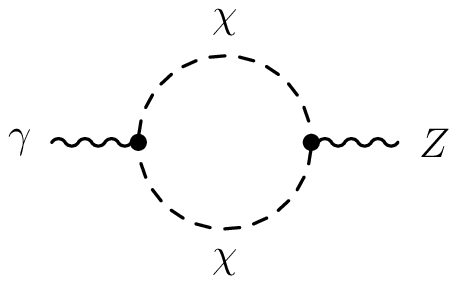}}
    \hspace{2cm}
  \subfigure[]
  {\label{gZverth}
  \includegraphics{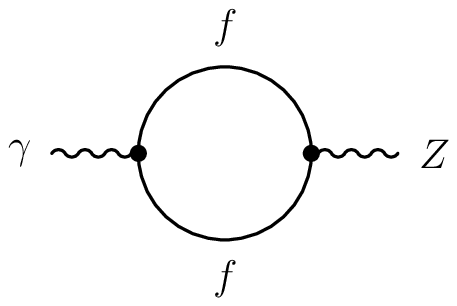}}
    \caption{\label{fig2}\subref{gZverta}-\subref{gZverth}
    $\gamma-Z$ self-energy diagrams. $f$ denotes
    the electron, muon, $\tau$-lepton and $u$, $c$, $t$, $d$, $s$ and $b$
    quarks (the charge of ghosts is indicated by the symbols $\oplus$ and $\ominus$).}
\end{center}
\end{figure}
 Consider \cite{DvoStuPRD04_DvoStuJETP04} the full set of one-loop
Feynman diagrams contributing to the Dirac massive neutrino
electromagnetic vertex function in the framework of the Standard
Model  supplied with the $SU(2)$-singlet right-handed neutrino in the
general $R_{\xi}$ gauge. The vertex function $\Lambda_\mu (q)$, in
the one-loop approach, contains contributions given by two types of
diagrams: the proper vertices (Fig.~\ref{fig1f}) and the $\gamma - Z$
self-energy diagrams (Fig.~\ref{fig4_0}).

The direct calculation \cite{DvoStuPRD04_DvoStuJETP04} of the massive neutrino electromagnetic vertex function, taking
into account all of the diagrams (Fig.~\ref{fig1f} and Fig.~\ref{fig2}), reveals that each of the Feynman diagrams gives
nonzero contribution to the term proportional to $\gamma_{\mu}\gamma_{5}$. These contributions are not
vanishing even at $q^{2}=0$. Therefore in addition to the usual four terms in (\ref{vert_func}) an extra term
proportional to $\gamma_{\mu}\gamma_{5}$ appears and the corresponding additional form factor $f_5 (q^2)$ can be
introduced. This problem is related to the decomposition of the massive neutrino electromagnetic vertex function. The
calculation of the contributions of the proper vertex diagrams (Fig.~\ref{fig1f}) and $\gamma-Z$ self-energy diagrams
(Fig.~\ref{fig4_0}) for arbitrary gauge fixing parameter $\alpha=\frac{1}{\xi}$ and arbitrary mass parameter
$a=\frac{m_l^2}{M_W^2}$ shows that at least in the zeroth and first orders of the expansion over the small neutrino mass
parameter $b=\big(\frac{m_\nu}{M_W}\big)^2$ the corresponding ``charge'' $\phi=f_5 (q^2=0)$ is zero. The cancellation of
contributions from the proper vertex and self-energy diagrams to the form factor $f_5 (q^2)$ at $q^2\neq0$,
\begin{equation}
  f_{5}(q^{2})=f_{5}^{(\gamma-Z)}(q^{2})+
  f_{5}^{(\mathrm{prop. vert.})}(q^{2})=0,
\end{equation}
was also shown \cite{DvoStuPRD04_DvoStuJETP04} for arbitrary mass parameters
$a$ and $b$ in the `t Hooft-Feynman gauge $\alpha=1$.

For a {\it massive} Dirac neutrino, by performing the direct
calculations \cite{DvoStuPRD04_DvoStuJETP04} of the complete set of
one-loop diagrams it is established that the neutrino vertex function
consists of only three electromagnetic form factors (in the case of a
model with $CP$ conservation). Closed integral expressions are found
for electric, magnetic, and anapole form factors of a {\it massive}
neutrino. On this basis, the electric charge (the value of the
electric form factor at zero momentum transfer), magnetic moment, and
anapole moment of a {\it massive} neutrino have been derived. It has
been shown by means of direct calculations for the case of a {\it
massive} neutrino  that the electric charge is independent of the
gauge parameters and is equal to zero, the magnetic moment is finite
and does not depend on the choice of gauge.

\subsection{Neutrino electric charge}

It is usually believed  \cite{BerRudPR63} that the neutrino electric
charge is zero. This is often thought to be attributed to gauge
invariance and anomaly cancellation constraints imposed in the
Standard Model . In the Standard Model  of $SU(2)_L \times U(1)_Y$
electroweak interactions it is possible to get \cite{BabMohPRD89,
FooJosLewVolMPL90_FooLewVolJPG93} a general proof that neutrinos are
electrically neutral. The electric charges of particles in this model
are related to the $SU(2)_L$ and $U(1)_Y$ eigenvalues by (see
Table~\ref{003})
\begin{equation}
Q=I_{3} + \frac{Y}{2}.
\end{equation}
 In the Standard Model  without right-handed neutrinos
$\nu_{R}$ the triangle anomalies cancellation constraints (the
requirement of renormalizability) lead to certain relations among
particles hypercharges $Y$, that are enough to fix all $Y$, so that
hypercharges, and consequently electric charges, are quantized
\cite{FooJosLewVolMPL90_FooLewVolJPG93}.
In this case, neutrinos are electrically neutral.


The direct calculation of the neutrino charge in the Standard Model
under the assumption of a vanishing neutrino mass in different gauges
and with use of different methods is presented in
\cite{BarGasLauNP72, BegMarRudPRD78_MarSirPRD80_SakPRD81,
LucRosZepPRD84_85, Car_RosBerVid_ZepEPJC00}. For the flavor massive
Dirac neutrino the one-loop contributions to the charge, in the
context of the minimal extension of the Standard Model  within the
general $R_{\xi}$ gauge, were considered in
\cite{DvoStuPRD04_DvoStuJETP04}. By these direct calculations within
the mentioned above theoretical frameworks it is proven that at least
at one-loop level approximation neutrino electric charge is gauge
independent and vanish.

However, if the neutrino has a mass, the statement that a neutrino
electric charge is zero is not so evident as it meets the eye. It is
not entirely assured that the electric charge should be quantized (see
\cite{Raf_book96_RafPR99} and references therein). We recall here
that the problem of charge quantization has been always a mystery
within quantum electrodynamics \cite{DavCamBaiPRD91}. The absence
of an algebraic quantization of the charge eigenvalues in
electrodynamics led to the proposal \cite{DirPRSL31} of a possible
topological explanation leading to magnetic monopoles.

The strict requirements for charge quantization may also disappear in
extensions of the standard  $SU(2)_L \times U(1)_Y$ electroweak interaction
models if right-handed neutrinos $\nu_R$ with $Y\neq0$ are included. In
this case the uniqueness of particles hypercharges $Y$ is lost (hypercharges
are no more fixed) and in the absence of hypercharge quantization the electric
charge gets ``dequantized'' \cite{FooJosLewVolMPL90_FooLewVolJPG93}. As a
result, neutrinos may become electrically millicharged particles.

  In
general, the situation with charge quantization is different for
Dirac and Majorana neutrinos. As it was shown in \cite{BabMohPRD89},
charge dequantization for Dirac neutrinos occurs in the extended
Standard Model  with right-handed neutrinos $\nu_R$ and also in a
wide class of models that contain an explicit $U(1)$ symmetry. On the
contrary, if the neutrino is a Majorana particle, the arbitrariness
of hypercharges in this kind of models is lost, leading to electric
charge quantization and hence to neutrino neutrality
\cite{BabMohPRD89}.

Finally, while there are other Standard Model  extensions
(superstrings, GUT's etc) that provide enforcing of charge
quantization, there are also models (for instance, with a ``mirror
sector'' \cite{HolPLB86}) that predict the existence of new particles
of arbitrary mass and small (unquantized) electric charge, in which
neutrino can be a millicharged particle.

The most severe experimental constraints on the electric charge of the neutrino
\begin{equation} q_\nu\leq \times 10^{-21} e,
\end{equation}
are obtained assuming electric charge conservation in neutron
beta decay $n\rightarrow p+e^-+\nu_e$,  from the neutrality of matter (from
the measurements of the total charge $q_{p}+q_{e}$)
\cite{MarMorPLB84} and from the neutrality of the neutron itself
\cite{BauGahKalMamPRD88}. Constraints from direct accelerator
searches, charged leptons anomalous magnetic moments, stellar
astrophysics and primordial nucleosynthesis are in general less
stringent \cite{DavCamBaiPRD91, BabVolPRD92}:
\begin{equation} q_\nu\leq \times 10^{-6} - 10^{-17} e.
\end{equation}
A detailed discussion of different constraints on the neutrino
electric charge can be found in \cite{Raf_book96_RafPR99}.

\subsection{Neutrino charge radius}\label{NCR}

Even if the electric charge of a neutrino is vanishing, the electric form factor
$f_Q(q^2)$ can still contain nontrivial information about neutrino static
properties. A neutral particle can be characterized by a superposition of two
charge distributions of opposite signs so that the particle's form
factor $f_Q(q^2)$ can be non zero for $q^2\neq0$. The application of
this notion to neutrinos has a long-standing history and is puzzling.
In the case of an electrically neutral neutrino, one usually
introduces the mean charge radius, which is determined by the second
term in the expansion of the neutrino charge form factor $f_Q (q^2)$
in series of powers of $q^2$,
\begin{equation}\label{nu_cha_rad}
f_{Q}(q^2)=f_{Q}(0)+q^2\frac{df_{Q}(q^2)}{dq^2}_{\mid q^2=0}+\ ... \
\ \ .
\end{equation}
The definition of the neutrino charge radius follows an analogy
 with the elastic
electron scattering off a static spherically symmetric charged
distribution of density $\rho(r)$ ($r=|{\bf x}|$),  for which the
differential cross section is determined \cite{Ng_prepr92_NC94,
HalMar1984, GreReiQED03} by the point particle cross section
$\frac{d\sigma}{d\Omega}_{\mid_{point}}$ ,
\begin{equation}
\frac{d\sigma}{d\Omega}=\frac{d\sigma}{d\Omega}_{\mid_{point}} |f(q^2)|^2 ,
\end{equation}
where the correspondent form factor $f(q^2)$ in the so-called {\it
Breit frame}, in which $q_0=0$, can be expressed as
\begin{equation}
f(q^2)=\int \rho (r)e^{i{\bf q}{\bf x}}d^3x=4\pi \int dr r^2
\rho(r)\frac {\sin (qr)}{qr},
\end{equation}
here $q=|{\bf q}|$. Thus, one has
\begin{equation}
\frac{df_{Q}}{dq^2}=\int \rho(r)\frac{qr\cos (qr) - \sin
(qr)}{2q^{3/2}r}d^3 x.
\end{equation}
In the case of small $q$, we have $\lim _{q^2\rightarrow 0} \frac{qr\cos
(qr) - \sin (qr)}{2q^{3/2}r}= -\frac{r^2}{6}$ and
\begin{equation}
f(q^2)=1- |{\bf q}|^2\frac{\langle r^2\rangle}{6}+\ ... \ .
\end{equation}
Therefore, the neutrino charge radius (in fact, it is the charge
radius squared) is usually defined by
\begin{equation}
{\langle r_{\nu}^2\rangle}=-{6}\frac{df_{Q}(q^2)}{dq^2}{\mid_{ q^2=0}}.
\end{equation}
Since the neutrino charge density is not a positively defined quantity,
${\langle r_{\nu}^2\rangle}$ can be negative.

\begin{figure}
  \centering
  \includegraphics[scale=1]{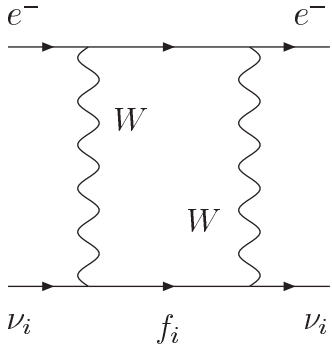}
  \caption{\label{Diagramm}Contribution of the W box diagram to the
  scattering process $\nu_l+l'\rightarrow\nu_l+l'$.}
  \end{figure}

Just in one of the first studies \cite{BarGasLauNP72}, it was claimed
that in the Standard Model  and in the unitary gauge the neutrino
charge radius is ultraviolet-divergent and so it is not a physical
quantity. A recent direct one-loop calculation
\cite{DvoStuPRD04_DvoStuJETP04} of proper vertices (Fig.~\ref{fig1f})
and $\gamma - Z$ self-energy (Fig.~\ref{fig4_0}) contributions to the
neutrino charge radius performed in a general $R_\xi $ gauge  for a
massive Dirac neutrino gave also a divergent result. However, it was
shown \cite{LeePRD72}, using the unitary gauge, that by including in
addition to the usual terms also contributions from diagrams of the
neutrino-lepton neutral current scattering ($Z$ boson diagrams), it
is possible to obtain for the neutrino charge radius a gauge
dependent but finite quantity. Later on it was also shown
\cite{LeeShrPRD77} that in order to define the neutrino charge radius
as a physical quantity one has also to consider box diagrams (see
Fig.~\ref{Diagramm}),
which contribute to the scattering process $\nu_l+l'\rightarrow
\nu_l+l'$, and that in combination with contributions from the proper
diagrams it is possible to obtain a finite and gauge-independent
value for the neutrino charge radius. In this way, the neutrino
electroweak radius was introduced \cite{LucRosZepPRD84_85} and
an additional set of diagrams that give contribution to its value was
discussed in \cite{DegMarSirPRD89}. Finally, in a series of recent
papers \cite{BerCabPapVidPRD00_BerPapVidPRL02_NPB04} the neutrino
electroweak radius as a physical observable has been introduced.
In the correspondent calculations, performed in the one-loop
approximation including additional terms from the $\gamma-Z$ boson
mixing and the box diagrams involving $W$ and $Z$ bosons, the
following gauge-invariant result for the neutrino charge radius have
been obtained:
\begin{equation}
{\langle
r_{\nu_i}^2\rangle}=\frac{G_F}{4\sqrt{2}\pi^2}\Big[3-2\log\big(\frac{m_i^2}
{m^2_W}\big)\Big]
\end{equation}
(where $m_W$ and $m_i$ are the $W$ boson and lepton masses,
$i=e,\mu,\tau$). This result, however, revived the discussion
\cite{FujShrPRD04, PapBerBisVidEPJ04_NPB05}  on the definition of
the neutrino charge radius. Numerically, for the electron neutrino electroweak
radius it yields
 \cite{BerCabPapVidPRD00_BerPapVidPRL02_NPB04}
\begin{equation}
{\langle r_{\nu_e}^2\rangle}=4 \times 10^{-33} \, \text{cm}^2,
\label{r01}
\end{equation}
which is very close to the numerical estimation obtained much earlier in
\cite{LucRosZepPRD84_85}.

Note that the neutrino charge radius can be considered as an
effective scale of the particle's ``size'', which should influence
physical processes such as, for instance, neutrino scattering off
electron (the differential cross section is given in
Eq.~(\ref{d_sigma_SM}) below). To incorporate the neutrino charge
radius contribution in the cross section, the following substitution
\cite{GraGriPLB86_VogEngPRD89_HagMasHaiKimZP94} can be used:
\begin {equation}\label{g_V_ch_rad}
g_V\rightarrow \frac{1}{2}+2\sin^2 \theta_W + \frac{2}{3}m^2_W
{\langle r_{\nu_e}^2\rangle}\sin^2 \theta_{W}.
\end{equation}

It is interesting to compare the theoretical results for the neutrino charge
radius with some available experimental bounds
\cite{GriMasMPL87_PRD89_AllPRD91_MouPulRalPLB92}:
from primordial nucleosynthesis,
\begin{equation}
\langle r_{\nu_e}^2\rangle <7 \times 10^{-33} \, \text{cm}^2,
\label{r02}
\end{equation}
from SN 1987A,
\begin {equation}
\langle r_{\nu_e}^2\rangle <2 \times 10^{-33} \, \text{cm}^2,
\label{r03}
\end{equation}
from neutrino neutral-current reactions,
\begin {equation}
-2.74\times 10^{-32} \, \text{cm}^2 <\langle r_{\nu_e}^2\rangle<4.88 \times 10^{-33} \, \text{cm}^2,
\label{r04}
\end{equation}
from solar experiments (Kamiokande II),
\begin{equation}
\langle r_{\nu_e}^2\rangle <2 \times 10^{-32} \, \text{cm}^2.
\label{r05}
\end{equation}
 Recently, a
new constraint have been obtained \cite{BarMirRas_07074319} from a new
evaluation of the weak mixing angle $\sin^2 \theta_W$ by a combined fit of all
electron neutrino elastic scattering data,
\begin {equation}
-1.3\times 10^{-32} \, \text{cm}^2 <\langle r_{\nu_e}^2\rangle<3.32 \times 10^{-32} \, \text{cm}^2.
\label{r06}
\end{equation}

Comparing the theoretical value in Eq.~(\ref{r01})
with the experimental limits in Eqs.~(\ref{r02})--(\ref{r06}),
one can see that they differ at most by one
order of magnitude.
Therefore, one may expect that the experimental
accuracy will soon reach the value needed to probe the neutrino
effective charge radius.

It is obvious that the effects of new physics beyond the Standard
Model can also contribute to the neutrino charge radius. In this
concern, a recent work \cite{NovRosSaTos08054177} should be
mentioned, where the anomalous $WW\gamma$ vertex contribution to the
neutrino effective charge radius has been studied and the value for
the correspondent additional contribution of
\begin {equation}
|\langle r_{\nu_e}^2\rangle|\leq 10^{-34} \, \text{cm}^2
\end{equation}
was obtained. Note that this is only one order of magnitude lower
than the expected value of the charge radius in the Standard Model .

A detailed discussion on possibility to constrain the $\nu_{\tau}$
and $\nu_{\mu}$ charge radii from astrophysical and cosmological
observations and from the terrestrial experiments can be found in
\cite{HirNarResPRD03}.

\subsection{Neutrino anapole moment}
\label{anap_mom}

 The anapole form factor is the most mysterious and
ambiguous among the neutrino form factors. The notion of an
anapole moment for a Dirac particle was introduced in \cite{ZelSPJETP57} for
a $T$-invariant interaction which, however, is not invariant under
$P$ and $C$ transformations.

To understand the physical meaning of the anapole form factor, as
well as the meaning of other form factors, it is instructive to
couple the correspondent term of the current to an external
electromagnetic field (given by a potential $A_\mu$), to derive
the corresponding Dirac equation of motion for a neutrino field $\psi$ of mass $m$,
and
finally to obtain the interaction energy with a static electromagnetic
field in the nonrelativistic limit. From this perspective, it is
straightforward to understand that the charge form factor $f_Q({q^2})$ at
$q^2=0$ is the electric charge, $f_Q({q^2})=Q$ \cite{GreReiQED03,
ItzZub_book80}. Similarly, $\mu=f_M(0)$ and $\epsilon=if_E(0)$ are
the dipole magnetic and electric moments, respectively. In the
nonrelativistic approximation, from the anapole term of the neutrino
current (see Eqs.~(\ref{matr_elem}) and (\ref{vert_func})), it is possible
to obtain \cite{DubKuzIJMPA98} the interaction energy
\begin{equation}
{\it H}_{int}\propto f_A (0)\big( {{\bm \sigma}}\cdot  curl \ {\bf B} - {\dot
{\bf E}}\big),
\end{equation}
which corresponds to a $T$-invariant toroidal (anapole) interaction of the
neutrino that does not conserve the $P$ and $C$ parities. This interaction defines
the axial-vector interaction with an external electromagnetic field. The
poloidal currents on a torus can be considered as a geometrical model for the
anapole \cite{BukDubKuzPAN98}.

The direct calculation \cite{DvoStuPRD04_DvoStuJETP04} of the
corresponding vertex contributions (the diagrams in Figs.~\ref{fig1f}
and \ref{fig4_0}) to the massive Dirac neutrino anapole moment gives
an infinite and gauge-dependent result.  The same behavior of the
charged leptons anapole moments has been demonstrated in
\cite{CzyKolZraKhrCJP88}. Note that even in the case of massless
neutrinos this is not a trivial task to obtain the anapole moment as
a physical quantity (see section \ref{NCR}). Here we also recall that
for the massless case the neutrino anapole moment is connected to the
derivative of the electric charge form factor $f_Q(q^2)$ with respect
to $q^2$ at $q^2=0$, which is the charge radius, by the relation
\begin{equation}
a_{\nu}=f_A(0)=\frac{1}{6}\langle r^2_{\nu}\rangle.
\end{equation}
This relation is obtained within the Standard Model  and in general
it is model dependent. As it has been shown in
\cite{NovRosSaTos08054177}, the same relation between a massless
Dirac neutrino anapole moment and charge radius  in the context of an
effective Yang-Mills theory which includes a general
$SU_L(2)$-invariant Lorentz tensor structure of nonrenormalizable
type for the $WW\gamma$ vertex is also fulfilled. This relation is
obtained within the Standard Model  and in general it is model
dependent. As it has been shown in \cite{NovRosSaTos08054177}, the
same relation between a massless Dirac neutrino anapole moment and
charge radius  in the context of an effective Yang-Mills theory which
includes a general $SU_L(2)$-invariant Lorentz tensor structure of
nonrenormalizable type for the $WW\gamma$ vertex is also fulfilled.

As it was discussed in \cite{DubKuzIJMPA98}, since the anapole form
factor does not correspond to a multipole distribution, the anapole
moment has a quite intricate classical analog. A more convenient and
transparent characteristic, the toroidal dipole moment, was proposed
instead for the description of $T$-invariant interactions. In this
case, the electromagnetic vertex of a neutrino can be rewritten in an
alternative multipole (toroidal) parameterization. In some sense this
parameterization has a more transparent and clear physical
interpretation, because it provides a one-to-one correspondence
between the multipole moments and the corresponding form factors. In
one-loop calculations \cite{DubKuzIJMPA98} of the toroidal (and
anapole) moment of a massive and massless Majorana neutrino (the
diagrams in Figs.~\ref{fig1f} and \ref{fig4_0} contribute) it was
shown that its value does not depend significantly on the neutrino
mass (through the parameters $\frac{m^2_{\nu_i}}{m^2_{W}}$) and is of
the order of
\begin{equation}
\tau_{\nu}=f_{A}(q^2)_{q^2=0}\propto 10^{-33}-10^{-34} \,
\text{cm}^2,
\end{equation}
depending on the values of the quark masses that propagate in the
loop diagrams of Fig.~\ref{fig2}.

Note that the anapole form factors can contribute to the neutrino
vertex function in both the diagonal and and off-diagonal cases. The
anapole and the toroidal parameterizations coincide in the case when
the current is diagonal on the neutrino initial and final masses.

To conclude this section, it should be mentioned that the anapole interactions
of a Majorana as well as a Dirac neutrino are expected to contribute to the total
cross section of neutrino elastic scattering off electrons, quarks and nuclei.
Due to the fact that the anapole interaction conserves helicity, its
contribution to the cross section is similar to that of the neutrino charge
radius. In principle, these contributions can be probed in low-energy
scattering experiments in the future.

\subsection{Neutrino magnetic and electric dipole moments}

The neutrino dipole magnetic and electric form factors (and the
corresponding magnetic and electric dipole moments) are theoretically
the most well studied and understood among the form factors. They
also attract a reasonable attention from experimentalists, although
the neutrino magnetic moment predicted in the Standard Model  is
proportional to the neutrino mass and therefore is many orders of
magnitude smaller than the present experimental limits obtained in
terrestrial experiments.

As it has been mentioned before, the first calculations of the neutrino dipole
moments within a minimal extension of the Weinberg-Salam model (with nonzero
neutrino mass and with a right-handed neutrino $\nu_{R}$) were performed
\cite{MarSanPLB77, LeeShrPRD77, FujShrPRL80, PetSNP77} by evaluating the
radiative diagrams (a) and (d) shown in Fig.~\ref{fig1f}. The explicit
evaluation of the one-loop contributions to the neutrino dipole moments in the
leading approximation over the small parameters
$b_i=\frac{m_{i}^{2}}{M_{W}^{2}}$ (here $m_i$ are the neutrino masses,
$i=1,2,3$), that in addition exactly accounts for the dependence on the small
parameters $a_l= \frac{m_{l}^{2}}{M_{W}^{2}}$ ($l=e,\ \mu , \ \tau$), yields,
for Dirac neutrinos \cite{PalWolPRD82,PetSNP77},
\begin{equation}\label{m_e_mom_i_j}
\begin{array}{c}
    \mu^{D}_{ij}\\
    \epsilon^{D}_{ij}
  \end{array}\Bigg  \}=\frac{e G_F m_{i}}{8\sqrt {2} \pi ^2}
  \Big(1\pm \frac{m_j}{m_i}\Big)\sum_{l=\ e, \ \mu, \
  \tau}f(a_l)U_{lj}U^{\ast}_{li},
\end{equation}
where
\begin{equation}\label{f_a_l}
f(a_l)=\frac{3}{4}\Big[1+\frac{1}{1-a_l}-2\frac{a_l}{(1-a_l)^2}-2\frac{a_l^2}{(1-a_l)^3}
\ln a_l\Big].
\end{equation}
All the charged lepton parameters $a_l$ are small. In the limit
$a_l\ll 1$ one has
\begin{equation}\label{f_appr}
f(a_l)\approx \frac{3}{2}\Big(1 - \frac{1}{2}a_l\Big).
\end{equation}
From Eqs.~(\ref{m_e_mom_i_j}) and (\ref{f_appr}), the diagonal magnetic moment
of Dirac neutrinos are given by
\cite{MarSanPLB77, LeeShrPRD77, FujShrPRL80}
\begin{equation}\label{nu_mu_D_ii}
    \mu^{D}_{ii}=\frac{3e G_F m_{i}}{8\sqrt {2} \pi ^2}
  \Big(1 - \frac{1}{2} \sum_{l=\ e, \mu, \tau}a_l\mid U_{li}\mid^{2}\Big).
\end{equation}

Several important features of this result should be mentioned. The
magnetic moment of a Dirac neutrino is proportional to the neutrino
mass and for a massless Dirac neutrino in the Standard Model  (in the
absence of right-handed charged currents) the magnetic moment is
zero. The magnetic moment of a massive Dirac neutrino, at the leading
order in $a_l$, is independent of the neutrino mixing matrix and also
independent of the values of the charged lepton masses. The numerical
value of the Dirac neutrino magnetic moment, as it follows from
Eq.~(\ref{nu_mu_D_ii}), is
\begin{equation}\label{mu_3_10_19}
    \mu^{D}_{ii}\approx 3.2\times 10^{-19}
  \Big(\frac{m_i}{1 \, \text{eV}}\Big) \mu_{B}.
\end{equation}
taking into account the existing constraints on neutrino masses,
this value is several orders of magnitude smaller than the present experimental
limits (see Section~\ref{sec3.6} for a
further discussion on the experimental constraints on magnetic moments).

From Eq.~(\ref{m_e_mom_i_j}) it can be clearly seen that in the
Standard Model  the static (diagonal) electric dipole moment of a
Dirac neutrino vanishes, $\epsilon^{D}_{ii}= 0$. Dirac neutrinos may
have nonzero diagonal electric moments in theories where $CP$
invariance is violated. For a Majorana neutrino both the diagonal
magnetic and electric moments are zero,
$\mu^{M}_{ii}=\epsilon^{M}_{ii}=0$.

Let us discuss the neutrino transition moments, which are
given by (\ref{m_e_mom_i_j}) for $i\neq j$.
If we again use the first two terms in the expansion (\ref{f_appr}) of the
function $f(a_l)$ and we insert the leading term in Eq.~(\ref{m_e_mom_i_j}), we get a
vanishing result. This happens because the neutrino mixing matrix $U_{li}$ is
unitary and its rows and columns are orthogonal vectors. Therefore, the
nonvanishing contribution comes only from the second term in the expansion of
$f(a_l)$, which contains the additional small factor $a_l=
\frac{m_{l}^{2}}{M_{W}^{2}}$.
For the Dirac neutrino magnetic and electric
transition moments, it is possible to obtain, rearranging the terms in
Eq.~(\ref{m_e_mom_i_j}),
\begin{equation}\label{m_e_mom_i__not_j}
\begin{array}{c}
    \mu^{D}_{ij}\\
    \epsilon^{D}_{ij}
  \end{array}\Bigg  \}=\frac{3e G_F m_{i}}{32\sqrt {2} \pi ^2}
  \Big(1\pm \frac{m_j}{m_i}\Big)
  \sum_{l=\ e, \ \mu, \ \tau}\Big(\frac{m_l}{m_W}\Big)^2U_{lj}U^{\ast}_{li}.
\end{equation}
Thus, they are reasonably suppressed with respect to the Dirac neutrino magnetic moment
(\ref{nu_mu_D_ii}) in the diagonal case ($i=j$). For convenience, numerically
the Dirac transition moments can be expressed as follows (see, for instance,
\cite{Raf_book96_RafPR99})
\begin{equation}
\begin{array}{c}
    \mu^D_{ij}\\
    \epsilon^D_{ij}
  \end{array}\Bigg  \}=4\times 10^{-23} \mu_{B}\Big(\frac{m_i\pm m_j}{1\ \text{eV}}\Big)
  \sum_{l=\ e, \ \mu, \ \tau}\Big(\frac{m_l}{m_\tau}\Big)^2U_{lj}U^{\ast}_{li}.
\end{equation}
The above-mentioned suppression by a factor of at least $a_l=
\frac{m_{l}^{2}}{M_{W}^{2}}$ is due to the well-known Glashow-Iliopoulos-Maiani
cancellation ($GIM$ mechanism) \cite{GlaIliMaiPRD70}. Note that in the diagonal
case ($i=j$) the leading term in the expression for the Dirac neutrino magnetic
moment is not zero, because the sum in Eq.~(\ref{m_e_mom_i_j}) is equal to unity.

Also Majorana neutrinos can  have nonvanishing transition magnetic and
electric moments. In this case, additional Feynman diagrams should be
considered, which also contribute to the dipole moments (for a
detailed discussion see, for instance, \cite{ShrNP82,PalWolPRD82}).
It is possible to show that, depending on the relative $CP$ phase of
the two neutrinos $\nu_i$ and $\nu_j$, one of the two options is
realized: $\mu^{M}_{ij}=2\mu^{D}_{ij}$ and $\epsilon^{M}_{ij}=0$, or
$\mu^{M}_{ij}=0$ and $\epsilon^{M}_{ij}=2\epsilon^{D}_{ij}$.

In recent studies, the value of a massive Dirac neutrino diagonal
magnetic moment was obtained in a one-loop approximation in the
Standard Model , accounting for the dependence on the neutrino mass
parameter $b_i=\frac{m_{i}^{2}}{M_{W}^{2}}$
\cite{Car_RosBerVid_ZepEPJC00} and accounting for the exact
dependence on both mass parameters  $b_i$ and  $a_l=
\frac{m_{l}^{2}}{M_{W}^{2}}$ \cite{DvoStuPRD04_DvoStuJETP04}. The
calculations of the neutrino magnetic moment which take into account
exactly the dependence on the masses of all particles can be useful
in the case of a heavy neutrino with a mass compared or even
exceeding the values of other known particle masses. Note that the
$LEP$ data require that the number of light neutrinos coupled to the
$Z$ boson is exactly three. Therefore, any additional active neutrino
must be heavier than $\frac{M_Z}{2}$. In general, such a possibility
is not excluded. That is the reason to consider the neutrino magnetic
moment for various ranges of particles masses. The value of the
neutrino magnetic moment for a light neutrino with mass $m_\nu\ll
m_\ell\ll M_W$ that was obtained in
\cite{Car_RosBerVid_ZepEPJC00,DvoStuPRD04_DvoStuJETP04},
\begin{equation}
    \mu_{\nu}=
    {\frac{eG_{F}}{4\pi^{2}\sqrt{2}}}m_{\nu}
    {\frac{3}{4(1-a_l)^{3}}}
    (2-7a_l+6a_l^{2}-2a_l^{2}\ln a_l-a_l^{3}),
  \end{equation}
reproduces the main term in Eq.~(\ref{nu_mu_D_ii}), i.e. the result derived in
\cite{MarSanPLB77, LeeShrPRD77, FujShrPRL80}.
The authors of Ref.~\cite{DvoStuPRD04_DvoStuJETP04} obtained
for an intermediate values of the neutrino mass,
$m_\ell\ll m_\nu\ll M_W$,
   \begin{equation}
    \mu_{\nu}={\frac{3eG_{F}}{8\pi^{2}\sqrt{2}}}m_{\nu}
    \left\{
    1+{\frac{5}{18}}b
    \right\},
  \end{equation}
and for a heavy neutrino, $m_\ell\ll M_W\ll m_\nu$,
  \begin{equation}
    \mu={\frac{eG_{F}}{8\pi^{2}\sqrt{2}}}m_{\nu}.
  \end{equation}
Note that in all the cases considered , the Dirac neutrino magnetic
moment is proportional to the neutrino mass. This is an expected
result, because the calculations have been performed within the
Standard Model .

In this concern, a question arises: ``Is a neutrino magnetic moment
always proportional to the neutrino mass?''. The answer is ``No''. For
example, much larger value for the Dirac neutrino magnetic moment can
be obtained in $SU_L (2) \times SU_R (2) \times U(1)$ left-right
symmetric models (see, for instance, \cite{KimPRD76,MarSanPLB77,Czakon:1998rf} and
the first paper in \cite{BegMarRudPRD78_MarSirPRD80_SakPRD81}) with
direct right-handed neutrino interactions.  The intermediate
gauge bosons mass states $W_1$ and $W_2$ have, respectively, predominant
left-handed and right-handed coupling, since
\begin{equation}
W_1=W_{L}\cos \xi - W_R \sin \xi ,
\end{equation}
\begin{equation}
W_2=W_{L}\sin \xi + W_R \cos \xi,
\end{equation}
where $\xi$ here is a mixing angle and the fields $W_L$ and $W_R$ have pure
$V\pm A$ interactions. The magnetic moment of a
neutrino $\nu_l$ calculated in this model is
\begin{equation}\label{mu_L_R}
\mu_{\nu _{l}}=\frac{eG_F}{2\sqrt{2}\pi ^2}
\Big[m_l\Big(1-\frac{m_{W_1}^2}{m_{W_2}^2}\Big)
\sin2\xi+\frac{3}{4}m_{\nu_{l}}\Big(1+\frac{m_{W_1}^2}{m_{W_2}^2}\Big)\Big],
\end{equation}
where the term proportional to the charged lepton mass $m_l$ is
due to the left-right mixing. This term can exceed the second term
in (\ref{mu_L_R}), which is proportional to the neutrino mass
$m_{\nu_{l}}$.

\subsection{Experimental limits on neutrino magnetic moment}
\label{sec3.6}

The most sensitive and established method for the experimental investigation of
the neutrino magnetic moment is provided by direct laboratory
measurements of electron neutrino(antineutrino)-electron scattering
at low energies in solar, accelerator and reactor experiments. A
detailed description of different experiments can be found in
\cite{WogLiMPLA05, BedStar_eaPAN07}.

Extensive experimental studies of the neutrino magnetic moment,
performed during many years, are stimulated by the hope to observe a
value much larger than the prediction (\ref{mu_3_10_19}) of the
minimally extended Standard Model  (with nonzero neutrino masses). It
would be a clear indication of new physics beyond the extended
Standard Model . For example, in calculations \cite{MohNgYuPRD04} of
the magnetic moment contribution to $\bar \nu_e$-$e$ scattering in a
class of extra-dimension models it was shown that the contribution to
the cross section can be comparable with the corresponding one for
the case in which there are no extra dimensions and the neutrino
magnetic moment is of order $\mu_{\nu} \sim 10^{-10} \mu_{B}$. Future
higher precision reactor experiments can therefore be used to provide
new constraints on a class of large extra-dimension theories.

The cross section for electron neutrino (antineutrino) scattering on electrons
can be written \cite{VogEngPRD89} (see also
\cite{BedStar_eaPAN07,WogLiMPLA05,Bal06}) as a sum of the Standard Model
contribution and the neutrino magnetic moment contribution:
\begin{equation}\label{d_sigma}
\frac{d\sigma}{dt}=\Big(\frac{d\sigma}{dt}\Big)_{SM}+
\Big(\frac{d\sigma}{dt}\Big)_{\mu}.
\end{equation}
The Standard Model  contribution in Eq.~(\ref{d_sigma}) is
\begin{equation}\label{d_sigma_SM}
\Big(\frac{d\sigma}{dt}\Big)_{SM}=\frac{G^2_F m_e}{2\pi}\Bigg[(g_V + g_A)^2 +
(g_V - g_A)^2\Big(1-\frac{T}{E_\nu}\Big)^2 + (g_A^2 -
g_V^2)\frac{m_eT}{E^2_\nu}\Bigg],
\end{equation}
where $E_\nu$ is the initial neutrino energy and $T$ is the electron recoil
energy, which is measured in experiments. The coupling constants $g_V$ and $g_A$
are
\begin{equation}
g_V=\Bigg  \{\begin{array}{l}
    2\sin ^2 \theta _W +\frac{1}{2}, \ \  $for$ \ \ \nu_e ,\\
    2\sin ^2 \theta _W -\frac{1}{2},  \ \  $for$ \ \ \nu_\mu, \nu_\tau ,
  \end{array} \ \ \ \ \ \ \
g_A=\Bigg  \{\begin{array}{rl}
  \frac{1}{2}, \ \  & $for$ \ \ \nu_e ,\\
 -\frac{1}{2},  \ \ & $for$ \ \ \nu_\mu, \nu_\tau .
  \end{array}
  \end{equation}
In the case of antineutrinos, the substitution $g_A\rightarrow -g_A$ should be
made. As it has been already mentioned, the neutrino charge radius  can also
contribute to the cross-section with the corresponding change
of $g_V$ in Eq.~(\ref{g_V_ch_rad}).

The neutrino magnetic moment contribution to the cross section is
\begin{equation}\label{d_sigma_mu}
\Big(\frac{d\sigma}{dt}\Big)_{\mu}=\frac{\pi \alpha ^2_{em}}{m^2_e}
\Big(\frac{1-{T}/{E_\nu}}{T}\Big)\Big(\frac {\mu_\nu}{\mu_B}\Big)^2.
\end{equation}
Note that the magnetic moment contribution to the cross section
changes the helicity of the neutrino, contrary to the Standard Model
contribution and also the possible contribution from the neutrino
charge radius. Therefore, for relativistic neutrino energies the
interference between $\Big(\frac{d\sigma}{dt}\Big)_{SM}$ and
$\Big(\frac{d\sigma}{dt}\Big)_{\mu}$ is a negligible effect in the
total cross section (\ref{d_sigma}).
\begin{figure}
  \centering
  \includegraphics[scale=1]{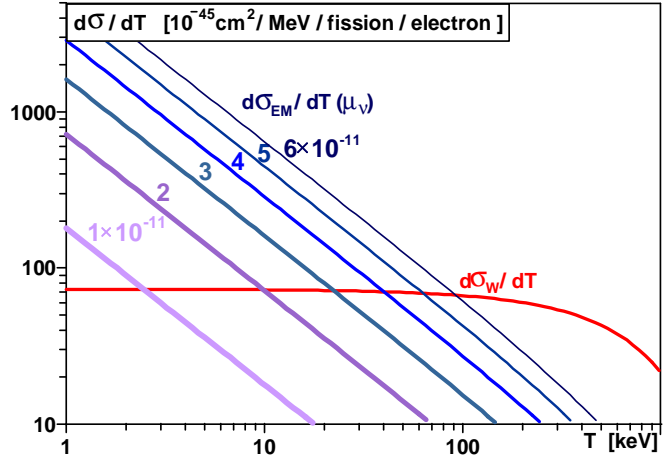}
  \caption{\label{gemma_f1}Standard Model  weak (W) and magnetic moment electromagnetic (EM)
  contributions to the cross section for several values of the neutrino
  magnetic moment
  \cite{BedStar_eaPAN07}. }
  \end{figure}
The two terms $\Big(\frac{d\sigma}{dt}\Big)_{SM}$ and
$\Big(\frac{d\sigma}{dt}\Big)_{\mu}$ exhibit a quite a different dependence on
the experimentally observable electron recoil energy $T$. The dependence of
these two terms on $T$ is shown \cite{BedStar_eaPAN07} in Fig.~\ref{gemma_f1}
for six fixed values of the neutrino magnetic moment, $\mu_{\nu}^{(N)}=N\times
10^{-11}\mu_{B}, \ \ N=1,2,3,4,5,6 $. The cross sections are averaged over the
typical antineutrino reactor spectrum (see also \cite{VogEngPRD89}). It is easy
to see that the lower the measured recoil energy is, the smaller neutrino
magnetic moment values are probed in the experiment. From
Eqs.~(\ref{d_sigma_SM}) and (\ref{d_sigma_mu}), it follows that
$\Big(\frac{d\sigma}{dt}\Big)_{\mu}$ exceeds
$\Big(\frac{d\sigma}{dt}\Big)_{SM}$ for
\begin{equation}
T<\frac{\pi^2\alpha^2}{G_F ^2m_e ^3}\Big(\frac {\mu_\nu}{\mu_B}\Big)^2.
\end{equation}

The constraints on the neutrino magnetic moment in direct
laboratory experiments have been obtained so far from the lack of any observable distortion
of the recoil electron energy spectrum. Experiments of this type have
started more than 30 years ago at the Savannah River Laboratory where
the ${\bar \nu}$-$e$ scattering process was studied
for the first time \cite{ReiGurSobPRL76}.
The upper limit on the
magnetic moment $\mu_\nu \leq (2\div 4) \times 10^{-10}\mu_{B}$ was
derived in Ref.~\cite{VogEngPRD89}. The results of experiments at the
Krasnoyarsk and Rovno reactors are,
respectively, $\mu_\nu \leq (2.4) \times 10^{-10}\mu_{B}$ and
$\mu_\nu \leq (1.9) \times 10^{-10}\mu_{B}$ \cite{VidJETPL92_DerJETPL93}. The analysis of the
recoil electron spectrum in the SuperKamiokande experiment
 gives $\mu_\nu \leq (1.1) \times
10^{-10}\mu_{B}$ \cite{Liu_ea_PRL04}. In reactor experiments carried
recently, the following upper bounds have been obtained: $\mu_\nu
\leq 9.0 \times 10^{-11}\mu_{B}$ ($MUNU$ \cite{Dar_eaPLB05}),
$\mu_\nu \leq 7.4 \times 10^{-11}\mu_{B}$ ($TEXONO$
\cite{Wong_PRD07_75_012001}), $\mu_\nu \leq 5.8 \times
10^{-11}\mu_{B}$ ($GEMMA$ \cite{BedStar_eaPAN07}\footnote{The
 new stringent constraint on the level $\mu_\nu \leq 3.2 \times
10^{-11}\mu_{B}$ has been also obtained within recently performed analysis
\cite{BedSta13LomCon}. }). The  limit $\mu_\nu \leq 5.4 \times 10^{-11}\mu_{B}$
has been recently obtained in the $Borexino$ solar neutrino scattering
experiment \cite{BOREXINO_08}.

An upper limit on the neutrino magnetic moment $\mu_\nu \leq 8.5
\times 10^{-11}\mu_{B}$ has been found in an independent analysis
 of the first release of the
$Borexino$ experiment data performed in \cite{MonPicPul08012643}. It
was also shown that with reasonable assumptions on the oscillation
probability this limit translates into the upper limits on the magnetic
moments of the muon and $\tau$ neutrinos
$\mu_{\nu _\mu} \leq 1.5 \times 10^{-10}\mu_{B}$ and $\mu_{\nu _\tau}
\leq 1.9 \times 10^{-10}\mu_{B}$. The limit on $\mu_{\nu
_\tau}$ is three order of magnitude stronger than that quoted by the
Particle Data Group \cite{ParDataGroup06}.

An interesting new possibility for providing more stringent
constraints on the neutrino magnetic moment from $\bar \nu_e$-$e$
scattering experiments was discussed in \cite{BerPapPassPLB05} on the
basis of an observation \cite{SegBerBotPRD94} that the ``dynamical
zeros" appear in the Standard Model  contribution to the scattering
cross section.

It should be mentioned that what is measured in experiments is an
effective magnetic moment $\mu^{exp}_{e}$ whose value is a rather
complicated function of the magnetic (transition) moments $\mu_{i
j}$. In addition, the dipole electric (transition) moments, if these
quantities do not vanish,  can also contribute to $\mu^{exp}_{e}$.

The magnetic moments $\mu_{i j}$, in the presence of mixing between
different neutrino states, are associated with the neutrino mass
eigenstates $\nu_i$.
The effective interaction Lagrangian which describes the coupling
of neutrinos with the electromagnetic field is given by
\begin{equation}\label{Lagr_sigma_F}
L_{int}=\frac{1}{2}{\bar \psi}_i \sigma_{\alpha \beta}(\mu
_{ij}+\epsilon_{ij} \gamma_{5}) \psi_j F^{\alpha \beta}+ h.c..
\end{equation}
One can see that also the electric (transition) moments
$\epsilon_{ij}$ contribute to the coupling. In the laboratory
neutrino scattering experiments on a neutrino magnetic moment the
recoil energy of an electron, coupled to a neutrino flavor state, is
the only measured quantity. That is why in order to extract from the
experimental data information on the neutrino magnetic moment it is
important to consider \cite{BeaVogPRL99} the interplay between
magnetic moment and mixing effects. Thus, the measured value
$\mu_{exp}$ depends on the composition of the neutrino beam at the
detector. The flavor composition of the initial beam of electron
neutrinos is changing with the distance $L$ from its source according
to
\begin{equation}
| \nu_{e}(L)\rangle = \sum _{i} U_{ei} e^{-iE_{i}L} |\nu_{i}\rangle,
\end{equation}
where $E_{i}$ is the neutrino energy. Therefore the electromagnetic
contribution to the scattering process amplitude is composed of terms
like
\begin{equation} A_{j} \sim \sum_{i} U_{ei}
e^{-iE_{i}L}\mu_{ji}.
\end{equation}
In the magnetic scattering different mass eigenstates contribute
incoherently,
\begin{equation}
\Big(\frac{d\sigma}{dt}\Big)_{\mu} \sim \sum_{j}\Big| \sum_{i} U_{ei}
e^{-iE_{i}L}\mu_{ji} \Big| ^{2}.
\end{equation}
Therefore, the effective value of the neutrino magnetic moment
measured in scattering experiments is
\begin{equation}\label{mu_exp}
\mu_{exp}^{2} = \mu_{\nu}^{2}(\nu_{l},L,E_{\nu})=\sum_{j}\Big|
\sum_{i} U_{li} e^{-iE_{i}L}\mu_{ji} \Big| ^{2}.
\end{equation}
If a neutrino has nonvanishing electric (transition) moments,
the substitution
\begin{equation}\label{mu_mu_epsilon}
 \mu _{ij}\rightarrow |\mu _{ij} -
\epsilon _{ij}| \end{equation}
should be made in Eq.~(\ref{mu_exp})
\cite{Raf_book96_RafPR99}. Therefore, in the case of Dirac neutrinos
a destructive interference between the magnetic and electric moments
is possible. In the case of Majorana neutrinos, only magnetic or
electric transition moments contribute to the cross section if $CP$
is not violated.

The general expression for $\mu_{\nu}^{2}(\nu_{l},L,E_{\nu})$ can be
simplified \cite{BeaVogPRL99} in several important cases. For
instance, for Dirac neutrinos with only diagonal magnetic moments
$\mu _{ij}=\mu _{i}\delta _{ij}$ we have
\begin{equation}\label{mu_exp_Dir}
\mu_{\nu}^{2}(\nu_{e},L,E_{\nu})\rightarrow
(\mu^{D}_{e})^2=\sum_{i}|U_{ei}| ^{2} |\mu_{i}|^2.
\end{equation}
Since in Eq.~(\ref{mu_exp_Dir}) there is no dependence on the distance $L$ and
the neutrino energy, the magnetic cross section is
characterized by the initial neutrino flavor rather than by the
composition of mass states in the detector. In this case, measurements of
all ``flavor'' magnetic moments and mixing parameters, in
principle, allow the extraction of the fundamental moments $\mu_{i}$.

In the case of Majorana neutrinos, assuming that only two mass
eigenstates are important, we have
\begin{equation}\label{mu_exp_Maj}
\mu_{\nu}^{2}(\nu_{e},L,E_{\nu})\rightarrow
(\mu^{M}_{e})^2=|\mu_{12}|^2(|U_{e1}| ^{2} +|U_{e2}|
^{2})=|\mu_{12}|^2,
\end{equation}
which is independent on $L$, as well as on the neutrino energy and
mixing. Note that the global fit \cite{Grim_ea_NPB03_Tor04} of the
magnetic moment data from the reactor and solar neutrino experiments
for the Majorana neutrinos produces limits on the neutrino transition
moments $\mu_{23},\ \mu _{31}, \ \mu _{12} < 1.8 \times
10^{-10}\mu_B$.

\subsection{Model independent bounds on magnetic moments of Dirac and
Majorana neutrinos}

As it was already mentioned before, there is a gap of many orders of
magnitude between the present experimental limits $\sim
10^{-11}\mu_{B}$ on neutrino magnetic moments (discussed in the
previous subsection) and the prediction (\ref{mu_3_10_19}) of the
minimal extension of the Standard Model . At the same time, the
experimental constraints have improved by only one order of magnitude
during a period of about twenty years since the first limit on
$\mu_{\nu}$ was obtained \cite{VogEngPRD89} from neutrino scattering
data. That is why experimental studies of $\mu_{\nu}$ are in a
reasonable extent stimulated by a hope that new physics beyond the
minimally extended Standard Model  might give much stronger
contributions to $\mu_{\nu}$. One of the examples in which it is
possible to avoid the neutrino magnetic moment being proportional to
a (small) neutrino mass, that would in principle make a neutrino
magnetic moment accessible for experimental observations, is realized
in the left-right symmetric models considered before.

Other interesting possibilities of obtaining neutrino magnetic
moments lager than the prediction (\ref{mu_3_10_19}) of the minimal
extension the Standard Model  have been considered recently. In this
concern, we note that it was proposed in \cite{MohNgYuPRD04} to probe
a class of large extra dimensions models with future reactors
searches for neutrino magnetic moments. The results obtained within
the Minimal Supersymmetric Standard Model with $R$-parity violating
interactions \cite{GozKamSimFaePRD06IJMP06} show that the Majorana
transition magnetic moment might be significantly above the scale of
(\ref{mu_3_10_19}).

 Considering the generic problem with
large neutrino magnetic moment, one can write down \cite{PalIJMP92,
BelCirRamVogWisPRL05_BelGorRamVogWanPLB06_BelIJMPA07} a naive
relationship between the size of $\mu_{\nu}$ and the neutrino mass
$m_{\nu}$. Suppose that a large neutrino magnetic moment is generated
by physics beyond a minimal extension of the Standard Model  at an
energy scale characterized by $\Lambda$. For a generic diagram
corresponding to this contribution to $\mu_{\nu}$, one can again use
the Feynman graph in Fig.~\ref{fig0}; the shaded circle in this case
denotes effects of new physics beyond the Standard Model. The
contribution of this diagram to the magnetic moment is
\begin{equation}\label{mu_Lambda}
\mu_{\nu} \sim \frac{eG}{\Lambda},
\end{equation}
where $e$ is the electric charge and $G$ is a combination of coupling
constants and loop factors. The same diagram of Fig.~13
but
without the photon line gives a new physics contribution to the
neutrino mass
\begin{equation}\label{m_Lambda}
\delta m_{\nu} \sim G\Lambda.
\end{equation}
Combining the estimates (\ref{mu_Lambda}) and (\ref{m_Lambda}), one
can get the relation
\begin{equation}\label{mu_Lambda1}
\delta m_{\nu} \sim \frac{\Lambda ^2}{2m_e}\frac{\mu_{\nu}}{\mu_B}=
\frac{\mu_{\nu}}{10^{-18}\mu_B}\Big(\frac{\Lambda}{1 \, \text{TeV}}\Big)^2\
\text{eV}
\end{equation}
between the one-loop contribution to the neutrino mass and the neutrino
magnetic moment. The ${\Lambda}^2$ dependence in Eq.~(\ref{mu_Lambda1}) was
also discussed in \cite{VolSJNP88,BarFreZeePRL90}.

It follows that, generally, in theoretical models that predict large
values for the neutrino magnetic moment, simultaneously large
contributions to the neutrino mass arise. Therefore, a particular
fine tuning is needed to get a large value for the neutrino magnetic
moment while keeping the neutrino mass within experimental bounds.
One of the possibilities \cite{VolSJNP88} is based on the idea of
suppressing the ratio $m_{\nu}/\mu_{\nu}$ with a symmetry: if a
$SU(2)_{\nu}$ symmetry is an exact symmetry of the Lagrangian of a
model, because of different symmetry properties of the mass and
magnetic moment even a massless neutrino can have a nonzero magnetic
moment. If, as it happens in a realistic model, the $SU(2)_{\nu}$
symmetry is broken and if this breaking is small, the ratio
$m_{\nu}/\mu_{\nu}$ is also small, giving a natural way to obtain a
magnetic moment on the order of $\sim 10^{-11}\mu_B$ without
contradictions with neutrino mass experimental constraints. Several
possibilities based on the general idea of \cite{VolSJNP88} were
considered in
\cite{LeMaPLB90BaMohPRD89GeRaPLB90EcGrNePLB90ChaKeuLipSenPRL91}.

Another idea of neutrino mass suppression without suppression of the
neutrino magnetic moment was discussed in \cite{BarFreZeePRL90}
within the Zee model \cite{ZeePLB80}, which is based on the Standard
Model  gauge group $SU(2)_{L}\times U(1)_{Y}$ and contains at least
three Higgs doublets and a charged field which is a singlet of
$SU(2)_{L}$. For this kind of models there is a suppression of the
neutrino mass diagram, while the magnetic moment diagram is not
suppressed.

It is possible to show with more general and rigorous considerations
\cite{BelCirRamVogWisPRL05_BelGorRamVogWanPLB06_BelIJMPA07} that the
${\Lambda}^2$ dependence in Eq.~(\ref{mu_Lambda1}) arises from the quadratic
divergence  in the renormalization of the dimension-four neutrino mass
operator. A general and model-independent upper bound on the Dirac neutrino
magnetic moment, which can be generated by an effective theory beyond the
Standard Model, has been derived
\cite{BelCirRamVogWisPRL05_BelGorRamVogWanPLB06_BelIJMPA07} from the demand of
absence of fine-tuning of effective operator coefficients and from the current
experimental information on neutrino masses. A model with Dirac fermions,
scalars and gauge bosons that is valid below the scale $\Lambda$ and respects
the Standard Model  $SU(2)_L \times U(1)_Y$ symmetry was considered.
Integrating out the physics above the scale $\Lambda$, the following effective
Lagrangian that involves right-handed neutrinos $\nu_{R}$, lepton isodoublets
and the Higgs doublet can be obtained:
\begin{equation}\label{L_operator}
\mathcal{L} _{eff}=\sum _{n,j} \frac{\mathcal{C}^{n}_{j}(\mu)}{\Lambda
^{n-4}}\mathcal{O}_{j}^{(n)}(\mu)+ h.c.,
\end{equation}
where $\mu$ is the renormalization scale, $n\geq 4$ denotes the operator
dimension and $j$ runs over independent operators of a given dimension. At
$n=4$ a neutrino mass arises from the operator $\mathcal{O}^{(4)}_{1}={\bar
L}{\tilde \Phi}\nu_{R}$, where ${\tilde \Phi}=i\sigma_2 \Phi ^{*}$. In
addition, if the scale $\Lambda$ is not extremely large with respect to the
electroweak scale, then an important contribution to the neutrino mass can
arise also from the higher dimension operators. At this point it is important
to note that the combination of the $n=6$ operators appearing in the Lagrangian
(\ref{L_operator}) contains the magnetic moment operator ${\bar \nu}\sigma
_{\mu \nu} \nu F^{\mu \nu}$ and also generates the contribution $\delta
m_{\nu}$ to the neutrino mass (a detailed discussion of this item is given in
\cite{BelCirRamVogWisPRL05_BelGorRamVogWanPLB06_BelIJMPA07}).  Solving the
renormalization group equation from the scale $\Lambda$ to the electroweak
scale, one finds that the contributions to the neutrino magnetic moment and to
the neutrino mass are connected to each other by
\begin{equation}\label{mu_D_Bell}
|\mu_{\nu}^{D}|=\frac{16{\sqrt 2}G_Fm_e \delta m_{\nu}\sin ^{4}\theta
_{W}}{9\alpha ^2|f|\ln\big(\Lambda / {{\it v}}\big)}\mu _{B},
\end{equation}
where $\alpha$ is the fine structure constant, ${\it v}$ is the
vacuum expectation value of the Higgs doublet,
\begin{equation}\label{f_r}
f=1-r-\frac{2}{3}\tan ^2 \theta _{W} -\frac{1}{3}(1+r)\tan ^4 \theta
_{W},
\end{equation}
$r$ is a ratio of effective operator coefficients defined at the scale
$\Lambda$ and is of order unity without fine-tuning. If the neutrino magnetic
moment is generated by new physics at a scale $\Lambda \sim 1 \ \text{TeV}$ and
the corresponding contribution to the neutrino mass is $\delta m_{\nu} \lesssim
1\ \text{eV}$, then the bound $\mu_{\nu}\lesssim 10^{-14} \mu_B$ can be
obtained. This bound is several orders of magnitude stronger than the
constraints from reactor and solar neutrino scattering experiments discussed
before.

The model-independent limit on a Majorana neutrino transition magnetic moment
was also discussed in
\cite{BelCirRamVogWisPRL05_BelGorRamVogWanPLB06_BelIJMPA07}. However, the limit
in the Majorana case is much weaker than that in the Dirac case, because for a
Majorana neutrino the magnetic moment contribution to the mass is Yukawa
suppressed. The limit on $\mu ^{M}_{\nu}$ is also weaker than the present
experimental limits if $\mu ^{M}_{\nu}$ is generated by new physics at the
scale $\Lambda \sim 1 \ \text{TeV}$. An important conclusion of
\cite{BelCirRamVogWisPRL05_BelGorRamVogWanPLB06_BelIJMPA07}, based on
model-independent considerations of the contributions to $\mu_{\nu}$, is that
if a neutrino magnetic moment of order $\mu _{\nu} \geq 10^{-15} \mu _{B}$ were
observed in an experiment, it would give a proof that neutrinos are Majorana
rather than Dirac particles.

\section{Effects of neutrino electromagnetic properties}
\begin{figure}
  \centering
  \includegraphics[scale=0.6]{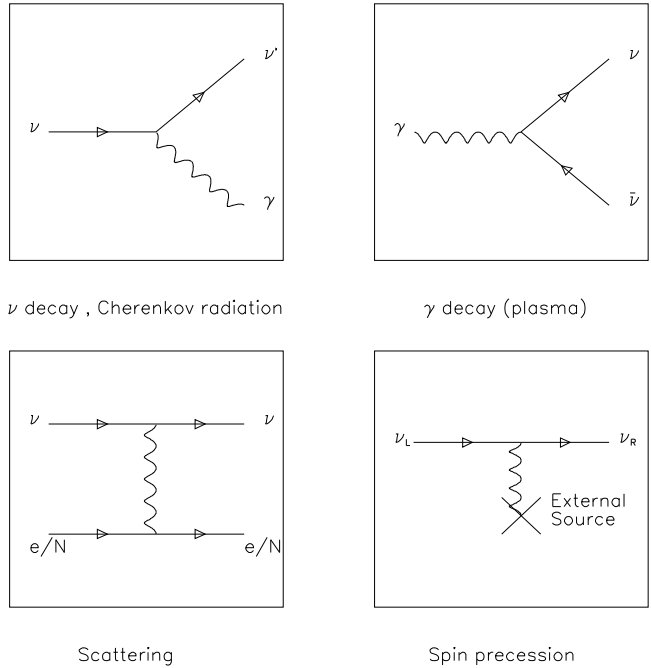}
  \caption{\label{processes} Schematic diagrams for neutrino-photon processes
  \cite{Raf_book96_RafPR99,WogLiMPLA05}.}
  \end{figure}
If a neutrino has non-trivial electromagnetic properties, notably
nonvanishing magnetic and electric (transition) dipole moments or
nonzero millicharge and charge radius, then a direct neutrino
coupling to photons is possible and several processes important for
applications exist. A set of typical and most important neutrino
electromagnetic processes involving the direct neutrino couplings
with photons is shown in Fig.~\ref{processes}. These processes are:
1) the neutrino radiative decay $\nu_{1}\rightarrow \nu_{2} +\gamma$,
neutrino Cherenkov radiation in an external environment (plasma
and/or electromagnetic fields), the spin light of neutrino, $SL\nu$,
in the presence of a medium; \ 2) the photon (plasmon) decay to a
neutrino-antineutrino pair in a plasma $\gamma \rightarrow \nu {\bar
\nu }$; \ 3) neutrino scattering off electrons (or nuclei), this
process has been already considered in detail in
Section~\ref{sec3.6}; \ 4) neutrino spin (spin-flavor) precession in
a magnetic field.

\subsection{Neutrino radiative decay and
other $\nu \rightarrow \nu +\gamma$ processes}

If the masses of neutrinos are not degenerate, the radiative decay of
a heavier neutrino $\nu_i$ into a lighter neutrino $\nu_{j}$ ($m_i>m_j$ ) with
emission of a photon,
\begin{equation}\label{nu_i_nu_j+gamma}
\nu_i \rightarrow \nu_j + \gamma,
\end{equation}
may proceed in vacuum \cite{MarSanPLB77,LeeShrPRD77, PetSNP77,
GolStePRD77, BilPetRMP87, ZacSmiYadFiz78, PalWolPRD82}. A discussion
of the possible role of the neutrino radiative decay in different
astrophysical and cosmological setting has been started in
\cite{DicPRD77_SatKobPTP77_StePRL80_KimBowJakPRL81_MelSciPRL81_DRGlaPRL80}.

 The neutrino radiative decay process is described by the effective Lagrangian
given in Eq.~(\ref{Lagr_sigma_F}). The corresponding typical one-loop
Feynman diagrams that contribute to the process in the standard
$SU(2)_L \times U(1)_Y$ model are similar to those shown in
Fig.15
if to consider the initial and final neutrinos as
different mass states. For the case of a Dirac neutrino, the decay
rate is found to be \cite{MarSanPLB77,LeeShrPRD77, PetSNP77,
GolStePRD77, BilPetRMP87, ZacSmiYadFiz78, PalWolPRD82}
\begin{equation}\label{Gamma_Dir_nu_i+nu_j+gamma}
\Gamma _{\nu^D _{i}\rightarrow \nu^D_{j}+\gamma}=\frac{\alpha G_F ^2}
{128 \pi ^4}\Big (\frac{m^2_i-m^2_j}{m_j}\Big )^3 (m^2_i+m^2_j)
  \Big |\sum_{l=\ e, \ \mu, \
  \tau}f(a_l)U_{lj}U^{\ast}_{li}\Big |^2,
\end{equation}
where $f(a_l)$ is given by Eq.~(\ref{f_a_l}). Recalling the results
 for the Dirac neutrino magnetic and electric
transition moments $\mu_{ij}$ and $\epsilon _{ij}$, given in
Eq.~(\ref{m_e_mom_i_j}), one may rewrite
Eq.~(\ref{Gamma_Dir_nu_i+nu_j+gamma}) in the following form (see, for
instance, \cite{Raf_book96_RafPR99}):
\begin{equation}\label{Gamma}
\Gamma _{\nu _{i}\rightarrow
\nu_{j}+\gamma}=\frac{|\mu_{ij}|^2+|\epsilon _{ij}^2|} {8 \pi}\Big
(\frac{m^2_i-m^2_j}{m_j}\Big )^3 .
\end{equation}
For degenerate neutrino masses ($m_i = m_j$), the process is
kinematically forbidden in vacuum.

Note that there are models ( see for instance \cite{PetPLB82}) in which the
neutrino radiative decay rate (as well as the magnetic moment discussed above)
of a non-standard Dirac neutrino are much larger than those predicted in
minimally extended Standard Model.

In the evaluation of the decay rate for Majorana neutrinos, two cases
should be considered that correspond to the two possible relative $CP$
phases of $\nu_i$ and $\nu_j $. If the Majorana neutrinos $\nu_i$ and
$\nu_j $ have the same $CP$ eigenvalues, the decay rate is
\begin{equation}\label{Gamma_Maj_nu_i+nu_j+gamma_e}
\Gamma _{\nu^M _{i}\rightarrow \nu^M_{j}+\gamma}=\frac{\alpha G_F ^2}
{64 \pi ^4}\Big (\frac{m^2_i-m^2_j}{m_j}\Big )^3 (m^2_i-m^2_j)
  \Big |\sum_{l=\ e, \ \mu, \
  \tau}f(a_l)U_{lj}U^{\ast}_{li}\Big |^2.
\end{equation}
In this case the decay process is induced purely by the neutrino
electric transition dipole moment, because $\mu _{ij}=0$. If the
Majorana neutrinos have opposite $CP$ eigenvalues, the decay
rate is
\begin{equation}\label{Gamma_Maj_nu_i+nu_j+gamma_mu}
\Gamma _{\nu^M _{i}\rightarrow \nu ^M_{j}+\gamma}=\frac{\alpha G_F
^2} {64 \pi ^4}\Big (\frac{m^2_i-m^2_j}{m_j}\Big )^3 (m^2_i+m^2_j)
  \Big |\sum_{l=\ e, \ \mu, \
  \tau}f(a_l)U_{lj}U^{\ast}_{li}\Big |^2.
\end{equation}
In this case, the transition is purely of magnetic dipole type ($\epsilon
_{ij}=0$).

For numerical estimations it is convenient to express Eq.~(\ref{Gamma})
in the following form:
\begin{equation}\label{Gamma_num}
\Gamma _{\nu _{i}\rightarrow \nu_{j}+\gamma}=5.3 \times \Big
(\frac{\mu_{eff}}{\mu_B}\Big)^2\Big (\frac{m^2_i-m^2_j}{m_j^2}\Big )^3 \big
(\frac{m_i}{1\ \text{eV}}\big)^3 s^{-1},
\end{equation}
with the effective neutrino magnetic moment $\mu_{eff}=\sqrt
{|\mu_{ij}|^2+|\epsilon _{ij}^2|}$. The expression (\ref{Gamma_num}) is
valid for both Dirac and Majorana neutrinos. In the case of
Majorana neutrinos, only $\mu_{ij}$ or $\epsilon _{ij}$
contributes, depending on the relative $CP$ phase of the neutrino states.
Note that there is no destructive interference between $\mu_{ij}$ and
$\epsilon _{ij}$ in Eq.~(\ref{Gamma_num}), contrary to the neutrino
scattering off electrons cross section given in Eqs.~(\ref{d_sigma_mu})
and (\ref{mu_mu_epsilon}).

The neutrino radiative decay can be constrained by the absence of
decay photons in reactor $\bar \nu_e$ and solar $\nu _e$ fluxes.
The limits on $\mu_{eff}$ that are obtained from these
considerations are much weaker than those obtained from neutrino
scattering terrestrial experiments. Stronger constraints on
$\mu_{eff}$ (though still weaker than those mentioned above)  are
obtained from the neutrino decay limit set by SN 1987A and also from
the limit on the cosmic microwave background radiation distortions.
These limits can be expressed as (see \cite{Raf_book96_RafPR99} and
references therein)
\begin{equation}
\frac{\mu_{\rm eff}}{\mu_{\rm B}}
=
\left\{\begin{array}{lcl}
0.9{\times}10^{-1}\big(\frac{\text{eV}}{m_\nu}\big)^2 & \ \ & \text{Reactor ($\bar\nu_e$)},\\
0.5{\times}10^{-5}\big(\frac{\text{eV}}{m_\nu}\big)^2 &     & \text{Sun ($\nu_e$)}, \\
1.5{\times}10^{-8}\big(\frac{\text{eV}}{m_\nu}\big)^2 &     & \text{SN 1987A (all flavors)}, \\
1.0{\times}10^{-11}\big(\frac{\text{eV}}{m_\nu}\big)^{9/4} && \text{Cosmic background (all flavors)}.
  \end{array}\right.
  \end{equation}
A detailed discussion (and corresponding references) on astrophysical
constraints on the neutrino magnetic and electric transition moments,
summarized in Fig.~21,
can be found in
\cite{Raf_book96_RafPR99}.
\begin{figure}
  \centering
  \includegraphics[scale=.3]{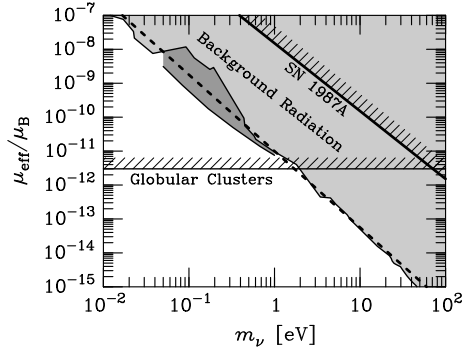}
  \caption{\label{Fig8}Astrophysical limits on neutrino transition moments \cite{Raf_book96_RafPR99}.}
\end{figure}

For completeness, we would like to mention that other processes
characterized by the same signature of Eq.~(\ref{nu_i_nu_j+gamma})
have been considered previously (for a review of the literature see
\cite{IoaRafPRD97, LobStuPLB03,StuPAN04_StuPAN07,StuAFB06}):

 i) the photon radiation by a massless neutrino
$(\nu_{i} \rightarrow \nu_{j} + \gamma,\; i=j)$ due to the vacuum
polarization loop diagram  in the presence of an external magnetic field
\cite{GalNik72_Sko76};

ii) the photon radiation by a massive neutrino with nonvanishing
magnetic moment in constant magnetic and electromagnetic wave fields
\cite{BorZhuTer88_Sko91};

iii) the Cherenkov radiation due to the nonvanishing neutrino
magnetic moment in an homogeneous and infinitely extended medium, which
is only possible if the speed of the neutrino is larger than the speed of
light in the medium \cite{Rad75_GriNeu93};

iv) the transition radiation due to a nonvanishing neutrino magnetic
moment which would be produced when the neutrino crosses the
interface of two media with different refractive indices
\cite{SakKur94-95_GriNeu95};

v) the Cherenkov radiation of a massless neutrino due to its induced
charge in a medium \cite{OraSemSmoJETP86_OliNiePalPLB96}\footnote
{Note that the neutrino electromagnetic properties are in general
affected by the external environment. In particular, a neutrino can
acquire an electric charge in magnetized matter
\cite{OraSemSmoJETP86_OliNiePalPLB96} and the neutrino magnetic
moment depends on the strength of external electromagnetic fields
\cite{BorZhuKurTerSJNP85_DAN88_MasPerelRMF02_EgoLikStu99}. A recent
study of the neutrino electromagnetic vertex in magnetized matter
can be found in \cite{NiePRD03}. See also \cite{StuPAN04_StuPAN07}
for a review of neutrino interactions in external electromagnetic
fields.}

vi) the Cherenkov radiation of massive and massless  neutrinos in
a magnetized medium \cite{MohSam96, IoaRafPRD97}.

vii) the neutrino radiative decay $(\nu_{i} \rightarrow \nu_{j} +
\gamma, \; i\not=j)$ in external fields and media (see
\cite{GiuKLPRD91GvoMikVas92Sko95ZhuEmiGri96KacWun97TerEmi03} and
references therein).

Recently, another mechanism of electromagnetic radiation by a massive
neutrino in presence of matter (termed the spin light of neutrino,
$SL\nu$), has been proposed \cite{LobStuPLB03}. The $SL\nu$ is an
electromagnetic radiation that can be emitted by a massive neutrino
due to the neutrino magnetic or electric (transition) moments when
the particle moves in the background matter. Within a
quasi-classical treatment, the existence of the $SL\nu$ was first
studied \cite{LobStuPLB03,LobStuPLB04_DvoGriStuIJMP05} on the basis
of the developed Lorentz invariant approach to the neutrino spin
evolution that implies the use of the generalized
Bargmann-Michel-Telegdi equation \cite{EgoLobStuPLB00_LobStuPLB01}.

Within the developed Lorentz invariant approach, it is also possible to find
the solution of the neutrino spin evolution problem for a general case when
the neutrino is subjected to general types of non-derivative interactions with
external fields \cite{DvoStuJHEP02} (see also \cite{BeGrNaPRD99}). These
interactions are given by the Lagrangian
\begin{equation}
-{\cal L}=g_{s}s(x){\bar \nu}\nu+ g_{p}{\pi}(x) {\bar \nu}\gamma^{5}\nu+
g_{v}V^{\mu}(x){\bar \nu}\gamma_{\mu}\nu+ g_{a}A^{\mu}(x){\bar
\nu}\gamma_{\mu}\gamma^{5}\nu+ {{g_{t}}\over{2}}T^{\mu\nu}{\bar
\nu}\sigma_{\mu\nu}\nu+ {{g^{\prime}_{t}}\over{2}} \Pi^{\mu\nu}{\bar
\nu}\sigma_{\mu\nu}\gamma_{5}\nu,
 \end{equation} where $s, \pi,
V^{\mu}=(V^{0}, {\bf V}), A^{\mu}=(A^{0}, {\bf A}), T_{\mu\nu}=({\bf
a}, {\bf b}), \Pi_{\mu\nu}=({\bm c}, {\bm d})$ are the scalar,
pseudoscalar, vector, axial-vector, tensor and pseudotensor fields,
respectively. For the corresponding spin evolution equation it has
been found
\begin{equation}\label{S_eq_gen}
\begin{array}{c}  \displaystyle {{d{\bf S} \over dt}}=  2g_{a}\left\{ A^{0}[{\bf S}
\times{\bm \beta}]- {{({\bf A}{\bm \beta})[{\bf S} \times{\bm
\beta}]}\over{1+{\gamma}^{-1}}} - {1 \over \gamma}[{\bf S}\times{\bf A}]
\right\}
 \\ \displaystyle +2g_{t}\left\{ [{\bf S}\times{\bf b}]-
{{({\bm \beta}{\bf b})[{\bf S}\times{\bm \beta}]}\over{1+{\gamma}^{-1}}} +
[{\bf S}\times[{\bf a}\times{\bm \beta}]] \right\} \\
\displaystyle + 2ig^{\prime}_{t}\left\{ [{\bf S}\times{\bf c}]- {{({\bm
\beta}{\bf c})[{\bf S}\times{\bm \beta}]}\over{1+{\gamma}^{-1}}}- [{\bf
S}\times[{\bf d}\times{\bm \beta}]] \right\}.
\end{array}
\end{equation}
This is a rather general equation for the neutrino spin evolution that can be
also used for the description of neutrino spin oscillations in different
environments, such as moving and polarized matter with external electromagnetic
fields (see \cite{StuPAN04_StuPAN07} and references therein). The $SL\nu$ in
gravitational fields has been studied
(see the second paper of \cite{LobStuPLB04_DvoGriStuIJMP05}), for the first time,
on the basis of a neutrino spin
evolution equation (\ref{S_eq_gen}).
Some general aspects of
chiral dynamics for a neutrino non-minimally coupled with an external
magnetic field have been discussed in \cite{BerJPA06}.

\begin{figure}[h]
\begin{center}
{  \includegraphics[scale=.3]{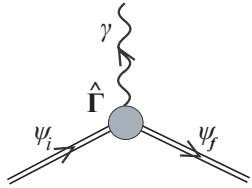}}
    \caption{\label{diagram}
    The spin light of neutrino ($SL\nu$) radiation diagram. }
  \end{center}
\end{figure}

It should be mentioned that the $SL\nu$ in matter is a new mechanism of
electromagnetic radiation that cannot be considered as the neutrino Cherenkov
radiation in matter mentioned above, because it can exist even when the emitted
photon refractive index is equal to unity. The $SL\nu$ radiation is due to
radiation of the neutrino by its own, rather than radiation of the background
particles. As it was clear from the very beginning \cite{LobStuPLB03}, the
$SL\nu$ is a quantum phenomenon by its nature. The quantum theory of this
radiation has been elaborated
\cite{StuTerPLB05GriStuTerPLB05GriStuTerG&C05GriStuTerPAN06} (see also
\cite{LobPLB05LobDAN05}) within a development \cite{StuAFB06,StuJPA_06_08,
StuJPA_08,GriStuTerPAN09} of a quite powerful method that implies the use of
the exact solutions of the modified Dirac equation for the neutrino wave
function in matter. The corresponding Feynman diagram of the $SL\nu$ processes
is shown in Fig.~\ref{diagram},
where the neutrino initial ($\psi_{i}$) and final ($\psi_{f}$) states
(shown by ``broad lines'') are exact solutions of the corresponding
Dirac equations accounting exactly for the interaction with matter.
The amplitude of the $SL\nu$ process is
given by
\begin{equation}\label{amplitude}
\begin{array}{c} \displaystyle
  S_{f i}=-\mu \sqrt{4\pi}\int d^{4} x {\bar \psi}_{f}(x)
  ({\hat {\bf \Gamma}}{\bf e}^{*})\frac{e^{ikx}}{\sqrt{2\omega L^{3}}}
   \psi_{i}(x), \ \ \ \
     \hat {\bf \Gamma}=i\omega\big\{\big[{\bf \Sigma} \times
  {\bm \varkappa}\big]+i\gamma^{5}{\bf \Sigma}\big\},
\end{array}
\end{equation}
where $\mu$ is the neutrino magnetic moment, $k^{\mu}=(\omega,{\bf
k})$ and ${\bf e}^{*}$ are the photon momentum and polarization
vectors, ${\bm \varkappa}={\bf k}/{\omega}$ is the unit vector
pointing in the possible direction of the emitted photon propagation.
From Eq.~(\ref{amplitude}) it follows that the $SL\nu$ rate and
radiation power are proportional to $\mu_{\nu}^2$, so that these
quantities are in general quite small. At the same time, for a wide
range of matter densities these characteristics of radiation are
increasing with the neutrino momentum. One may expect
\cite{LobStuPLB03,StuPAN04_StuPAN07,LobStuPLB04_DvoGriStuIJMP05,
StuTerPLB05GriStuTerPLB05GriStuTerG&C05GriStuTerPAN06,LobPLB05LobDAN05,StuAFB06,
StuJPA_06_08,StuJPA_08,GriStuTerPAN09,GriLobStuTer_Quarks_06,
GriSavStuTer_DSPIN_07} that this radiation can be produced by
high-energy neutrinos propagating in different astrophysical and
cosmological environments. In the most interesting  for possible
astrophysical and cosmology applications case of ultra-high energy
neutrinos, the average energy of the $SL\nu$ photons is one third of
the neutrino momentum, so that in principle the $SL\nu$ spectrum
spans up to the range peculiar of gamma-rays.

It should be emphasized that the $SL\nu$ mechanism of radiation (i.e.
the transition between neutrino states with equal masses) can only
become possible because of an external environment (plasma) influence
on neutrino states. A possible impact of the background plasma on the
$SL\nu$ radiation through the plasma influence on propagation of
$SL\nu$ photons has been first considered in
\cite{StuTerPLB05GriStuTerPLB05GriStuTerG&C05GriStuTerPAN06}. The
plasma effects for the $SL\nu$ were further studied in
\cite{KuzMikhMPL06IJMPA07} where the role of the $SL\nu$ plasmon mass
was discussed. In the case of ultra-high energy neutrino (i.e., in
the only case when the time scale of the process can be much less
than the age of the Universe) the $SL\nu$ rate of
\cite{KuzMikhMPL06IJMPA07} exactly reproduces the result obtained in
\cite{StuTerPLB05GriStuTerPLB05GriStuTerG&C05GriStuTerPAN06}. For a
more detailed discussion on the historical aspects of this issue see
\cite{StuJPA_08, GriSavStuTer_DSPIN_07, GriLobStuTer_Quarks_06}.

\subsection{Photon (plasmon) decay into neutrino-antineutrino pair}

The most interesting process, for the purpose of constraining
neutrino electromagnetic properties, is the photon (plasmon) decay
into a neutrino-antineutrino pair, $\gamma^{*} \rightarrow \nu + {\bar
\nu }$. This plasmon process becomes kinematically allowed in media,
because a photon with the dispersion relation $\omega_{\gamma} ^{2} +
{\bf k}_{\gamma}^2 >0$ roughly behaves as a particle with an
effective mass. Note that the $\gamma ^{*} \rightarrow \nu + {\bar
\nu }$ generated by the neutrino coupling to photons due to a
magnetic moment $\mu_{\nu}$ (and/or also due to a neutrino electric
millicharge $q_\nu$) was first considered in \cite{BerRudPR63} as a
new energy-loss channels of the Sun. From the requirement that new
energy-loss channels do not reasonably exceed the standard solar
model luminosity, the constraints are found \cite{Raf_book96_RafPR99} to
be $\mu_{\nu}\leq 4 \times 10^{-10} \mu _B$ and $q_{\nu}\leq 6 \times
10^{-14} e$.

The tightest astrophysical bound on a neutrino magnetic moment is provided by
observed properties of globular cluster stars. The plasmon decay $\gamma ^{*}
\rightarrow \nu + {\bar \nu }$ inside the star liberates the energy
$\omega_{\gamma}$ in the form of neutrinos that freely escape the stellar
environment. This nonstandard energy loss cools a red giant star so fast that
it can delay helium ignition. The energy-loss rate per unit volume due to the
plasmon decay process is
\begin{equation}
Q_{\gamma ^{*}\rightarrow \nu {\bar \nu}}=\frac{g}{(2
\pi)^3}\int\omega_{\gamma} f_{k_{\gamma}}\Gamma_{\gamma \rightarrow \nu {\bar
\nu}}d^3k_{\gamma},
\end{equation}
where $f_{k_\gamma}$ is the photon Bose-Einstein distribution function, $g=2$
is the number of polarization states and the decay rate is
\begin{equation}
\Gamma_{\gamma ^{*}\rightarrow \nu {\bar \nu}}=\frac{\mu_{\nu}^2}{24
\pi}\frac{(\omega_{\gamma} ^2 - k_{\gamma}^2)^2}{\omega_ \gamma}.
\end{equation}

For a sufficiently large neutrino magnetic moment, the plasmon decay rate
can be enhanced, inducing a significant delay of helium ignition.
From the lack of observational evidence of this effect, the following
limit has been found \cite{RafPRL90}:
\begin{equation}
\mu_{\nu}\leq 3 \times 10^{-12} \mu _B.
\end{equation}
This is the most stringent astrophysical constraint on a neutrino
magnetic moment, applicable to both Dirac and Majorana neutrinos. The
same limit applies for the neutrino magnetic transition moments as well
as for the electric (transition) moments.

\subsection{Neutrino spin-flavor precession in magnetic field} \label{041a} \nopagebreak

If neutrinos have magnetic moments, the spin can precess in a transverse magnetic field
\cite{Cisneros:1971nq,Voloshin:1986ty,Okun:1986na}. Considering for simplicity only one undetermined flavor, a neutrino
produced at $x=0$ is described at a distance $x$ by the state
\begin{equation}
| \nu(x) \rangle = \varphi_{L}(x) \, | \nu_{L} \rangle + \varphi_{R}(x) \, | \nu_{R} \rangle \,, \label{101}
\end{equation}
where $| \nu_{L} \rangle$ and $| \nu_{R} \rangle$ are, respectively, neutrino states with negative and positive
helicity, which are called left-handed and right-handed. The respective amplitudes $\varphi_{L}(x)$ and
$\varphi_{R}(x)$ should not be confused with chiral fields. Their evolution equation in a transverse magnetic field
$B_{\perp}(x)$ is given by
\begin{equation}
i \frac{ \text{d} }{ \text{d}x }
\begin{pmatrix}
\varphi_{L}(x)
\\
\varphi_{R}(x)
\end{pmatrix}
=
\begin{pmatrix}
0 & \mu B_{\perp}(x)
\\
\mu B_{\perp}(x) & 0
\end{pmatrix}
\begin{pmatrix}
\varphi_{L}(x)
\\
\varphi_{R}(x)
\end{pmatrix}
\,, \label{102}
\end{equation}
where $\mu$ is the magnetic moment (we consider a Dirac neutrino). This differential equation can be solved through the
transformation
\begin{equation}
\begin{pmatrix}
\varphi_{L}(x)
\\
\varphi_{R}(x)
\end{pmatrix}
= \frac{1}{\sqrt{2}}
\begin{pmatrix}
1 & 1
\\
-1 & 1
\end{pmatrix}
\begin{pmatrix}
\varphi_{-}(x)
\\
\varphi_{+}(x)
\end{pmatrix}
\,. \label{103}
\end{equation}
The new amplitudes $\varphi_{-}(x)$ and $\varphi_{+}(x)$ satisfy decoupled differential equations, whose solutions are
\begin{equation}
\varphi_{\mp}(x) = \exp\!\left[ \pm i \int_{0}^{x} \text{d}x' \, \mu \, B_{\perp}(x') \right] \varphi_{\mp}(0) \,.
\label{104}
\end{equation}
If we consider an initial left-handed neutrino, we have
\begin{equation}
\begin{pmatrix}
\varphi_{L}(0)
\\
\varphi_{R}(0)
\end{pmatrix}
=
\begin{pmatrix}
1
\\
0
\end{pmatrix}
\qquad \Longrightarrow \qquad
\begin{pmatrix}
\varphi_{-}(0)
\\
\varphi_{+}(0)
\end{pmatrix}
= \frac{1}{\sqrt{2}}
\begin{pmatrix}
1
\\
1
\end{pmatrix}
\,. \label{105}
\end{equation}
Then, the probability of $\nu_{L}\to\nu_{R}$ transitions is given by
\begin{equation}
P_{\nu_{L}\to\nu_{R}}(x) = |\varphi_{R}(x)|^2 = \sin^2\!\left( \int_{0}^{x} \text{d}x' \, \mu \, B_{\perp}(x') \right)
\,. \label{106}
\end{equation}
Note that the transition probability is independent from the neutrino energy (contrary to the case of flavor
oscillations) and the amplitude of the oscillating probability is unity. Hence, when the argument of the sine is equal
to $\pi/2$ there is complete $\nu_{L}\to\nu_{R}$ conversion.

The precession $\nu_{e_L}\to\nu_{e_R}$ in the magnetic field of the
Sun was considered in 1971 \cite{Cisneros:1971nq} as a possible
solution of the solar neutrino problem. If neutrinos are Dirac
particles, right-handed neutrinos are sterile and a
$\nu_{e_L}\to\nu_{e_R}$ conversion could explain the disappearance of
active solar $\nu_{e_L}$'s.

In 1986 it was realized \cite{Voloshin:1986ty,Okun:1986na} that the
matter effect during neutrino propagation inside of the sun
suppresses $\nu_{eL}\to\nu_{eR}$ transition by lifting the degeneracy
of $\nu_{e_L}$ and $\nu_{e_R}$. Indeed, taking into account matter
effects, the evolution equation~(\ref{102}) becomes
\begin{equation}
i \frac{ \text{d} }{ \text{d}x }
\begin{pmatrix}
\varphi_{L}(x)
\\
\varphi_{R}(x)
\end{pmatrix}
=
\begin{pmatrix}
V & \mu B_{\perp}(x)
\\
\mu B_{\perp}(x) & 0
\end{pmatrix}
\begin{pmatrix}
\varphi_{L}(x)
\\
\varphi_{R}(x)
\end{pmatrix}
\,, \label{107}
\end{equation}
with the appropriate potential $V$ which depends on the neutrino flavor, according to Eq.~(\ref{i025}). Again, we
consider a Dirac neutrino, which can have a magnetic moment. In the case of a constant matter density, this
differential equation can be solved analytically with the orthogonal transformation
\begin{equation}
\begin{pmatrix}
\varphi_{L}(x)
\\
\varphi_{R}(x)
\end{pmatrix}
=
\begin{pmatrix}
\cos\xi & \sin\xi
\\
- \sin\xi & \cos\xi
\end{pmatrix}
\begin{pmatrix}
\varphi_{-}(x)
\\
\varphi_{+}(x)
\end{pmatrix}
\,. \label{108}
\end{equation}
The angle $\xi$ is chosen in order to diagonalize the matrix operator in Eq.~(\ref{107}):
\begin{equation}
\sin 2 \xi = \frac{ 2 \mu B_{\perp} }{ \Delta{E}_{\text{M}} } \,,
\label{109}
\end{equation}
with the effective energy splitting in matter
\begin{equation}
\Delta{E}_{\text{M}} = \sqrt{ V^2 + \left( 2 \mu B_{\perp} \right)^2 } \,. \label{109a}
\end{equation}
The decoupled evolution of $\varphi_{\mp}(x)$ is given by
\begin{equation}
\varphi_{\mp}(x) = \exp\!\left[ - \frac{i}{2} \left( V \mp \Delta{E}_{\text{M}} \right) \right] \varphi_{\mp}(0) \,.
\label{110}
\end{equation}
For an initial left-handed neutrino,
\begin{equation}
\begin{pmatrix}
\varphi_{-}(0)
\\
\varphi_{+}(0)
\end{pmatrix}
=
\begin{pmatrix}
\cos\xi
\\
\sin\xi
\end{pmatrix}
\,, \label{111}
\end{equation}
leading to the oscillatory transition probability
\begin{equation}
P_{\nu_{L}\to\nu_{R}}(x) = |\varphi_{R}(x)|^2 = \sin ^2 2 \xi
\sin^2\!\left( \frac{1}{2} \, \Delta{E}_{\text{M}} x \right) \,.
\label{1061}
\end{equation}
Since in matter $ \Delta{E}_{\text{M}}> 2 \mu B_{\perp}$, the matter
effect suppresses the amplitude of $\nu_{L}\to\nu_{R}$ transitions.
However, these transitions are still independent from the neutrino
energy, which does not enter in the evolution equation (\ref{107}).

Once it was known, in 1986 \cite{Voloshin:1986ty,Okun:1986na}, that the matter
potential has the effect of suppressing $\nu_{L}\to\nu_{R}$ transitions because
it breaks the degeneracy of left-handed and right-handed states, it did not
take long to realize, in 1988 \cite{Akhmedov:1988nc,Lim:1988tk}, that the
matter potentials can cause resonant spin-flavor precession if different flavor
neutrinos have transition magnetic moments (spin-flavor precession in vacuum
was previously discussed in \cite{SchValPRD81}).

A detailed discussion on how to attack the solar neutrino problem
using a neutrino magnetic moment can be found in \cite{PulPRep92,
ShiSchRosDeaCommNPP93}.

Let us consider two neutrino flavors: $\nu_{e}$ and $\nu_{\mu}$. A neutrino produced at $x=0$ is described at a
distance $x$ by the state
\begin{equation}
| \nu(x) \rangle = \varphi_{e_L}(x) \, | \nu_{e_L} \rangle +
\varphi_{e_R}(x) \, | \nu_{e_R} \rangle + \varphi_{\mu _L}(x) \, |
\nu_{\mu _L} \rangle + \varphi_{\mu _R}(x) \, | \nu_{\mu _R} \rangle
\,, \label{201}
\end{equation}
which is the generalization of Eqs.~(\ref{state}) and (\ref{101}).

Considering Dirac neutrinos, which can have diagonal magnetic moments $\mu_{ee}$ and $\mu_{\mu\mu}$, as well as
transition magnetic moments $\mu_{e\mu}$ and $\mu_{\mu e}$, the evolution equation of the amplitudes, obtained by
combining Eqs.~(\ref{i064}) and (\ref{107}) and neglecting irrelevant diagonal terms, is
\begin{equation}
i \frac{ \text{d} }{ \text{d}x }
\begin{pmatrix}
\varphi_{e_L}(x)
\\
\varphi_{\mu _L}(x)
\\
\varphi_{e_R}(x)
\\
\varphi_{\mu _R}(x)
\end{pmatrix}
= \mathcal{H}
\begin{pmatrix}
\varphi_{e_L}(x)
\\
\varphi_{\mu _L}(x)
\\
\varphi_{e_R}(x)
\\
\varphi_{\mu _R}(x)
\end{pmatrix}
\,, \label{202}
\end{equation}
with the effective Hamiltonian matrix
\begin{equation}
\mathcal{H} =
\begin{pmatrix}
- \frac{ \Delta{m}^{2} }{ 4 E } \cos{2\vartheta} + V_{e} & \frac{ \Delta{m}^{2} }{ 4 E } \sin{2\vartheta} & \mu_{ee}
B_{\perp}(x) & \mu_{e\mu} B_{\perp}(x)
\\
\frac{ \Delta{m}^{2} }{ 4 E } \sin{2\vartheta} & \frac{ \Delta{m}^{2} }{ 4 E } \cos{2\vartheta} + V_{\mu} & \mu_{\mu e}
B_{\perp}(x) & \mu_{\mu\mu} B_{\perp}(x)
\\
\mu_{ee} B_{\perp}(x) & \mu_{\mu e} B_{\perp}(x) & - \frac{ \Delta{m}^{2} }{ 4 E } \cos{2\vartheta} & \frac{
\Delta{m}^{2} }{ 4 E } \sin{2\vartheta}
\\
\mu_{e\mu} B_{\perp}(x) & \mu_{\mu\mu} B_{\perp}(x) & \frac{ \Delta{m}^{2} }{ 4 E } \sin{2\vartheta} & \frac{
\Delta{m}^{2} }{ 4 E } \cos{2\vartheta}
\end{pmatrix}
\,. \label{203}
\end{equation}
The matter potential can generate resonances, when two diagonal
elements of $\mathcal{H}$ become equal. There are two resonances:
\begin{enumerate}
\item There is a resonance in the $\nu_{e_L}\leftrightarrows\nu_{\mu
_ R}$ channel for
\begin{equation}
V_{e} = \frac{ \Delta{m}^{2} }{ 2 E } \, \cos2\vartheta .\label{211}
\end{equation}
The density at which this resonance occurs is not the same as that of the MSW resonance, given by Eq.~(\ref{i074}),
because of the neutral-current contribution to $V_{e}=V_{\text{CC}}+V_{\text{NC}}$. The location of this resonance
depends on both $N_{e}$ and $N_{n}$.
\item There is a resonance in the $\nu_{\mu
_L}\leftrightarrows\nu_{e_R}$ channel for
\begin{equation}
V_{\mu} = - \frac{ \Delta{m}^{2} }{ 2 E } \, \cos2\vartheta .
\label{212}
\end{equation}
If $\cos2\vartheta>0$, this resonance is possible in normal matter, since the sign of $V_{\mu}=V_{\text{NC}}$ is
negative, as one can see from Eq.~(\ref{i017}).
\end{enumerate}
In practice the effect of these resonances could be the disappearance
of active $\nu_{e_L}$ or $\nu_{\mu_L}$ into sterile right-handed
states.

Let us consider now the more interesting case of Majorana neutrinos, which presents two fundamental differences with
respect to the Dirac case:
\begin{enumerate}
\renewcommand{\labelenumi}{\theenumi}
\renewcommand{\theenumi}{(\Alph{enumi})}
\item Majorana neutrinos can have only a transition magnetic moment $\mu_{e\mu}=-\mu_{\mu e}$.
\item The right-handed states are not sterile, but interact as right-handed Dirac antineutrinos.
\end{enumerate}
The evolution equation of the amplitudes is given by Eq.~(\ref{202}) with the effective Hamiltonian matrix
\begin{equation}
\mathcal{H} =
\begin{pmatrix}
- \frac{ \Delta{m}^{2} }{ 4 E } \cos{2\vartheta} + V_{e} & \frac{ \Delta{m}^{2} }{ 4 E } \sin{2\vartheta} & 0 &
\mu_{e\mu} B_{\perp}(x)
\\
\frac{ \Delta{m}^{2} }{ 4 E } \sin{2\vartheta} & \frac{ \Delta{m}^{2} }{ 4 E } \cos{2\vartheta} + V_{\mu} & -
\mu_{e\mu} B_{\perp}(x) & 0
\\
0 & - \mu_{e\mu} B_{\perp}(x) & - \frac{ \Delta{m}^{2} }{ 4 E } \cos{2\vartheta} - V_{e} & \frac{ \Delta{m}^{2} }{ 4 E
} \sin{2\vartheta}
\\
\mu_{e\mu} B_{\perp}(x) & 0 & \frac{ \Delta{m}^{2} }{ 4 E } \sin{2\vartheta} & \frac{ \Delta{m}^{2} }{ 4 E }
\cos{2\vartheta} - V_{\mu}
\end{pmatrix}
\,. \label{303}
\end{equation}
Again, there are two resonances:
\begin{enumerate}
\item There is a resonance in the $\nu_{e_L}\leftrightarrows\nu_{\mu
_ R}$ channel for
\begin{equation}
V_{\text{CC}}+2V_{\text{NC}} = \frac{ \Delta{m}^{2} }{ 2 E } \, \cos2\vartheta \,. \label{311}
\end{equation}
\item There is a resonance in the $\nu_{\mu
_L}\leftrightarrows\nu_{e_R}$ channel for
\begin{equation}
V_{\text{CC}}+2V_{\text{NC}} = - \frac{ \Delta{m}^{2} }{ 2 E } \, \cos2\vartheta \,. \label{312}
\end{equation}
\end{enumerate}
The location of both resonances depend on both $N_{e}$ and $N_{n}$. If $\cos2\vartheta>0$, only the first resonance can
occur in normal matter, where $ N_{n} \simeq N_{e}/6 $. A realization of the second resonance requires a large neutron
number density, as that in a neutron star.

The neutrino spin oscillations in a transverse magnetic field with a
possible rotation of the field-strength vector in a plane orthogonal
to the neutrino-propagation direction (such rotating fields may exist
in the convective zone of the Sun) have been considered in
\cite{VidWudPLB90_SmiPLB91_AkhPetSmiPRD93,LikStuJETP95}. The effect
of the magnetic-field rotation may substantially shift the resonance
point of neutrino oscillations. Neutrino spin oscillations in
electromagnetic fields with other different configurations, including
a longitudinal magnetic field and the field of an electromagnetic
wave, were first examined in \cite{AkhKhlMPL88SJNP88} and
\cite{EgoLobStuPLB00_LobStuPLB01,DvoStuPAN01_DvoStuPAN04}.

It is possible to formulate a criterion \cite{LikStuJETP95} for
finding out if the neutrino spin (spin-flavor)
precession is significant for given neutrino and background medium properties.
The probability of oscillatory transitions between two neutrino states
$\nu_{\alpha L}\leftrightarrows\nu_{\beta R}$ can be expressed in terms of the
elements of the effective
Hamiltonian matrices (\ref{203}) and (\ref{303}) as
\begin{equation}
P_{\nu_{\alpha _L} \leftrightarrows \nu_{\beta _R}}=\sin ^2 \vartheta
_{eff} \sin ^2 \frac{x\pi}{L_{eff}},
\end{equation}
where
\begin{equation}
\sin ^2 \vartheta _{eff}=
\frac{4\mathcal{H}^2_{\alpha\beta}}{4\mathcal{H}^2_{\alpha\beta}+(\mathcal{H}_{\beta\beta}-\mathcal{H}_{\alpha\alpha})^2},
\ \text {and} \ \ L_{eff}=\frac {2\pi}{\sqrt
{4\mathcal{H}_{\alpha\beta}+(\mathcal{H}_{\beta\beta}-\mathcal{H}_{\alpha\alpha})^2}}.
\end{equation}
The transition probability can be of order unity
if the following two conditions hold simultaneously: 1) the amplitude
of the transition probability must be ``far'' from zero (at least
$\sin ^2 \vartheta _{eff}>1/2$), 2) the neutrino path length in a
medium with a magnetic field should be longer than half the effective
length of oscillations $L_{eff}$. In accordance with this criterion,
it is possible to introduce the critical strength of a magnetic field
$B_{cr}$ which determines the region of field values $B_{\perp}>
B_{cr}$ at which the probability amplitude is not small ($\sin ^2
\vartheta_{eff} > 1/2$):
\begin{equation}\label{B_cr}
B_{cr}=\frac {1}{2 {\tilde \mu}}\sqrt
{(\mathcal{H}_{\beta\beta}-\mathcal{H}_{\alpha\alpha})^2},
\end{equation}
where $\tilde \mu $ is $\mu _{ee}$, $\mu _{\mu \mu}$, $\mu _{e\mu}$, or
$\mu_{\mu e}$ depending on the type of neutrino transition process in question.

Consider, for instance,  the case of
$\nu_{e_L}\leftrightarrows\nu_{\mu _R}$
transitions between Majorana neutrinos. From Eqs.~(\ref{B_cr}) and (\ref{303}), it
follows \cite{LikStuJETP95}  that
\begin{equation}B_{cr}= \left|\frac{1}{2\tilde\mu}
\Big(\frac{\Delta m_{\nu}^2}{2E}A- \sqrt{2}G_F N_{eff}\Big) \right|,
\label{B_cr2}
\end{equation}
where $A=\cos 2\vartheta$ and $N_{eff}=N_e
-N_n$. For getting numerical estimates of $B_{cr}$ it is convenient
to re-write Eq.~(\ref{B_cr2}) in the following form:
\begin{equation}
B_{cr}\approx43 \frac{\mu_{B}}{\tilde \mu}\Bigg|A \Big(\frac{\Delta m^2
_{\nu}}{1 \ \text{eV}^2}\Big)\Big(\frac{1 \ \text{MeV}}{E_{\nu}}\Big)-
2.5\times 10^{-31} \frac{N_{eff}}{1 \ cm^{-3}} \Bigg| \ Gauss.
\end{equation}

An interesting feature of the evolution equation~(\ref{202}) in the case of
Majorana neutrinos is that the interplay of spin precession and flavor
oscillations can generate $\nu_{e_L}\to\nu_{e_R}$ transitions
\cite{Akhmedov:1991uk}. Since $\nu_{e_R}$ interacts as right-handed Dirac
antineutrinos, it is often denoted by $\bar\nu_{e_R}$, or only $\bar\nu_{e}$,
and called ``electron antineutrino''. This state can be detected in through the
inverse $\beta$-decay reaction
\begin{equation}
\bar\nu_{e} + p \to n + e^{+} \,, \label{g096}
\end{equation}
having a threshold $ E_{\text{th}} = 1.8 \, \text{MeV} $.

The possibility of $\nu_{e_L}\to\bar\nu_{e_R}$ transitions generated
by spin-flavor precession is particularly interesting for solar
neutrinos, which experience matter effects in the interior of the Sun
in the presence of the solar magnetic field. In 2002, the
Super-Kamiokande Collaboration established for the flux of solar
$\bar\nu_{e}$'s an upper limit of 0.8\%, at 90\% C.L., of the
Standard Solar Model neutrino flux in the range of energy from 8 to
20 MeV \cite{hep-ex/0212067}. This limit was improved in 2003 by the
KamLAND Collaboration to $ 2.8 \times 10^{-4} $ in the energy range
8.3 -- 14.8 MeV \cite{hep-ex/0310047}. The implications of this limit
for the spin-flavor precession of solar neutrinos have been studied
in several papers
\cite{hep-ph/0311014,hep-ph/0406066,hep-ph/0504185,hep-ph/0505165,0810.1037},
taking into account the dominant $\nu_{e}\to\nu_{\mu},\nu_{\tau}$
transitions due to neutrino oscillations (see the brief review in
Section~\ref{053}). Considering turbulent solar magnetic field models
in which $\nu_{eL}\to\bar\nu_{eR}$ transitions are strongly enhanced,
the authors of Refs.~\cite{hep-ph/0311014,hep-ph/0406066} obtained
the interesting limit
\begin{equation}
\mu_{ea} < \text{few} \times 10^{-12} \, \mu_{\text{B}} \,, \label{331}
\end{equation}
where $\mu_{ea}$ is the transition magnetic moment between $\nu_{e}$
and
$\nu_{a}=\cos\vartheta_{23}\nu_{\mu}-\sin\vartheta_{23}\nu_{\tau}$.
This limit has been, however, criticized in
Ref.~\cite{hep-ph/0505165}.

The spin-flavor mechanism was considered \cite{PulChaRag12LomCon} in order to
describe time variations of solar-neutrino fluxes in gallium experiments. The
effect of a nonzero neutrino magnetic moment is also of interest in connection
with the analysis of helioseismological observations \cite{CouTurKosAPJ03}.

The idea that the neutrino magnetic moment may solve the supernova problem,
i.e. that the neutrino spin-flip transitions in a magnetic field provide an
efficient mechanism of energy transfer from a protoneutron star, was first
discussed in \cite{DarIAS_rep_1987} and then investigated
 \cite{NusRep87GolAhaAleNusPRD88LatCooPRL88} in some detail. The possibility of a
loss of up to half of the active left-handed neutrinos because of their transition
to sterile right-handed neutrinos in strong magnetic fields at the boundary of
the neutron star (the so-called boundary effect) was considered in
\cite{LikStuJETP95}.

In conclusion, we would like to point out \cite{StuNPB09} that there
is a huge gap of many orders of magnitude between the present limits
$\propto 10^{-(11\div 14)}\mu_{B}$ on a neutrino magnetic moment
$\mu_{\nu}$ and the prediction of a minimal extension of the Standard
Model. Therefore, if any direct experimental confirmation of non-zero
neutrino magnetic moment were obtained within a reasonable time in
the future, it would open a window to new physics.

\raggedright

\section{Acknowledgments}
The authors are thankful to Jose Bernab\'{e}u, Anatoly Borisov, Alexander
Grigoriev, Max\-im Dvornikov, Jo\~{a}o Pulido, Timur Rashba, Alexander
Starostin and Alexei Ternov  for useful discussions on various problems
connected to the subject of the paper. One of the authors (A.S.) is thankful to
the INFN Section of Turin for support during his stay in Turin where the main part
of the paper was written.

\end{document}